\documentclass{elsarticle}

\makeatletter
\def\ps@pprintTitle{%
  \let\@oddhead\@empty
  \let\@evenhead\@empty
  \let\@oddfoot\@empty
  \let\@evenfoot\@oddfoot}
\makeatother

\usepackage{graphicx} 
\usepackage{hyperref} 

\usepackage{amsmath} 
\numberwithin{equation}{section}

\usepackage{amssymb}

\usepackage{tabularx} 
\usepackage{booktabs} 
\usepackage{caption} 
\usepackage{subcaption} 

\usepackage{tikz}
\usepackage{adjustbox}

\usepackage{xcolor} 
\usepackage{multicol}

\begin{document}

\newcommand{\refEQ}[1]{Eq.(#1)}
\newcommand{\refFIG}[1]{Fig.(#1)}

\setlength{\abovedisplayskip}{0pt}
\setlength{\belowdisplayskip}{0pt}
\setlength{\abovedisplayshortskip}{0pt}
\setlength{\belowdisplayshortskip}{0pt}

%
%
\newcommand{\varx}{\boldsymbol{x}}

%
%
\newcommand{\varElec}{\mathbf{E}}
\newcommand{\varBmag}{\mathbf{B}}
\newcommand{\varDelec}{\mathbf{D}}
\newcommand{\varHmag}{\mathbf{H}}
\newcommand{\varJcur}{\mathbf{J}}
\newcommand{\varMvec}{\mathbf{A}}

\newcommand{\varEcharge}{\rho}
\newcommand{\varPermitZero}{\epsilon_0}
\newcommand{\varPermeaZero}{\mu_0}
\newcommand{\varEcond}{\sigma}
\newcommand{\varPermit}{\epsilon}
\newcommand{\varPermea}{\mu}
\newcommand{\varFcoil}{\Bar{\omega}}

%
%

\newcommand{\varCelsius}{$^\circ\text{C}$}
\newcommand{\varHflux}{q}
\newcommand{\varHsource}{Q}

\newcommand{\varTemp}{T}
\newcommand{\varTempEle}{\varTemp_e}
\newcommand{\varTempI}{\varTemp_i}
\newcommand{\varTempDisc}{\boldsymbol{\varTemp}}

\newcommand{\varDens}{\rho}
\newcommand{\varHeatC}{C_p}
\newcommand{\varTcond}{\boldsymbol{C}}

%
%

\newcommand{\varEspace}{\varOmega}
\newcommand{\varSigmaAlgebra}{\mathcal{F}}
\newcommand{\varProbMeas}{\mathbb{P}}
\newcommand{\varBorel}{\mathcal{B}}
\newcommand{\varProbSpace}{(\varEspace, \varSigmaAlgebra, \varProbMeas )}

\newcommand{\vardiag}{\mathfrak{d i a g}}
\newcommand{\varDiag}{\text{Diag+}}
\newcommand{\varso}{\mathfrak{s o}}
\newcommand{\varSO}{\mathrm{SO}}
\newcommand{\varReal}{\mathbb{R}}
\newcommand{\varDomain}{\mathcal{G}}
\newcommand{\varSymPlus}{\text{Sym}^\text{+}}
\newcommand{\varSym}{\text{Sym}}
\newcommand{\varSobo}{H^1}

%
%
\newcommand{\varEvent}{\omega}
\newcommand{\varRV}{\kappa(\varEvent)}
\newcommand{\varRfield}{\kappa(x,\varEvent)}
\newcommand{\varRVmean}{\Bar{\kappa}}
\newcommand{\varRVstd}{\Tilde{\kappa}}
\newcommand{\varVMFref}{\boldsymbol{\mu}}
\newcommand{\varVMFcon}{\eta}

\newcommand{\varTensor}{\boldsymbol{C}}
\newcommand{\varTensorG}{\mathbb{C}}
\newcommand{\varTensorM}{\overline{\boldsymbol{C}}}

\newcommand{\varEigVal}{\boldsymbol{\Lambda}}
\newcommand{\varEigVec}{\boldsymbol{Q}}
\newcommand{\varTensorDet}{\overline{\varTensor}}
\newcommand{\varEigValDet}{\overline{\varEigVal}}
\newcommand{\varEigVecDet}{\overline{\varEigVec}}
%
%
\newcommand{\varWeights}{\mathbf{\theta}}
\newcommand{\varLope}{A}
\newcommand{\varLidx}{k}
\newcommand{\varL}{l}
\newcommand{\varLcur}{\varL_{\varLidx}}
\newcommand{\varLpre}{\varL_{\varLidx-1}}
\newcommand{\varLn}{n}

\setlength{\abovedisplayskip}{5pt}
\setlength{\belowdisplayskip}{5pt}
\setlength{\abovedisplayshortskip}{5pt}
\setlength{\belowdisplayshortskip}{5pt}

\begin{frontmatter}

\title{Constitutive Manifold Neural Networks}

\author{Wouter J. Schuttert
, Mohammed Iqbal Abdul Rasheed, Bojana Rosi\'{c}}     

\address{Applied Mechanics and Data Analysis chair\\ Faculty of Engineering Technology\\ 
Drienerlolaan 5, 7522 NB Enschede\\University of Twente}

\begin{abstract}
Anisotropic material properties, such as the thermal conductivities of engineering composites, exhibit variability due to inherent material heterogeneity and manufacturing-related uncertainties. Mathematically, these properties are modeled as symmetric positive definite (SPD) tensors, which reside on a curved Riemannian manifold. Extending this description to a stochastic framework requires preserving both the SPD structure and the underlying spatial symmetries of the tensors.
From a numerical standpoint, this is achieved through the spectral decomposition of tensors, which enables the parameterization of uncertainties into scale (strength) and rotation (orientation) components. These components are modeled as independent random variables on a manifold structure derived from the maximum entropy principle.
To quantify the impact of strength and orientation uncertainties on the thermal behaviour of the composite, the stochastic material tensor must be propagated through a physics-based forward model, described by an elliptic partial differential equation (PDE). This process necessitates computationally efficient surrogate models, for which a feedforward neural network (FNN) is employed.
However, conventional FNN architectures are not well-suited for SPD tensors, as directly using tensor components as input features fails to preserve their underlying geometric structure, often leading to suboptimal performance. To address this issue, we introduce the Constitutive Manifold Neural Network (CMNN), which incorporates input layers that map SPD tensors from the curved manifold to the local tangent space—a flat vector space—thus preserving the statistical and geometric information in the dataset.
A case study involving steady-state heat conduction with stochastic anisotropic conductivity demonstrates that geometry-preserving neural network significantly enhances learning performance compared to conventional multilayer perceptrons (MLPs). These findings underscore the importance of manifold-aware methods when working with tensor-valued data in engineering applications.
\end{abstract}

\end{frontmatter}


\section{Introduction}

\noindent
In advanced materials science, many materials exhibit anisotropic behaviour, where physical properties vary with direction, such as in composites, biological tissues, or soils \cite{corwin_field-scale_2020,ahmadi_moghaddam_stochastic_2019, saturnino_principled_2019}. This directional behaviour is inherently captured by mathematical objects called tensors, which describe how key material quantities change with spatial direction \cite{kolecki_introduction_2002}. Therefore, advances in physics-based modeling increasingly depend on methods that incorporate both the geometric structure and the statistical variability of anisotropic material tensors—accounting for their symmetry, positive definiteness, and stochastic nature. Failure to preserve the nature of material tensors, such as those representing conductivity, can result in unphysical behavior, numerical instability, or invalid simulation outputs—ultimately compromising the reliability of the design or analysis \cite{siengchin_review_2023, leijsen_electrical_2021, chen_digital_2022}.

An important class of material tensors is the second-order symmetric and positive definite (SPD) tensor, or SPD matrix, which encodes quantities such as thermal conductivity, electrical conductivity, and magnetic permeability \cite{itskov_tensor_2015}. These properties often vary between specimens due to manufacturing defects or natural phenomena. As a result, accurately predicting material behaviour requires modelling tensor-valued quantities as uncertain \cite{corwin_field-scale_2020, schuttert_modeling_2024}. For isotropic second-order properties, this uncertainty can be captured using a scalar-valued random variable. In contrast, for anisotropic materials, the full tensor must be modelled as a random variable. In this work, we use a parametric stochastic model for second-order SPD tensors recently introduced by Shivanand et al. \cite{shivanand_stochastic_2024} where the positive definiteness is preserved via the exponential map. The method also separates variation in material properties and orientation through spectral decomposition, allowing for a physics-based interpretation of the tensor-valued random variable.


Incorporating uncertain material properties into physics-based models --- particularly those involving coupled physical processes or multiple spatial scales --- often incurs significant computational cost due to the complexity of solving the associated PDEs. These PDEs typically involve a large number of parametric degrees of freedom, making the computation of quantities of interest (QoIs) via stochastic integration methods --- such as Monte Carlo, quasi-Monte Carlo, or deterministic quadrature --- impractical \cite{zhang_modern_2021, lataniotis_data-driven_2019, schurer_comparison_2003}.

One way to reduce this burden is to replace the expensive simulator with a surrogate model that approximates the parameter-QoI map \cite{mahdavi_kriging_2023, sudret_surrogate_2017, yin_solving_2023}. Among the available approaches, neural networks and related deep learning architectures \cite{abdar_review_2021} are attractive because they can represent functions in a scalable manner \cite{tripathy_deep_2018}. These networks are usually trained on labelled input–output pairs that cover a prescribed input space, and are usually not geometrically aware with respect to the input space.
In materials science, where SPD tensors are ubiquitous, constitutive neural networks (CNNs) have been developed to capture stress–strain relations \cite{linka_constitutive_2021, xu_learning_2021, linka_new_2023}. These models often exploit mechanics-based invariants by tailoring the inputs, outputs or internal architecture. More recently, symmetry-enforcing networks \cite{garanger_symmetry-enforcing_2023} employ eigenvalue activations and rotation-invariant matrices to model symmetric strain behaviour. However, these networks do not encode the full manifold structure into the architecture. 


Neural networks that explicitly incorporate Riemannian geometry \cite{pennec_riemannian_2004, pennec_riemannian_2006, buchfink_symplectic_2023} have demonstrated superior approximation accuracy. For example, SPDNet \cite{huang_riemannian_2016} preserves the SPD structure by treating weight matrices as points on a manifold and applying Euclidean activation functions via the logarithmic map. Subsequent research has extended classical Euclidean concepts to the manifold setting, including input normalization through the Fréchet mean \cite{brooks_riemannian_2019}, residual connections to mitigate vanishing gradients \cite{katsman_riemannian_2023}, and convolutional operators based on the weighted Fréchet mean \cite{chakraborty_cnn_2018}. However, despite these advances, the majority of such methods have been developed primarily for large covariance matrices in computer vision and pattern recognition contexts, often overlooking critical physical constraints essential in materials science applications—such as the interpretation of eigenvectors as encoding material orientation \cite{schuttert_anisotropy_2024}.

This study investigates surrogate modelling for physics-based simulations that involve uncertain SPD material tensors. The goal is to build a Consitutive Manifold Neural Network (CMNN) architecture that preserves the geometric structure of SPD material data through built-in manifold-aware input layers. A case study on a cube-shaped domain with stochastic thermal conductivity and prescribed boundary conditions compares the newly introduced CMNN network with a conventional multilayer perceptron (MLP), and SPDNet. Each model employs a distinct mapping from the SPD manifold to Euclidean space. By evaluating their performance under different forms of uncertainty, we show that geometry-aware networks improve approximation accuracy and training stability.

The paper is organized as follows. Section 2 introduces the essential physics and the modeling framework for stochastic parametric tensors. Section 3 presents the constitutive manifold network architecture. In Section 4, a case study is conducted involving three distinct types of uncertainty: scaling, orientation, and their combination. The neural networks described in Section 3 are then evaluated for their approximation performance. Finally, Section 5 summarizes the conclusions, discusses the limitations of the current work, and outlines potential directions for future research.

\section{Stochastic heat transfer}
\label{sec:2}


\noindent
Let be given a physical system undergoing heat conduction described by the following elliptic partial differential equation (PDE):
\begin{alignat}{2}
\label{eq:heat_pde}
-\nabla \cdot (\varTcond \nabla \varTemp(\varx)) &= \varHsource(\varx) &\quad &\text{a.e. in } \varDomain, \\
\intertext{with boundary conditions}
\varTemp(\varx) &= \varTemp_D(\varx) &\quad &\text{on } \Gamma_D \subseteq \partial \varDomain, \\
-\varTcond \nabla \varTemp(\varx) \cdot \vec{n} &= \varHsource_N(\varx) &\quad &\text{on } \Gamma_N \subseteq \partial \varDomain,
\end{alignat}
in which the bounded domain $\varDomain \subset \varReal^d$ resides in $d$-dimensional Euclidean space, with $d = 2$ or $d = 3$. Within this domain, the temperature at each point $\varx \in \varDomain$ is represented by $\varTemp(\varx)$, while the source term $\varHsource(\varx)$ accounts for thermal sources or sinks. On the Dirichlet boundary $\Gamma_D$, the temperature is prescribed by $\varTemp_D(\varx)$, while on the Neumann boundary $\Gamma_N$, the heat flux through the surface is specified as $\varHsource_N(\varx)$. The physical system is parametrised by a thermal conductivity tensor belonging to a space of SPD matrices, i.e.
\begin{equation}\label{eq:symplus}
    \varTensor \in \varSymPlus(d) := \left\{ \varTensor \in (\varReal^d \otimes \varReal^d) \mid \varTensor = \varTensor^T, \ \boldsymbol{z}^T \varTensor \boldsymbol{z} > 0, \ \forall \boldsymbol{z} \in \varReal^d \backslash \{ \boldsymbol{0} \} \right\},  
\end{equation}
here assumed to be spatially homogeneous and anisotropic. Positive definiteness ensures that energy conservation principles are maintained, while symmetry captures the balanced interactions within a material governed by Onsager's reciprocal relations \cite{onsager_reciprocal_1931}. Due to these properties, the conductivity tensor does not belong to a Euclidean space but an open convex cone within the space of symmetric matrices $\varSym(d)$. This open convex cone forms a Riemannian manifold, a curved space that locally resembles Euclidean space $\varReal^d$ in its tangent space. The local resemblance is formalised through continuous (homeomorphic) and differentiable (diffeomorphic) mappings between the manifold and its tangent space, enabling the use of calculus on the manifold \cite{lee_smooth_2003}.

In real-world applications, the conductivity tensor $\varTensor$ is typically unknown and must be estimated from experimental observations of the temperature field. Prior to obtaining any measurements, $\varTensor$ can be described a priori based on expert knowledge or prior assumptions. To reflect this lack of knowledge, $\varTensor$ is modeled as a symmetric positive definite (SPD) tensor-valued random variable representing epistemic uncertainty. Let $\varProbSpace$ be the probability space in which $\varEspace$ is the set of all elementary events $\varEvent$, $\varSigmaAlgebra$ is the sigma algebra, and $\varProbMeas$ is the probability measure. Then, $\varTensor(\varEvent)$ is modelled as a finite variance SPD-valued random variable in $L_2(\varProbSpace, \varSymPlus(d))$. As a consequence, the deterministic PDE in \refEQ{\ref{eq:heat_pde}} becomes stochastic
\begin{alignat}{2}
\label{eq:stochastic_heat_pde}
- \nabla \cdot \left( \varTcond(\omega) \nabla \varTemp(\varx, \omega) \right) &= \varHsource(\varx) &\quad &\text{a.e. in } \varDomain, \text{ and a.s. } \omega \in \varEspace, \\
\intertext{together with boundary conditions}
\varTemp(\varx, \varEvent) &= \varTemp_D(\varx) &\quad &\text{on } \Gamma_D \subseteq \partial \varDomain, \forall\varEvent \in \varEspace, \\
- \varTcond(\varEvent) \nabla \varTemp(\varx, \varEvent) \cdot \vec{n} &= \varHsource_N(\varx) &\quad &\text{on } \Gamma_N \subseteq \partial \varDomain,  \forall\varEvent \in \varEspace.
\end{alignat}
As a result, the temperature field $T(\boldsymbol{x}, \varEvent)$ becomes a finite variance random field in 
$L_2(\varProbSpace, \varSobo(\varDomain))$, where $\varSobo$ is the first-order Sobolev space. 

The primary objective of this study is to quantify the uncertainty in the temperature field $T(\varx, \varEvent)$ given a known stochastic description of $\varTensor(\varEvent)$. Since defining a probabilistic description of $\varTensor(\varEvent)$ is non-trivial due to its non-Euclidean (manifold) structure, we first introduce a suitable parametrization of the tensor-valued random variables, and subsequently outline the procedure for propagating this uncertainty to the temperature field $\varTemp(\varx, \varEvent)$.

\subsection{Stochastic anisotropic thermal conductivity tensor}
\noindent
To stochastically model the SPD tensor 
$\varTensor$, a probability distribution must be selected that satisfies several physical constraints, as outlined in \cite{shivanand_stochastic_2024}. First, the stochastic tensor must be SPD for all $\varEvent \in \varEspace$. Second, it must exhibit invariance under a subgroup $G \subset O(\mathbb{R}^d)$ of orthogonal transformations, satisfying $\boldsymbol{R}^T \varTensor(\varEvent) \boldsymbol{R} = \varTensor(\varEvent)$ for all $\boldsymbol{R} \in G$ and $\varEvent \in \varEspace$. Lastly, the mean tensor $\varTensorM := \mathbb{E}_{\vartheta}(\varTensor(\cdot))$ must remain invariant under a broader group $G_m \subseteq O(\mathbb{R}^d)$, including transformations such as inversion, scaling, and frame changes.

A model that satisfies these requirements is based on an eigenvalue decomposition approach, as introduced in \cite{shivanand_stochastic_2024}. The conductivity tensor 
$\varTensor$ is decomposed via its spectral representation into eigenvalues and eigenvectors, such that 
\begin{equation} \label{eq:spectral_decomp}
    \varTensor =  \varEigVec \, \varEigVal \,  \varEigVec^T,
\end{equation}
holds. Here, $\varEigVec \in \varSO(d)$ is an orthogonal matrix whose columns are the corresponding eigenvectors, and $\varEigVal \in \varDiag(d)$ represents eigenvalues in the space of positive-definite diagonal matrices, which is a Lie group under matrix multiplication. The eigenvectors $\varEigVec$ represent the principal directions of material symmetry, whereas the eigenvalues $\varEigVal$ reflect the material properties, such as conductivity along these principal directions. However, as these quantities lie on nonlinear manifolds, classical linear algebraic operations are not directly applicable \cite{gallier_differential_2020}. To address this, we represent the eigenvalues $\varEigVal$ and $\varEigVec$ in their respective Lie algebras $(\boldsymbol{Y}, \boldsymbol{W})$, in which $\boldsymbol{Y} \in \vardiag(d)$ is the space of diagonal matrices, and $\boldsymbol{W} \in \varso(d)$ is the space of skew-symmetric matrices. To map these Lie algebra elements back to the manifold, we employ the matrix exponential:
\begin{equation}\label{eq:tensor_mapping}
    (\boldsymbol{Y}, \boldsymbol{W}) \longmapsto(\boldsymbol{\Lambda}, \boldsymbol{Q})=(\exp \boldsymbol{Y}, \exp \boldsymbol{W}) \longmapsto \boldsymbol{C}=\boldsymbol{Q} \boldsymbol{\Lambda} \boldsymbol{Q}^T,
\end{equation} 
i.e.
\begin{equation}\label{eq:tensor_mapping_space}
\mathfrak{diag}(d) \times \mathfrak{s o}(d) \xrightarrow{\exp \times \exp } \operatorname{Diag}^{+}(d) \times \mathrm{SO}(d) \quad \longrightarrow \operatorname{Sym}^{+}(d). 
\end{equation} 
In this manner, stochastic variability can be incorporated in the material tensor by accounting for the uncertainty in both $\varEigVal(\varEvent)$ and $\varEigVec(\varEvent)$. Thus, regardless of the choice of distributions for these elements, the first and second constraints on the material tensor are inherently satisfied. To satisfy the third requirement, we adopt the scaling-rotation metric for the Fréchet mean as described in \cite{shivanand_stochastic_2024}, the generalised definition of the mean commonly used on the SPD manifold. For further details, the reader is referred to \cite{feragen_geometries_2017, shivanand_stochastic_2024}. 

\noindent Following this, the stochastic SPD tensor can be expressed as:
\begin{equation}\label{eq:random_both}
    \varTensor(\varEvent) = \exp(\boldsymbol{W}(\varEvent)) \; \overline{\varEigVec} \; \exp(\boldsymbol{Y}(\varEvent)) \; \overline{\varEigVec}^T \; \exp(\boldsymbol{W}(\varEvent)^T), \quad \forall \varEvent \in \varEspace,
\end{equation}
where $\exp(\boldsymbol{W}(\varEvent))$, taken around the reference eigenvector matrix $\overline{\varEigVec}$, models the random eigenvector matrix $\varEigVec(\varEvent)$. The random rotation matrix $\boldsymbol{W}(\varEvent)$ is modelled as
\begin{equation}\label{eq:stochastic_eigvec_1}
   \boldsymbol{W}(\varEvent) =  \textrm{skw}({\boldsymbol{w}})=\begin{bmatrix}
0 & -w_3(\varEvent)  & w_2(\varEvent) \\ 
w_3(\varEvent) & 0 & -w_1(\varEvent)\\ 
-w_2(\varEvent) &  w_1(\varEvent) & 0
\end{bmatrix}, \quad \boldsymbol{W} (\varEvent) = -\boldsymbol{W}^T(\varEvent)
\end{equation}
in which $\boldsymbol{w}(\varEvent)$ is the axis-angle representation or a minimal parameterization of a rotation near the identity. The vector \(\boldsymbol{w}(\varEvent)\) can also be decomposed according to
\begin{equation}\label{eq:stochastic_eigvec_2}
   \boldsymbol{w} = \phi(\varEvent) \boldsymbol{r}, \quad \boldsymbol{r} \cdot \boldsymbol{r}=\boldsymbol{1},
\end{equation}
into two components: a scalar random rotation angle \(\phi(\varEvent)\) and a unit vector \(\boldsymbol{r}\) that specifies the axis of rotation. Specifically, \(\boldsymbol{r}\) is a unit vector \(\left(\|\boldsymbol{r}\| = 1\right)\) representing the direction around which the rotation occurs, while \(\phi\) indicates the magnitude of the rotation about that axis.
This yields
\begin{equation}\label{eq:stochastic_eigvec}
   \boldsymbol{W}(\varEvent) = \phi(\varEvent) \boldsymbol{R}, \quad \boldsymbol{R} := \mathrm{skw}(\boldsymbol{r}),
\end{equation}
which is then suitable for the use of Rodrigues’ formula. To model the angle \(\phi(\varEvent)\) as random, one 
may choose a von Mises–Fisher distribution (vMF) \cite{mardia_directional_2000}, i.e.
\begin{equation}\label{eq:VMF}
    f(\phi; \varVMFref, \varVMFcon) = C_p(\varVMFcon)\exp(\varVMFcon \varVMFref^T\boldsymbol{\phi})
\end{equation}
where $\varVMFref$ is the mean vector, $\varVMFcon$ the concentration parameter (akin to the standard deviation of a normal distribution), and $C_p$ is a normalization constant which ensures the distribution lies on the unit sphere. The vMF is a generalization of a Gaussian distribution to the unit sphere and adheres to the principle of maximum entropy \cite{shannon_mathematical_1948, jaynes_information_1957}.
Similarly, the stochastic eigenvalue matrix $\boldsymbol{Y}(\varEvent)$, also chosen according to the principle of maximum entropy, is modelled as a Gaussian tensor-valued random variable:
\begin{equation}\label{eq:stochastic_eigval}
    \boldsymbol{Y}(\varEvent) =  \text{ diag} (\xi_i(\varEvent)) \quad \text{with} \quad \xi_i(\varEvent) \sim \mathcal{N}(\mu_i, \sigma_i^{2}), \quad i = 1, ...,d,
\end{equation}
where the subscript $i$ refers to the $i^{\text{th}}$ diagonal term of the eigenvalue matrix. Note that the exponential map of a Gaussian random variable corresponds to directly modelling a log-normal random variable at every diagonal element in the matrix $\varEigVal(\varEvent)$.

Together, the probabilistic models for scaling and orientation enable the specification of different symmetry classes by selecting the number of independent eigenvalues. For instance, isotropy corresponds to a single independent eigenvalue, while orthotropy involves three independent eigenvalues in $\mathbb{R}^3$. Additionally, by controlling the number of symmetry planes through the random rotation matrix, various types of material symmetries can be represented.

\subsection{Discretization and Approximation of the Temperature Field}
\noindent
By substituting the stochastic SPD model for the material tensor into the heat equation in \eqref{eq:stochastic_heat_pde}, one obtains a parametrized stochastic partial differential equation (SPDE). The resulting problem is defined in infinite-dimensional function spaces and must be discretized for numerical solution. For the spatial discretisation, we use the Finite Element Method (FEM) by choosing $\varDomain = \bigcup_{e} \mathcal{T}_e$, in which $\mathcal{T}_e$ denotes the corresponding elements. Within each element, the solution is approximated using a finite set of basis functions:
\begin{equation}\label{eq:FEM}
   \varTempDisc(\varx, \varEvent) = \sum_{i=1}^{M} \varTempI(\varTensor(\varEvent)) N_i(\varx),
\end{equation}
\noindent
here, the function $\varTempDisc(\varx, \varEvent)$ is expressed as a linear combination of basis functions $N_i(\varx)$, weighted by the stochastic coefficients $\varTempI(\varEvent)$. Since this formulation only discretises the spatial domain, the resulting system remains continuous in the probabilistic space and requires further stochastic discretisation. For this purpose, we employ the Monte Carlo (MC) sampling technique:
\begin{equation}\label{eq:MCFEM}
    \varTempDisc(\varx,\varEvent_j) = \sum_{i=1}^{M} \varTempI(\varTensor(\varEvent_j)) N_i(\varx), \; j  = 1 \;  \dots \; n,
\end{equation}
where, at each nodal point, we collect values of the temperature through $n$ samples, indicated by $\varEvent_j$. 

 MC estimation is a computationally costly technique when combined with FEM. To reduce the computational cost, we aim to utilise only part of the generated data to formulate a surrogate model. The objective is to approximate the quantity of interest—specifically, the temperature field and fluxes—through the parametric mapping
\begin{equation}\label{eq:func_approx}
    y(\varTensor(\varEvent)) = F(\varTensor(\varEvent); \varWeights),
\end{equation}
where
\begin{equation} \label{eq:function_map}
     F : \varSymPlus \to \mathbb{R}^m
\end{equation}
denotes a function from the space of symmetric positive definite tensors to an \( m \)-dimensional output space (e.g. the number of nodes in the finite element mesh), parameterized by \(\varWeights\).
Given a dataset of \( n \) Monte Carlo samples
\[
\mathcal{D} = \{(\varTensor^{(i)}, y^{(i)})\}_{i=1}^n,
\]
where \(\varTensor^{(i)}\) are realizations of the random tensor and \( y^{(i)} \) are the corresponding observed quantities of interest, the primary aim is to infer the parameters \(\varWeights\) such that \( F \) accurately predicts the temperature field.


\section{Constitutive Manifold Neural Networks}

\noindent  
In the previous equation, $F$ is assumed to be an abstract continuous and measurable mapping. For computational purposes, we choose a family of neural networks as the approximating functions, specifically a feedforward neural network architecture suited to the problem. Let $\boldsymbol{q}$ be a vectorised version of the tensor $\boldsymbol{C}$. Then, the QoI in \refEQ{\ref{eq:func_approx}} is approximated by
\begin{equation} \label{eq:NNfunction}
     \hat{y}(\boldsymbol{q})\approx H_{\varWeights}(\boldsymbol{q}) = (\varLope^L \circ f^{L}_a \ldots \circ f^{l_2}_a\circ \varLope^{l_2} \circ f^{l_1}_a\circ \varLope^{l_1}) (\boldsymbol{q}),
\end{equation} 
\noindent
in which
\begin{equation}\label{eq:layer_eq}
    \varLope^{\varLcur}(\boldsymbol{q}^{\varLcur}) = \mathbf{W}^{\varLcur} \boldsymbol{q}^{\varLcur} + \boldsymbol{b}^{\varLcur}
\end{equation}   
 represents the affine operation at depth $k$, with $k = l_1, l_2, \ldots, L$ being the layers, while $\boldsymbol{q}^{\varLcur} \in \mathbb{R}^{\varLn^{\varLcur}}$ is the input vector coming from layer $\varLpre$, $\varLn^{\varLcur}$ is the size of layer $k$ and $f^{\varLcur}_a $ is the nonlinear activation function of layer $\varLcur$. Together, the weights $\mathbf{W}^{l} \in \varReal^{\varLn^{\varLcur} \times \varLn^{\varLpre}}$ and biases $\boldsymbol{b}^{\varLcur} \in \varReal^{\varLcur}$ parameterise the linear operations in each layer. When combined with nonlinear activation functions, they enable the network to approximate the target function effectively.

While the previous approximation is commonly used in engineering practice, it is unfortunately inadequate in the context of stochastic symmetric positive-definite tensors. Specifically, it relies on mapping parameters--- which inherently lie on a nonlinear manifold--- to the quantity of interest (QoI) using classical algebraic operations called vectorisation or flattening. This approach fails to preserve the intrinsic geometric structure of the parameter space. While this embedding enables the use of standard neural network architectures, it imposes a Euclidean geometry on inherently non-Euclidean data \cite{arsigny_geometric_2007}, thereby potentially compromising the accuracy and validity of the approximation. Specifically, vectorisation induces the flat norm
\begin{equation}\label{eq:vector_metric}
    d_{\text{vec}}(\varTensor_1, \varTensor_2) = \|\varTensor_1 - \varTensor_2\|_F.
\end{equation}
which treats SPD tensors as elements of a flat Euclidean space, ignoring the geometry of the manifold. The nonlinear geometry of the manifold introduces additional nonlinearity into the data representation and undermines standard computations, such as those for the mean or variance, which are no longer meaningful under the previously given metric \cite{pennec_riemannian_2006}. 
Consequently, the distortions imposed by the flat metric in \refEQ{\ref{eq:vector_metric}} can severely limit the network’s ability to learn the desired functional relationships \cite{harandi_dimensionality_2016, ollivier_riemannian_2015,huang_riemannian_2016}.

To build the true map from the SPD tensor $\varTensor(\varEvent)$ to the QoI, we introduce an additional  input layer $B: \varSymPlus(d) \to \varReal^{\varLn^{l_0}}$ to the previously described neural network architecture in \refEQ{\ref{eq:NNfunction}}. This layer maps from the manifold of SPD tensors to an Euclidean space which serves as input to the hidden layers of the neural network, leading to the formulation of a Constitutive Manifold Neural Network (CMNN):
\begin{equation} \label{eq:CMNNfunction}
    \hat{y}(\varTensor) = F_{\varWeights}(\varTensor) = (\varLope^L \circ f^{L}_a \ldots \circ f^{l_2}_a\circ \varLope^{l_2} \circ f^{l_1}_a\circ \varLope^{l_1} \circ B) (\varTensor).
\end{equation}  
The network $F_{\varWeights}$ provides a parametric approximation of the target mapping defined in \refEQ{\ref{eq:function_map}}. In this work, we focus on introducing two alternative versions for $B$ instead of vectorization that respect the manifold structure, thereby potentially capturing the relationship between the stochastic material tensor and the random temperature field more effectively.

The first alternative is inspired by a widely used Riemannian metric for SPD tensors, the Log-Euclidean metric \cite{arsigny_geometric_2007}, defined as
\begin{equation}\label{eq:logeuclid}
d_{LE}(\varTensor_1, \varTensor_2) = ||\log(\varTensor_1) - \log(\varTensor_2)||_F,
\end{equation}
where the matrix logarithm $\log(\varTensor) = \varEigVec \log(\varEigVal) \varEigVec^\top$ maps tensors to the tangent space of symmetric matrices. This metric respects the manifold geometry, contrasting with the flat Euclidean metric induced by vectorization. To integrate the metric into a neural network, we utilise the log-eigenvalue operation (LogEig) as a geometry-aware input layer, first introduced in SPDNet \cite{huang_riemannian_2016}. The LogEig operation, i.e. the mapping
\begin{equation}\label{eq:LogEig} 
  B_{LogEig} : \varSymPlus(d) \xrightarrow{\log} \varSym(d)  \xrightarrow{flat} \mathbb{R}^{d(d+1)/2}. 
\end{equation}
maps the eigenvalues into the Lie algebra $\vardiag$ via a logarithmic map and reconstructs it in the linear space $\varSym(d)$. While computationally effective, this approach is known to introduce swelling, fattening, and shrinking effects when applied to SPD tensors \cite{schwartzman_random_2006, schwartzman_lognormal_2016}.

The second approach, inspired by the scaling-rotation metric, mitigates geometric distortions while remaining computationally efficient \cite{feragen_geometries_2017, shivanand_stochastic_2024}. The scaling-rotation metric is defined as
\begin{equation}\label{eq:scalerot} 
d_{sr}(\varTensor_1, \varTensor_2) = \sqrt{\,d_L(\varEigVal_1,\varEigVal_2)^2 + c\,d_R(\varEigVec_1,\varEigVec_2)^2} 
\end{equation}
\noindent
where
\begin{equation}\label{eq:scalerot2} d_L(\varEigVal_1,\varEigVal_2)=|\log(\varEigVal_1)-\log(\varEigVal_2)|_F, \quad
d_R(\varEigVec_1,\varEigVec_2)=|\log(\varEigVec_1^{\!\top}\varEigVec_2)|_F. 
\end{equation}
\noindent
This metric decomposes the distance into a scaling part, $d_L$, represented by the eigenvalues $\varEigVal_i$, and a rotation part, $d_R$, captured through the eigenvectors $\varEigVec_i$ of the SPD tensors. The parameter $c > 0$ balances these components by controlling the trade-off between scale and orientation alignment.

To leverage this metric in neural networks, we propose feeding the network separate strength and orientation information, referred to as the strength-angular (StrAng) layer. The StrAng layer decomposes the tensor into eigenvalues and eigenvectors, applies logarithmic scaling to the eigenvalues and eigenvectors, and flattens both components into a Euclidean vector through the following map:
\begin{alignat}{2}
\label{eq:StrAng}
B & := (\varrho^{\text{eig}} \circ \varrho^{\text{log}} \circ \varrho^{\text{flat}}) \\
\intertext{where}  
\varrho^{\text{eig}} & := \varSymPlus(d) \xrightarrow{eig} \varDiag(d) \times \varSO(d)\\
\varrho^{\text{log}} & := \varDiag(d) \times \varSO(d) \xrightarrow{(\log, \log)} \vardiag(d_1) \times \varso(d_2) \\
\varrho^{\text{flat}} & := \vardiag(d_1) \times \varso(d_2) \xrightarrow{flat} \mathbb{R}^{\,(d_1+d_2)}
\end{alignat}
\noindent
Here, $d_1$ denotes the number of independent eigenvalues, which are flattened into $d_1$ entries, while $d_2$ is the number of random angles flattened into $d_2$ entries, both are dependent on the type of symmetry and anisotropy class, i.e. $(d_1, d_2) = (3,3)$ for fully anisotropic materials or $(d_1,d_2)= (1,0)$ for isotropic materials. The map $\varrho^{\text{eig}}$ computes the eigen-decomposition and, as in the scaling–rotation metric, keeps the eigenvalues and eigenvectors separate. Using $\varrho^{\text{log}}$, we apply the matrix logarithm to map the eigenvalues from the open convex cone of positive-definite matrices to the flat tangent space. Similarly, the eigenvectors are mapped to their tangent space $\varso$ as angles. Finally, $\varrho^{\text{flat}}$ creates a representation in $\varReal$ from the eigenvalues and eigenvectors.

Regardless of choice for $B$, the weights and biases of the CMNN are trained by minimising 
 a standard mean squared error (MSE) loss over discretised output representations defined as 
\begin{equation}
\hat{\mathcal{L}}(\theta) = \frac{1}{n} \sum_{i=1}^n \left| \hat{y}(\varTensor_i; \theta) - y_i \right|^2,
\end{equation}
where $\{ (\mathbf{C}_i, y_i) \}_{i=1}^n$ are sampled input-output pairs from the data distribution.
For updating the weights and biases the backpropagation algorithm remains unchanged since the input transformations do not interfere with the internal architecture of the feedforward neural network. Note that one can equally utilize the physics-informed loss, or any other more advanced type of loss here.

Lastly, we require an unique spectral decomposition for the StrAng and LogEig layers, as the decomposition is inherently non-unique. Eigenvectors can change sign arbitrarily, and the ordering of the eigenvalues can vary, yet the decomposition remains valid. Despite this mathematical flexibility, such ambiguities can lead to inconsistencies when considering the physical interpretation of the material. To address this issue, we adopt a sorting algorithm that resolves both the sign and permutation ambiguities of eigenvectors and eigenvalues by aligning them with the reference tensor (e.g. the mean tensor) $\varTensorDet = \varEigVecDet \, \varEigValDet \, \varEigVecDet^T$. This alignment is achieved by solving the following minimisation problem
\begin{equation} \label{eq:decomp_min}
    (P^*, S^*) = \underset{P \in \mathcal{P}, \, S \in \mathcal{S}}{\operatorname{argmin}} \, d(P, S) \quad \text{where} \quad d = d_R(\boldsymbol{Q},\overline{\boldsymbol{Q}}),
\end{equation}
where $P \in \mathcal{P}$ represents one of the six possible permutations of the eigenvalue indices, $S \in \mathcal{S}$ represents one of the eight possible sign configurations for the eigenvector columns, and $\partial_R(P, S)$ is a cumulative distance measure between eigenvectors after applying the respective permutations and sign changes. The objective is to find the configuration $(P^*, S^*)$ that minimises $d(P, S)$ while ensuring eigenvectors align with $\overline{\varTensor}$. This alignment is robust for rotations up to a $45^\circ$ deviations from the initial eigenvectors, well beyond the typical variations observed in advanced materials.

\section{Case Study}

\begin{figure}[!b]               
\captionsetup{justification=centering,margin=0cm}
    \centering
    \begin{subfigure}[t]{0.35\textwidth}
        \includegraphics[width=\textwidth, trim=0cm 0cm 0cm 0cm,clip]{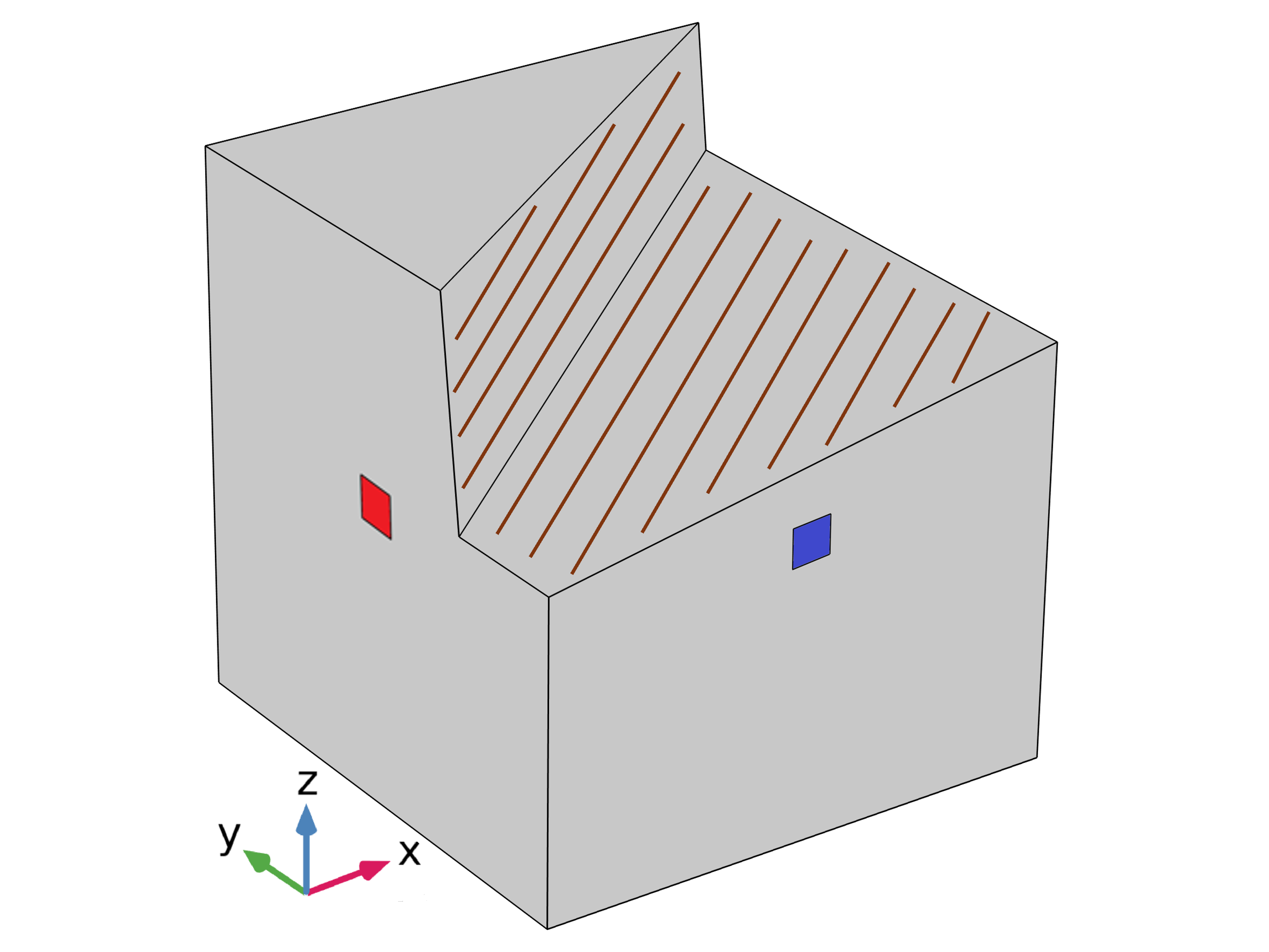}
        \caption{Cube geometry with patches and cut-out to show the fibres.}
        \label{fig:3Dcube}
    \end{subfigure}
    \hfill
     \begin{subfigure}[t]{0.31\textwidth}
        \includegraphics[width=\textwidth, trim=8cm 0cm 0cm 0cm,clip]{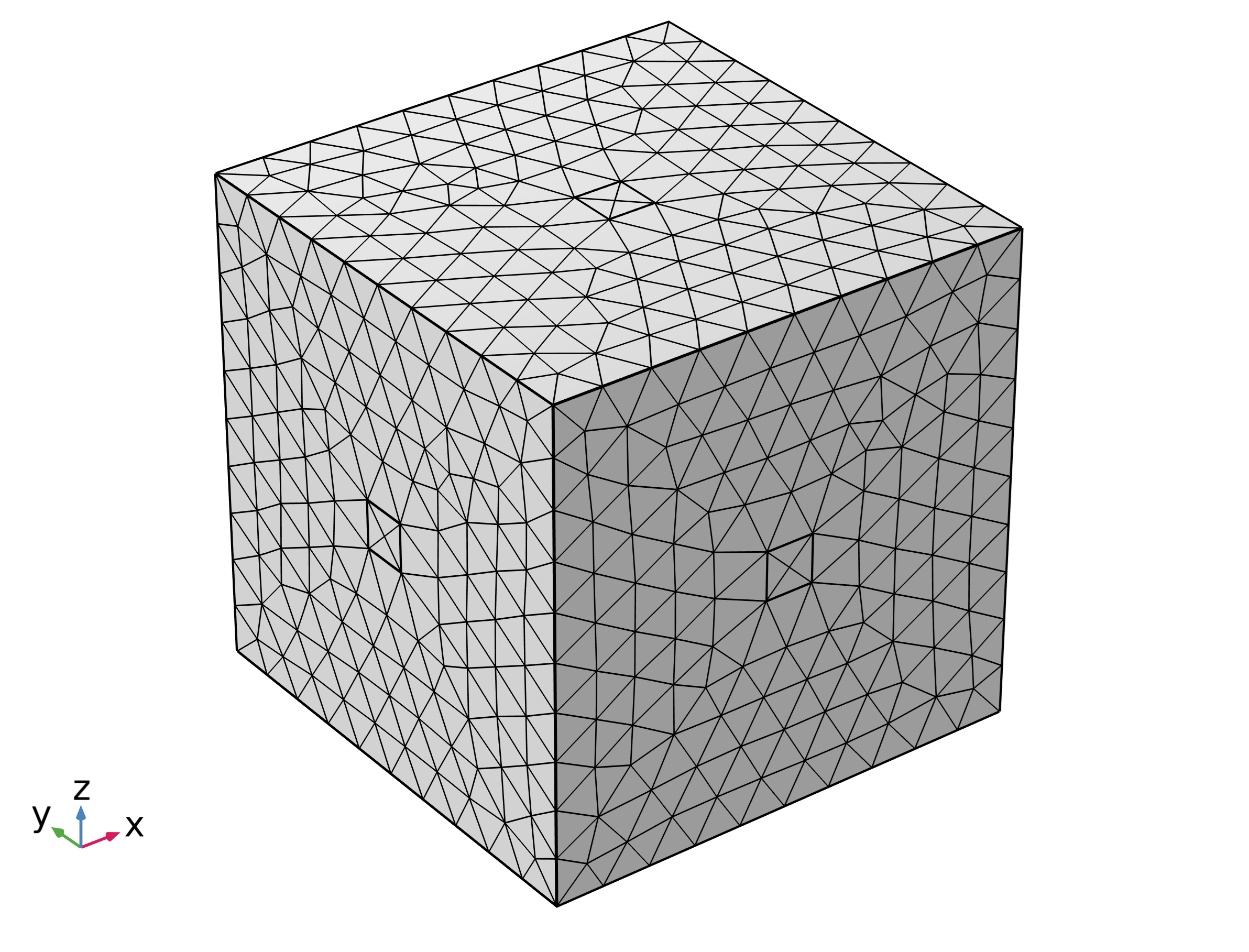}
        \caption{Finite element discretisation of the cube.}
        \label{fig:cube_discret}
    \end{subfigure}
    \hfill
    \begin{subfigure}[t]{0.32\textwidth}
        \includegraphics[width=\textwidth,  trim=7cm 0cm 0cm 2.5cm,clip]{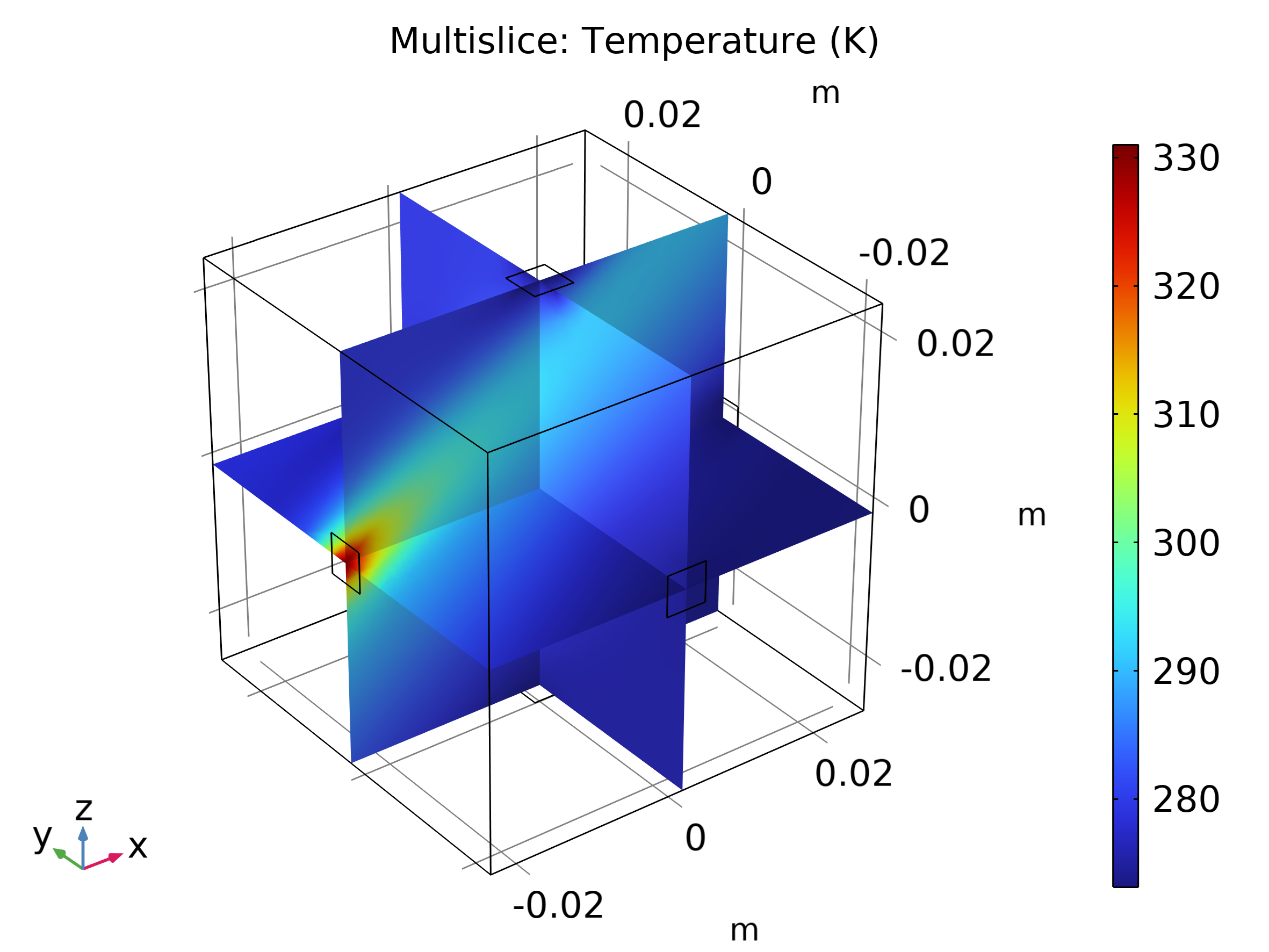}
        \caption{Multislice of the temperature field in the patched cube.}
        \label{fig:multislice_deterministic}
    \end{subfigure}
   
\vspace{-0.15cm}    
    \caption{Simulation domain and deterministic result.}
\vspace{-0.35cm}    
    \label{fig:subdomain}
\end{figure}
\noindent
To investigate the performance of the proposed neural networks in handling anisotropic SPD material data, we analyse steady-state heat conduction in a patched cube with anisotropic and uncertain thermal conductivity, governed by \refEQ{\ref{eq:stochastic_heat_pde}}. The patched cube is chosen for its direction-dependent geometry, consisting of six patches, each oriented along one of the Cartesian axes with normals aligned to the cube’s faces. The cube has a side length of $0.05$ m, with each patch measuring $0.005$ m. The geometry is shown in \refFIG{\ref{fig:3Dcube}}, where a red patch on one YZ face indicates a Neumann boundary condition, $Q_N = 50$ kW/m$^2$, while the other five patches (blue) are subject to a Dirichlet boundary condition, $T_D = 273.15$ K. The remaining boundary is insulated. The figure also features a cutout that shows the direction of the fibres within the cube. To parameterise the anisotropic thermal properties, we consider a deterministic thermal conductivity tensor representative of unidirectional stacked composites \cite{gouin_oshaughnessey_modeling_2016}. These composites are used in lightweight, load-bearing structures and exhibit distinct thermal properties in three principal directions: fibre, stacking, and transverse. The thermal conductivity tensor, its eigenvalues, and eigenvectors are given by:\begin{equation}\label{eq:initial_tensor}
    \bar{\varTcond} = \begin{bmatrix}
        11.24 & 5.18 & 1.73 \\
        5.18 & 3.49 & -0.356 \\
        1.73  & -0.356 & 1.78 \\
    \end{bmatrix} \text{where} \;
        \bar{\varEigVal}_\kappa  = \begin{bmatrix}
        14 \\
        0.11 \\
        2.4
    \end{bmatrix} \text{and} \;
    \bar{\varEigVec}_\kappa = \begin{bmatrix}
        0.892 & -0.416 & 0.174 \\
        0.436 & 0.700 & 0.565 \\
        0.113  & 0.580 & 0.807 \\
    \end{bmatrix}.
\end{equation} 
The first eigenvalue captures the high conductivity along the fibre direction, due to the fibre-dominant thermal properties of the material. The second eigenvalue reflects the lower conductivity in the out-of-plane stacking direction, influenced by polymer rich regions between tapes. The third eigenvalue represents the in-plane transverse conductivity, which is higher than in the stacking direction due to a more uniform fibre and polymer distribution. A dense rotation matrix, defined by Euler angles of $35^\circ$, $10^\circ$, and $25^\circ$, introduces pronounced interactions between material orientation and cube geometry, capturing complex anisotropic behaviour of the material.

To solve the deterministic case for $\varTemp(\varx)$, as described in \refEQ{\ref{eq:heat_pde}}, spatial discretisation is carried out using the approach outlined in \refEQ{\ref{eq:FEM}}. The simulation is performed in the commercial software COMSOL Multiphysics\textregistered{} \cite{noauthor_comsol_nodate}, where the geometry is discretised into approximately $16,000$ quadratic Lagrangian tetrahedral elements $e$, sharing about $3,000$ nodal points, as shown in \refFIG{\ref{fig:cube_discret}}. The Generalized Minimum Residual (GMRES) method is employed to solve the resulting system of equations.

\begin{figure}[t]               
    \captionsetup{justification=centering,margin=1cm}
    \centering
    \begin{subfigure}[t]{0.33\textwidth}
        \includegraphics[width=\textwidth, trim=2.3cm 0.5cm 2.7cm 0.8cm,clip]{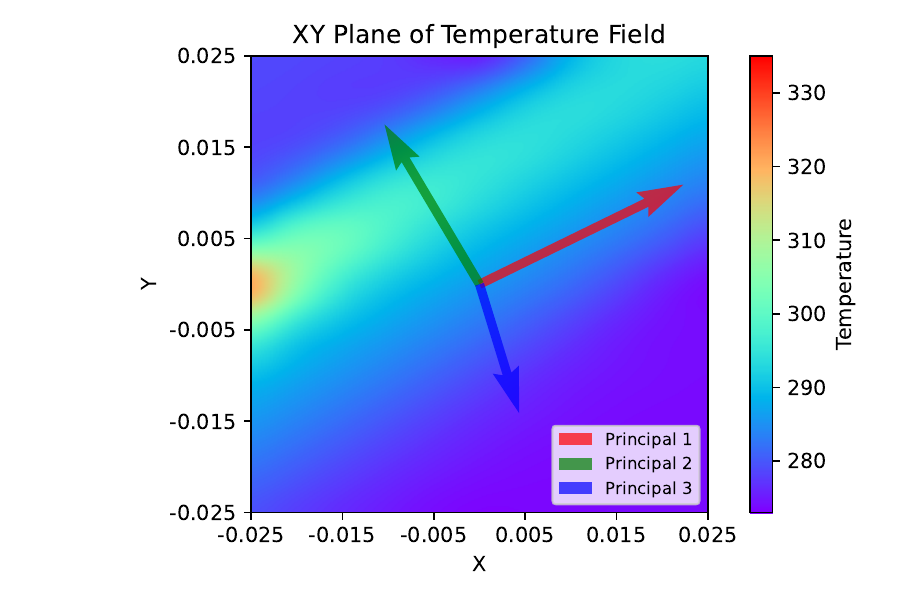}
        \caption{XY plane}
        \label{fig:XYplane}
    \end{subfigure}    
    \hfill
    \begin{subfigure}[t]{0.275\textwidth}        \includegraphics[width=\textwidth,trim=4.05cm 0.5cm 2.7cm 0.8cm,clip]{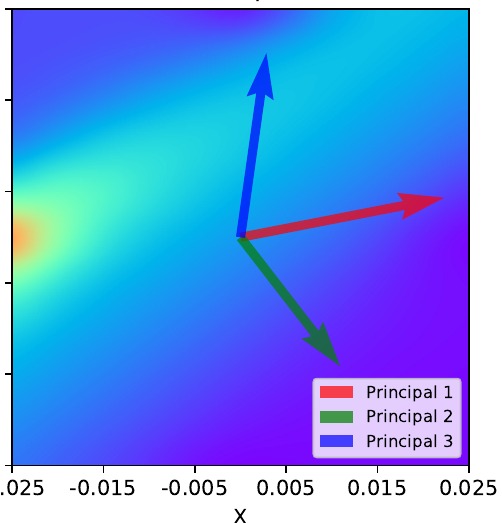}
        \caption{XZ plane}
        \label{fig:XZplane}
    \end{subfigure}    
    \vspace{-0.5cm}
    \hfill
    \begin{subfigure}[t]{0.35\textwidth}        \includegraphics[width=\textwidth,trim=4.05cm 0.5cm 0.5cm 0.8cm,clip]{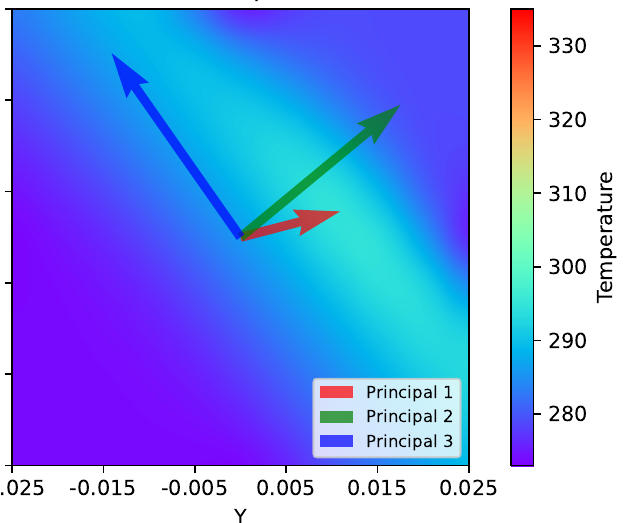}
        \caption{YZ plane}
        \label{fig:YZplane}
    \end{subfigure}
    \vspace{0.25cm}
    \caption{Deterministic temperature field for the multi-slice planes with projection of principal directions.}
    \label{fig:determnistic_plan}
\end{figure}

The deterministic solution, obtained using \refEQ{\ref{eq:initial_tensor}}, is shown in \refFIG{\ref{fig:multislice_deterministic}}. The results clearly demonstrate that the fibre direction significantly influences heat transport, directing it along principal paths. \refFIG{\ref{fig:determnistic_plan}} depicts the XY, XZ, and YZ planes, where boundary conditions create temperature gradients, with a noticeable drop near the patches where $T = T_b$. The highest temperature occurs at the heat source boundary, as expected. The projections of the 3D vectors onto the 2D planes visualise the principal directions of thermal conductivity, further revealing the fibre direction within the planes. Short vectors indicate directions with minimal in-plane projection, typically associated with lower effective conductivity in that plane. This deterministic reference sets the stage for evaluating the effect of uncertainty in the conductivity tensor on the temperature field. 

To evaluate the variability in the temperature field, we use $n = 500,000$ samples and compute the mean $\mu_{T}$ and standard deviation $\sigma_{T}$ using the Monte Carlo formulation in \refEQ{\ref{eq:MCFEM}}, as follows:
\begin{equation}\label{eq:MC_statistics}
\mu_{T} \approx \frac{1}{n} \sum_{j=1}^{n} \varTemp_i(\varEvent_j), \quad \sigma_{T} \approx \sqrt{\frac{1}{n-1} \sum_{j=1}^{n} ({\varTemp_i(\varEvent_j)} - \mu_{\varTemp})^2},
\end{equation}
for a single solution point $i$. To estimate the probability distribution function,  Kernel density estimation (KDE) is employed as a non-parametric method to approximate the probability density function $f(\varTemp_i)$ from data samples ${\varTemp_i(\varEvent_1), \varTemp_i(\varEvent_2), \dots, \varTemp_i(\varEvent_n)}$, defined  as
\begin{equation} \label{eq:KDE}
\hat{f}(\varTemp_i) = \frac{1}{
nh} \sum_{j=1}^n K\left(\frac{\varTemp_i - \varTemp_i(\varEvent_j)}{h}\right).
\end{equation}
Here, $K(u)$ is the kernel function, chosen as a Gaussian kernel $K(u) = \frac{1}{\sqrt{2\pi}} e^{-u^2 / 2}$. The bandwidth $h$ is computed using Scott’s rule: $h = \sigma_T \cdot n^{-1/5}$.

\subsection{Surrogate modelling of a patched cube with stochastic thermal conductivity}
\noindent
In the following, we compare the generalisation performance of the proposed MLP, SPDNet, and CMNN architectures using standard vectorization, \refEQ{\ref{eq:LogEig}}, and \refEQ{\ref{eq:StrAng}} as corresponding layers for $B$ in \refEQ{\ref{eq:CMNNfunction}}, respectively, when handling uncertain anisotropic material tensors. To facilitate the machine learning analysis, the temperature field is reduced to a coarser, equally-spaced grid of $11 \times 11 \times 11$ points over the cube, chosen to balance computational cost and data fidelity. For visualisation and further data processing, the temperature values at these grid points are interpolated to enhance the smoothness and readability of the visualised temperature distribution.

Each network is trained with $16,000$ samples, validated with $4,000$ samples, and tested on $30,000$ samples, which represent only a tenth of the reference Monte Carlo (MC) set. The hidden layers, layer sizes, activation functions, number of epochs, batch size, and learning rate are determined through Bayesian hyperparameter optimization using a tree-structured Parzen estimator \cite{akiba_optuna_2019}. For each network, training is repeated across 10 experiments with different initial sample configurations and network initializations.

To analyse the expected error at a grid point, we examine the sample-average normalised norms:
\begin{equation} \label{eq:L_norms} 
\ell_p = \mathbb{E_\omega} \left[ \frac{\|T(x,\omega_i)-\hat{T}(x,\omega_i)\|_p}{\|T(x,\omega_i)\|_p} \right]  
\end{equation} 
which provide insight into the expected error at a single grid point in the solution field. Specifically, for $p = 1$, this metric computes the expected relative difference between the predicted and finite element temperature fields, offering a measure of the average deviation. When $p = 2$, larger errors are emphasised, while for $p = \infty$, the worst-case scenario is produced. Norms are computed on a random temperature field, the mean temperature field and the standard deviation field.

Furthermore, to assess local errors across the entire temperature field, the absolute $\delta T_{a}$ and relative $T_{r}$ errors at each point are computed as:
\begin{equation}\label{eq:field_errors} 
\delta T_{a} = | \hat{T}_i - T_i |, \quad \delta T_{r} = \frac{| \hat{T}_i - T_i |}{| T_i |},
\end{equation} 
allowing for the visualisation of error distributions across an entire data slice using colour maps. Here, we restrict our analysis to the XY perspective, as \refFIG{\ref{fig:XYplane}} provides insight into the boundary conditions and the first principal direction.

Lastly, the Kullback–Leibler (KL) divergence quantifies the difference between the predicted and reference probability density functions. This metric evaluates how closely the predicted distributions align with the reference data:
\begin{equation}\label{eq:KLD} 
    D_\text{KL} = \sum_{j=1}^k P_j \log\left(\frac{P_j}{Q_j}\right), 
\end{equation}
where $P_j$ and $Q_j$ represent the probabilities from the reference and predicted histograms, respectively.

\subsubsection{Surrogate model for scaling uncertainty}


\begin{table}[b!]
\centering
\caption{Parameters of the log-normal random variables.}
\label{tab:lognormal_variables}
\begin{tabular}{lcc}
\hline
Variable & Mean ($\mu$) & Standard Deviation ($\sigma$) \\ \hline
$\lambda_1$           & 14.0                 & 0.8                                   \\
$\lambda_2$           & 0.11                 & 0.02                                   \\
$\lambda_3$           & 2.4                  & 0.27                                  \\ \hline
\end{tabular}
\end{table}

\begin{figure}[!t]               
    \captionsetup{justification=centering,margin=1cm}
    \centering
    \hfill
    \begin{subfigure}[t]{0.3\textwidth}
        \includegraphics[width=\textwidth, trim=2.3cm 0.5cm 0.7cm 0.9cm,clip]{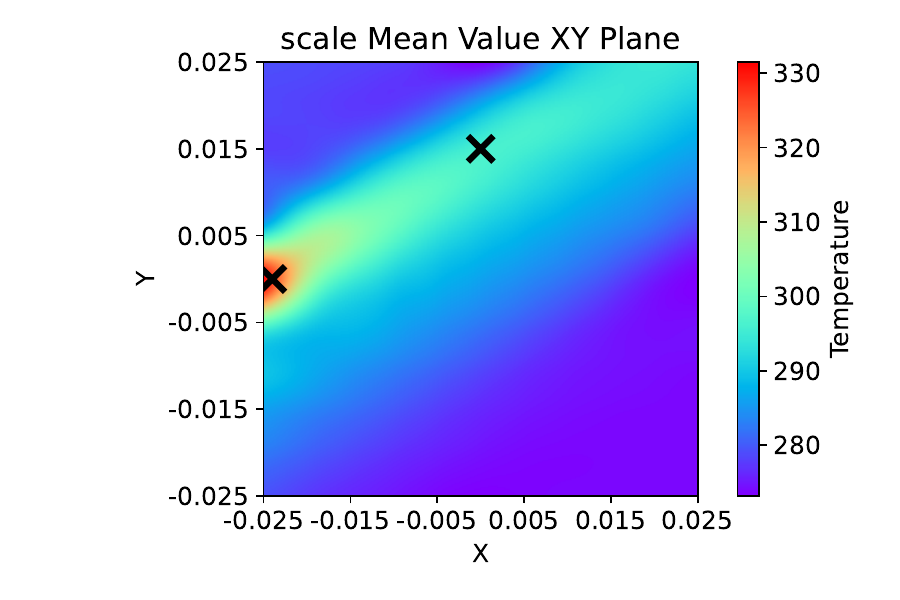}
        \caption{Mean temperature field (\varCelsius) with markers.}
        \label{fig:ref_mean_scl_YZ}
    \end{subfigure}    
    \hfill
    \begin{subfigure}[t]{0.3\textwidth}        
        \includegraphics[width=\textwidth, trim=2.3cm 0.5cm 0.7cm 0.9cm,clip]{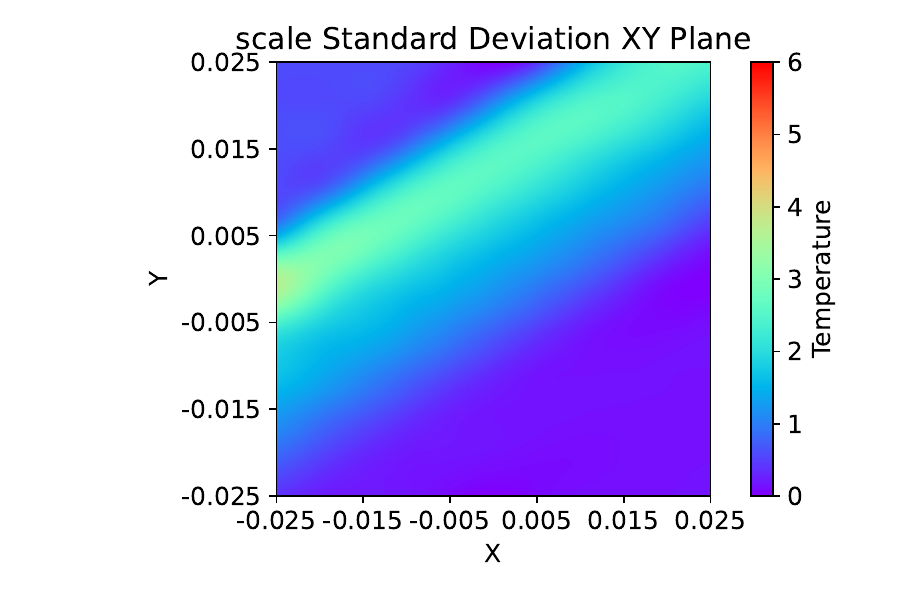}
        \caption{Standard deviation (\varCelsius) over the temperature field.}
        \label{fig:ref_std_scl_YZ}
    \end{subfigure} 
    \hfill    
    \begin{subfigure}[t]{0.33\textwidth}       
        \includegraphics[width=\textwidth, trim=0.3cm 0.2cm 0.7cm 1.3cm,clip]{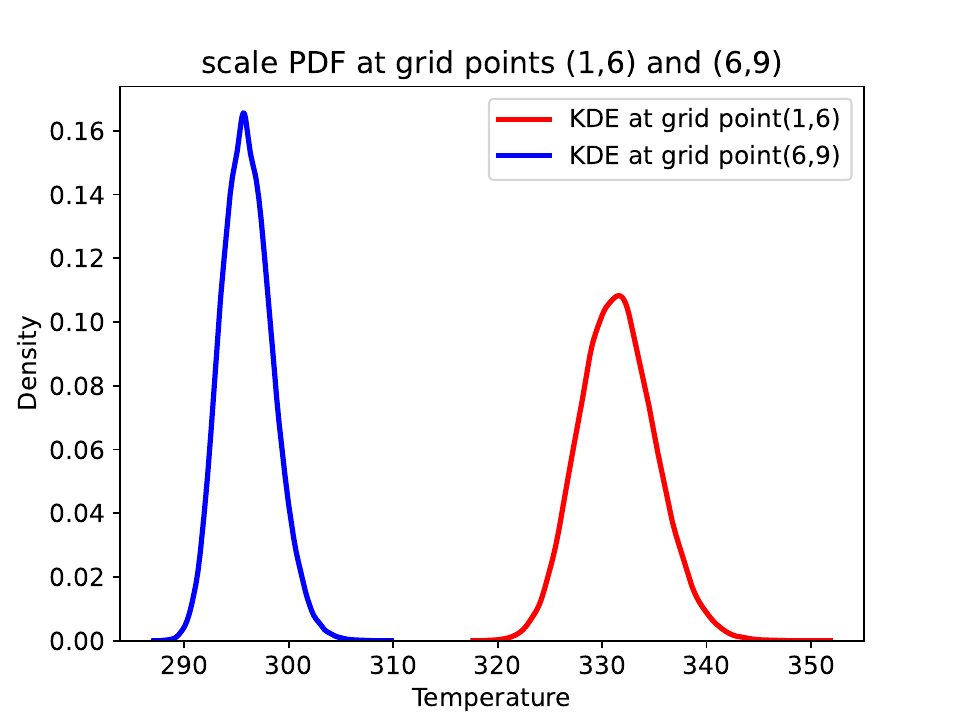}
        \caption{Probability density for the points marked on the mean plot.}
        \label{fig:ref_dist_scl_YZ}
    \end{subfigure} 
    \hfill    
    \caption{Temperature field statistics in the XY plane with the scaling uncertainty dataset.}
    \label{fig:ref_scl}
\end{figure}

\noindent
To evaluate networks' performance under scaling uncertainty, we define a tensor-valued random variable using \refEQ{\ref{eq:random_both}}, with three independent log-normal random variables. The mean values, $\mu_i$, are derived from \refEQ{\ref{eq:initial_tensor}}, while the standard deviations are chosen to reflect uncertainties observed in composites \cite{buser_characterisation_2022, grouve_simulating_2021}. Table \ref{tab:lognormal_variables} lists the mean and standard deviation for the log-normal variables.

\begin{figure}[!b]    
    \captionsetup{justification=centering}
    \centering
    \begin{subfigure}[t]{0.41\textwidth}
        \includegraphics[width=\textwidth,trim=0.35cm 0cm 0.25cm 1.3cm,clip]{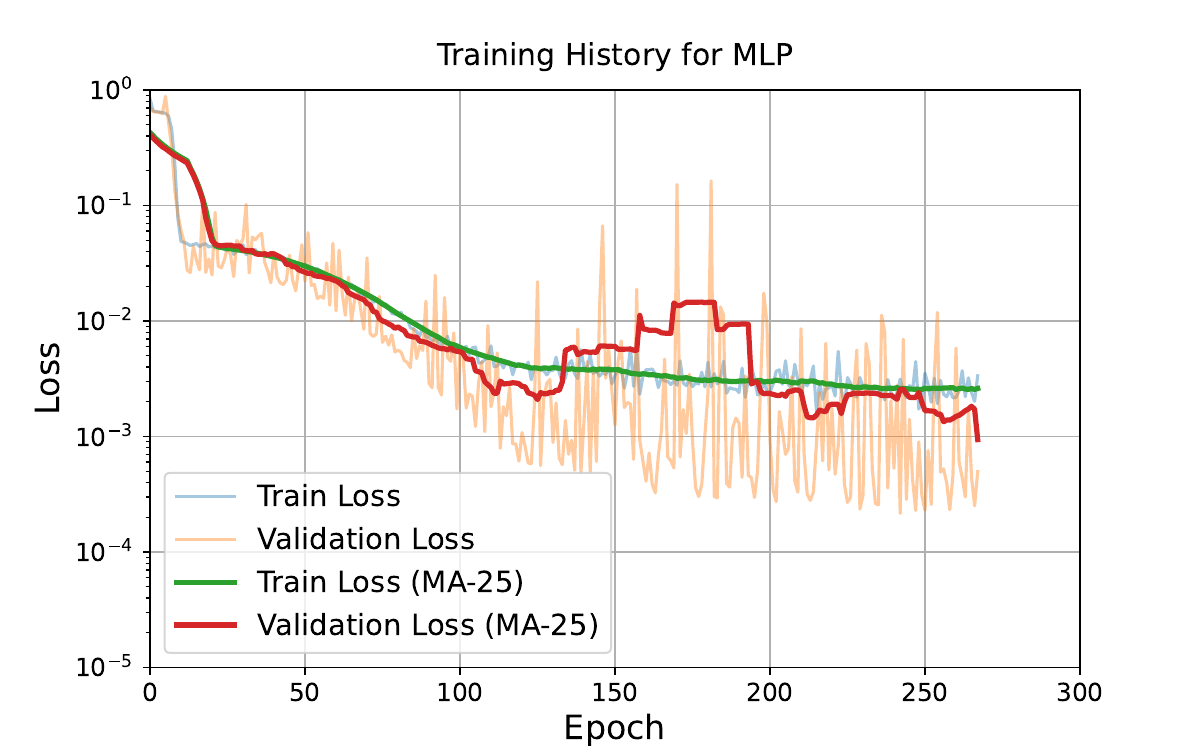}
        \caption{MLP}
        \label{fig:his_scl_MLP}
    \end{subfigure}
    \begin{subfigure}[t]{0.41\textwidth}
        \includegraphics[width=\textwidth,trim=0.35cm 0cm 0.25cm 1.3cm,clip]{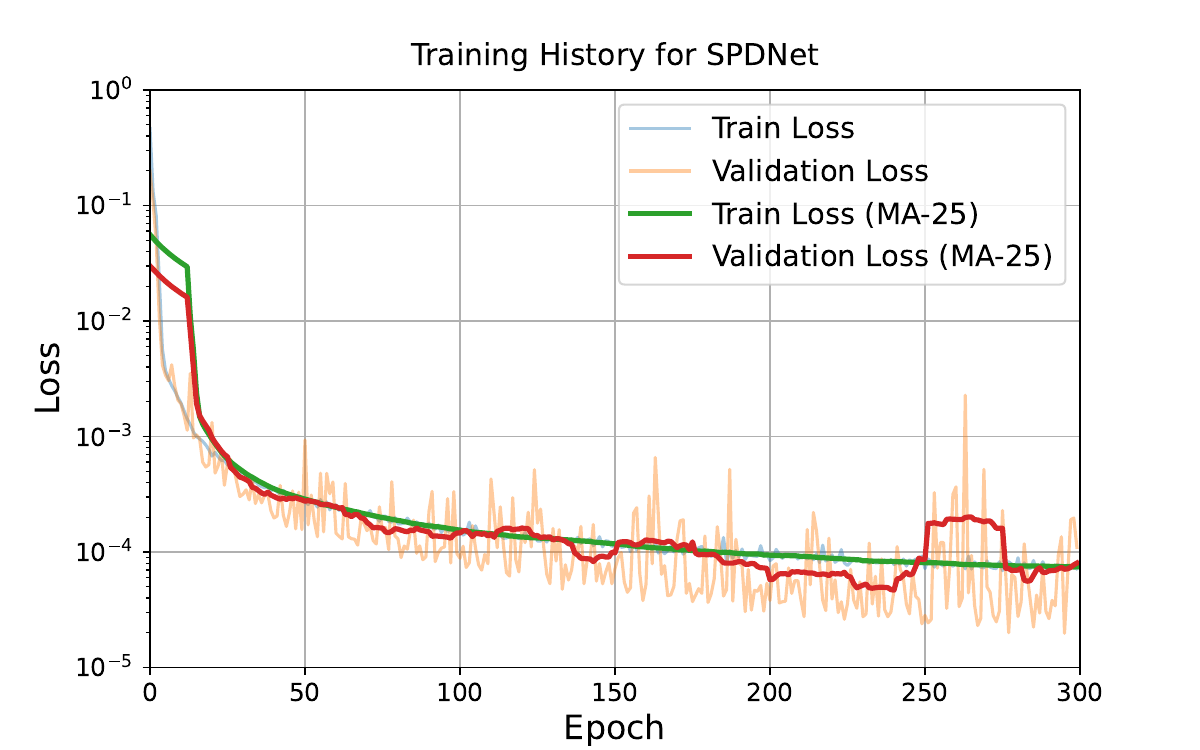}
        \caption{SPDNet}
        \label{fig:his_scl_SPD}
    \end{subfigure}
    \vspace{0.15cm}
    
    \begin{subfigure}[t]{0.41\textwidth}
        \includegraphics[width=\textwidth,trim=0.35cm 0cm 0.25cm 1.3cm,clip]{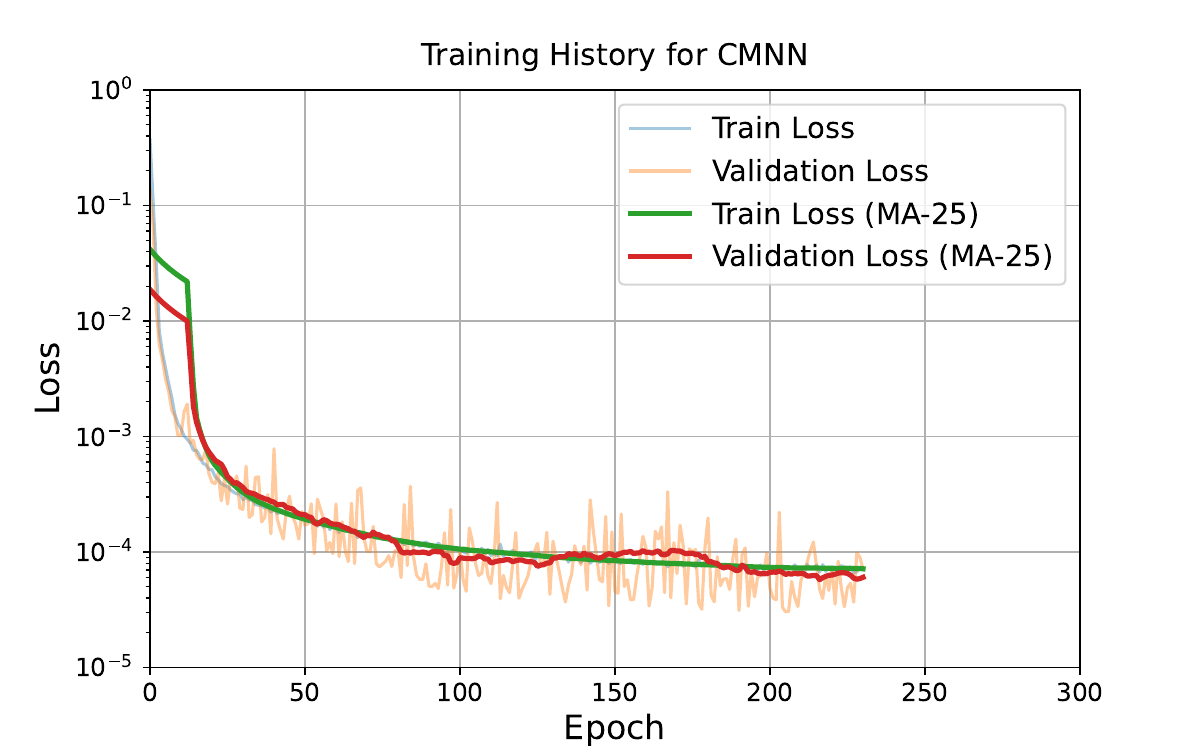}
        \caption{CMNN}
        \label{fig:his_scl_CMNN}
    \end{subfigure}    
    \caption{Training history of the neural networks with the scaling uncertainty dataset.}
    \vspace{-0.35cm}
    \label{fig:his_scl}
\end{figure}
The statistical characteristics of the temperature field under scaling uncertainty are depicted in \refFIG{\ref{fig:ref_scl}}. The mean temperature field, shown in \refFIG{\ref{fig:ref_mean_scl_YZ}}, is similar to that of the deterministic solution, as expected. The standard deviation of the temperature field, presented in \refFIG{\ref{fig:ref_std_scl_YZ}}, mirrors regions of high mean temperature, indicating that hotter regions exhibit greater variability. \refFIG{\ref{fig:ref_dist_scl_YZ}} shows the probability-density functions (PDFs) of the temperature at two selected points, marked in \refFIG{\ref{fig:ref_mean_scl_YZ}}. The PDF in the high-temperature region (red) exhibits a shifted mean compared to the cooler region (blue), consistent with heat conduction. Furthermore, both points display non-normal distributions seen in the shape of top of the curve, although their spreads and peakedness differ slightly.


To train the MLP, SPDNet, and CMNN, the optimised hyperparameters are 300 epochs, a batch size of 8, two hidden layers with 16 and 8 neurons, a Tanh activation function, the Adam optimizer, and a learning rate of $3 \times 10^{-3}$. Table \ref{tab:scl_results} compares the performance of the MLP, SPDNet, and CMNN architectures on the scaling dataset. The MLP architecture struggles to generalize, exhibiting a high validation loss ($2.16 \times 10^{-4}$). In contrast, the SPDNet and CMNN architectures demonstrate much lower validation losses ($0.20 \times 10^{-4}$ and $0.30 \times 10^{-4}$, respectively). SPDNet achieves the most consistent performance, with the smallest mean and standard deviation across repeated experiments, while CMNN delivers competitive results but with slightly higher variability. A similar trend is visible in Table \ref{tab:scl_norm_results}; notably, SPDNet outperforms the CMNN in estimating the standard deviation. Instabilities are further evident in the training histories depicted in \refFIG{\ref{fig:his_scl}}, where validation losses for the MLP (\refFIG{\ref{fig:his_scl_MLP}}) and, to a lesser extent, SPDNet (\refFIG{\ref{fig:his_scl_SPD}}) exhibit larger fluctuations compared to CMNN (\refFIG{\ref{fig:his_scl_CMNN}}). This behaviour suggests a tendency for the MLP and, to some extent, SPDNet to over-fit the training set rather than learn the input–output relationship effectively.

\begin{table}[!t]
\centering
\caption{Training metrics of the neural networks with the scaling uncertainty dataset. Data in order of magnitude of $ e^{-4}$.}
\begin{tabularx}{\textwidth}{l|XXX|XXX}
\toprule
Layer & Best Val. Loss & Test Loss & Train Loss & 
Mean Val. Loss (Std.) & Mean Test Loss (Std.) & Mean Train Loss (Std.) \\
\midrule
MLP & 2.16 & 4.71 & 31.40 & 3.13 (0.74) & 31.68(31.63) & 33.29 (8.75) \\
SPDNet & 0.20 & 1.08 & 0.73 & 0.43 (0.18) & 0.68 (0.27) & 1.09 (0.25) \\
CMNN & 0.30 & 0.68 & 0.71 & 0.47 (0.12) & 1.22 (0.92) & 1.16 (0.26) \\
\bottomrule
\end{tabularx}
\label{tab:scl_results}
\end{table}

\begin{table}[!t]
\centering
\caption{Normalised and per-sample norms of the neural networks with the scaling uncertainty dataset.}
\begin{tabularx}{\textwidth}{l|XXX|XXX|XXX}
\toprule
Model & L1 Sample $e^{-4}$ & L2 Sample $e^{-6}$ & L$\infty$ Sample $e^{-4}$ & 
        L1 Mean $e^{-4}$ & L2 Mean $e^{-6}$ & L$\infty$ Mean $e^{-4}$ & 
        L1 \; Std $e^{-4}$ & L2 \; Std $e^{-5}$ & L$\infty$ Std $e^{-3}$ \\
\midrule
MLP    & 1.52  & 6.33  & 8.49  & 1.87 & 7.17  & 7.20  & 4.42 & 1.64  & 3.69 \\
SPDNet & 1.64  & 6.55  & 7.75  & 0.38 & 1.51  & 5.37  & 1.92 & 0.74  & 0.80 \\
CMNN   & 0.78  & 3.33  & 6.69  & 0.44 & 1.94  & 2.60  & 3.58 & 1.32 & 2.03 \\
\bottomrule
\end{tabularx}
\label{tab:scl_norm_results}
\end{table}

\begin{figure}[h!]               
    \captionsetup{justification=centering,margin=0cm}
    \centering
    \vspace{-0.6cm}
    \begin{subfigure}[t]{0.28\textwidth}
        \includegraphics[width=\textwidth, trim=1.0cm 1.0cm 3.6cm 0.8cm,clip]{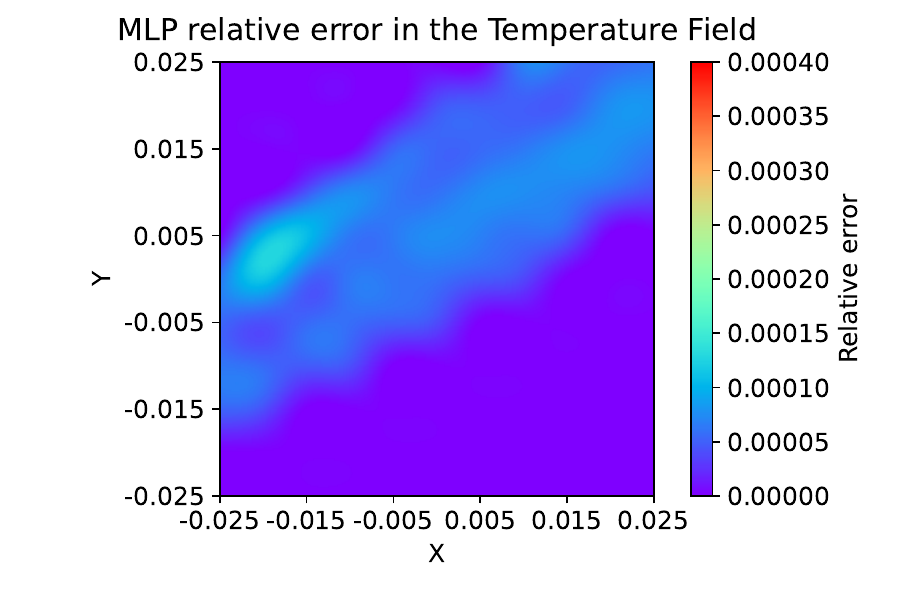}
        \caption{MLP relative error in a random sample.}
        \label{fig:MLP_rel_scl}
    \end{subfigure}    
    \hfill
    \begin{subfigure}[t]{0.25\textwidth}        
    \includegraphics[width=\textwidth,trim=2.3cm 1.0cm 3.6cm 0.8cm,clip]{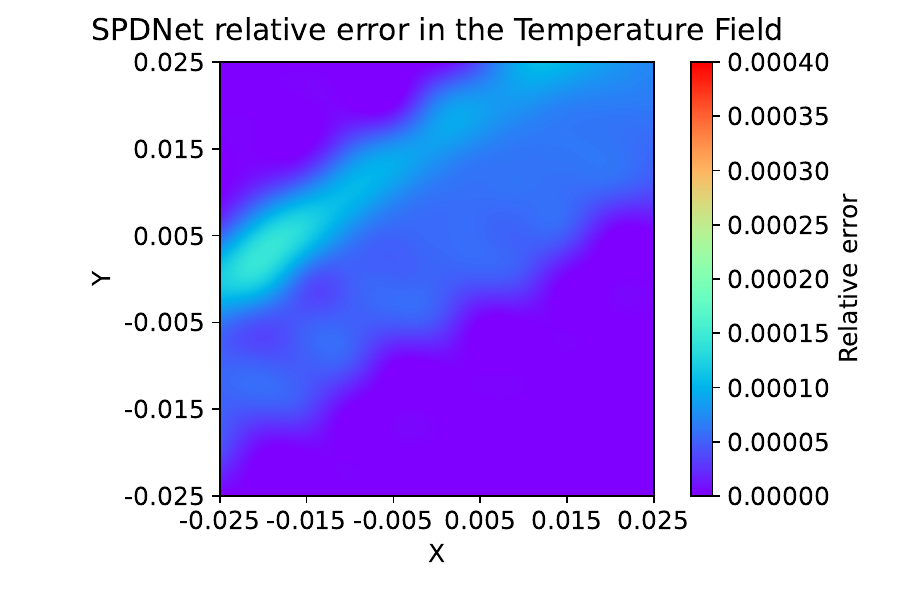}
        \caption{SPDNet relative error in a random sample.}
        \label{fig:SPD_rel_scl}
    \end{subfigure}
    \hfill
    \begin{subfigure}[t]{0.328\textwidth}        \includegraphics[width=\textwidth,trim=2.3cm 1.0cm 0.5cm 0.8cm,clip]{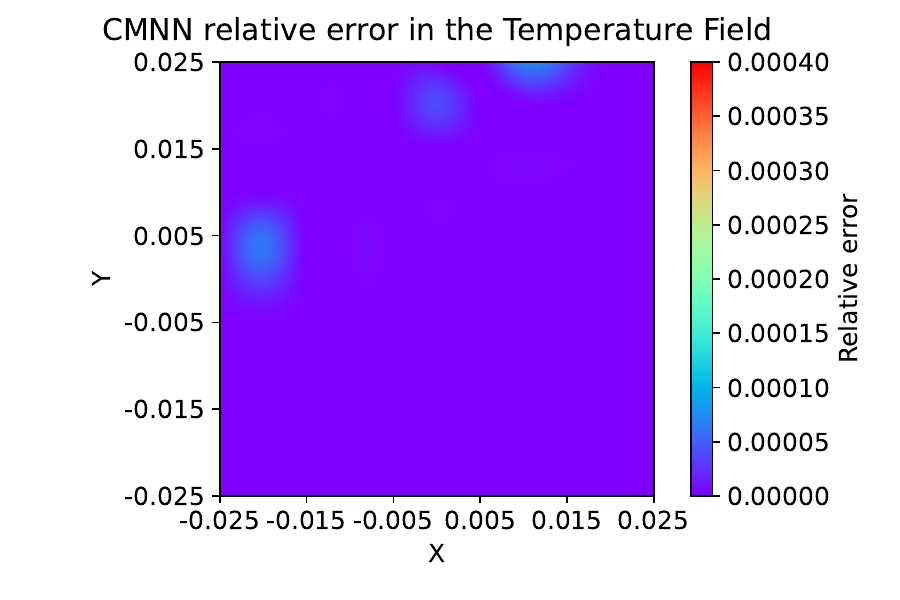}
        \caption{CMNN relative error in a random sample.}
        \label{fig:CMNN_rel_scl}
    \end{subfigure}    
    \hfill
    
    \begin{subfigure}[t]{0.295\textwidth}
        \includegraphics[width=\textwidth, trim=1.0cm 1.0cm 2.95cm 0.8cm,clip]{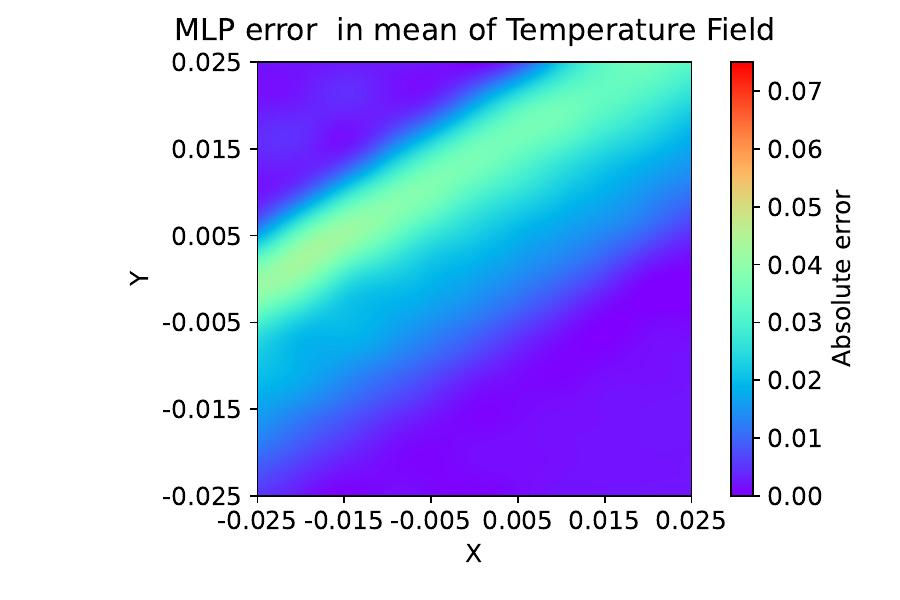}
        \caption{MLP absolute error in the mean.}
        \label{fig:MLP_mean_abs_scl}
    \end{subfigure}    
    \hfill
    \begin{subfigure}[t]{0.257\textwidth}        \includegraphics[width=\textwidth,trim=2.6cm 1.0cm 2.95cm 0.8cm,clip]{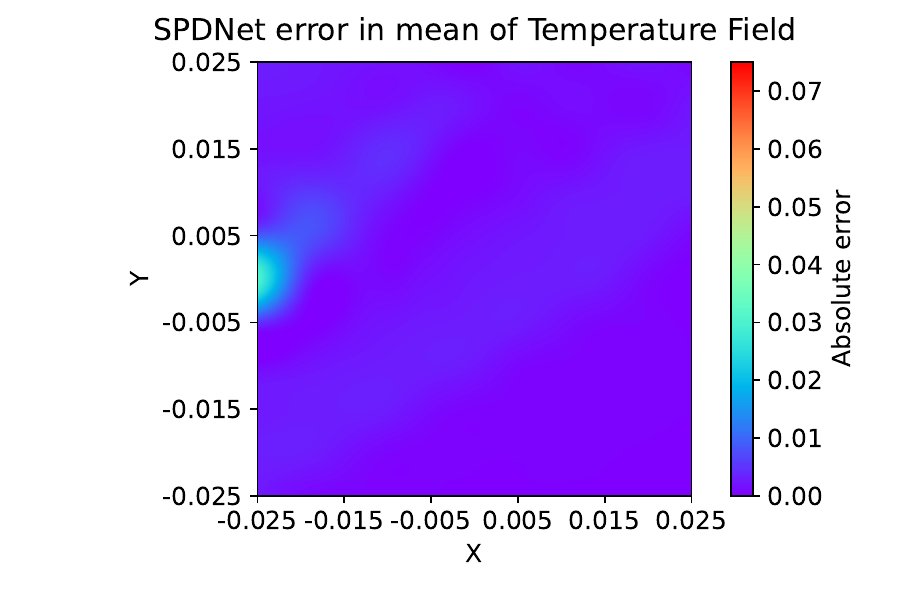}
        \caption{SPDNet absolute error in the mean.}
        \label{fig:SPD_mean_abs_scl}
    \end{subfigure}
    \hfill
    \begin{subfigure}[t]{0.328\textwidth}        \includegraphics[width=\textwidth,trim=2.6cm 1.0cm 0.5cm 0.8cm,clip]{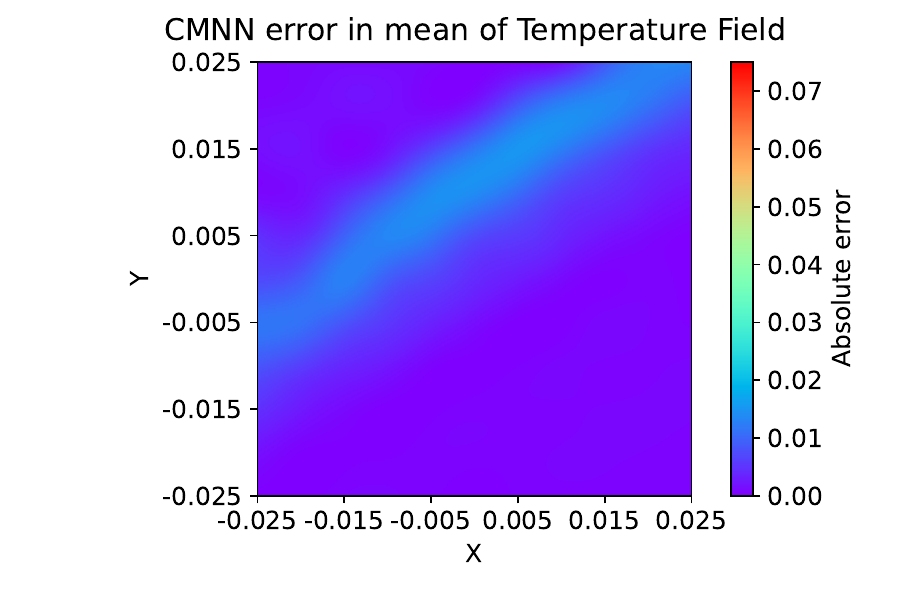}
        \caption{CMNN absolute error in the mean.}
        \label{fig:CMNN_mean_abs_scl}
    \end{subfigure}    
    \hfill
    
    \begin{subfigure}[t]{0.29\textwidth}
        \includegraphics[width=\textwidth, trim=1.0cm 1.0cm 3.6cm 0.8cm,clip]{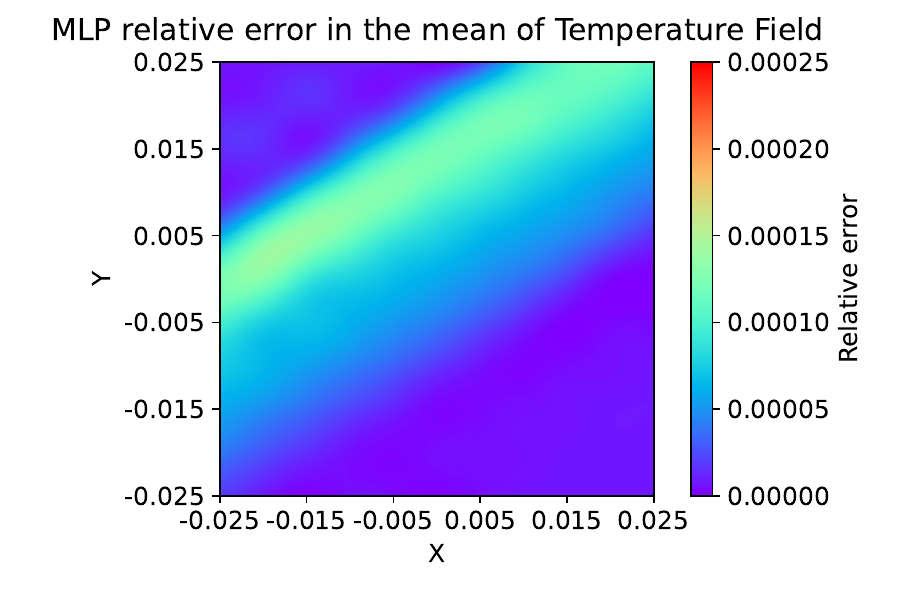}
        \caption{MLP relative error in the mean.}
        \label{fig:MLP_mean_rel_scl}
    \end{subfigure}    
    \hfill
    \begin{subfigure}[t]{0.257\textwidth}        \includegraphics[width=\textwidth,trim=2.0cm 1.0cm 3.6cm 0.8cm,clip]{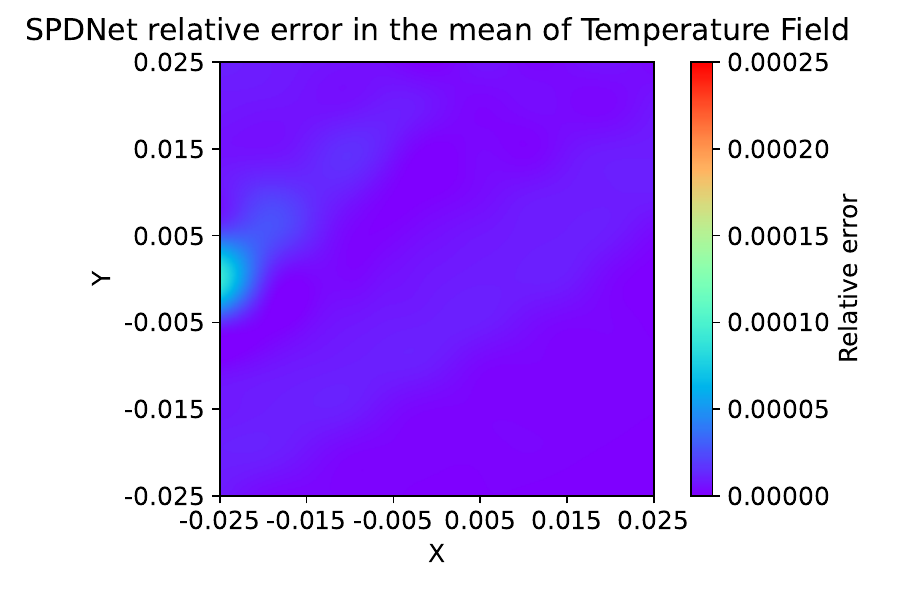}
        \caption{SPDNet relative error in the mean.}
        \label{fig:SPD_mean_rel_scl}
    \end{subfigure}
    \hfill
    \begin{subfigure}[t]{0.328\textwidth}        \includegraphics[width=\textwidth,trim=2.0cm 1.0cm 0.5cm 0.8cm,clip]{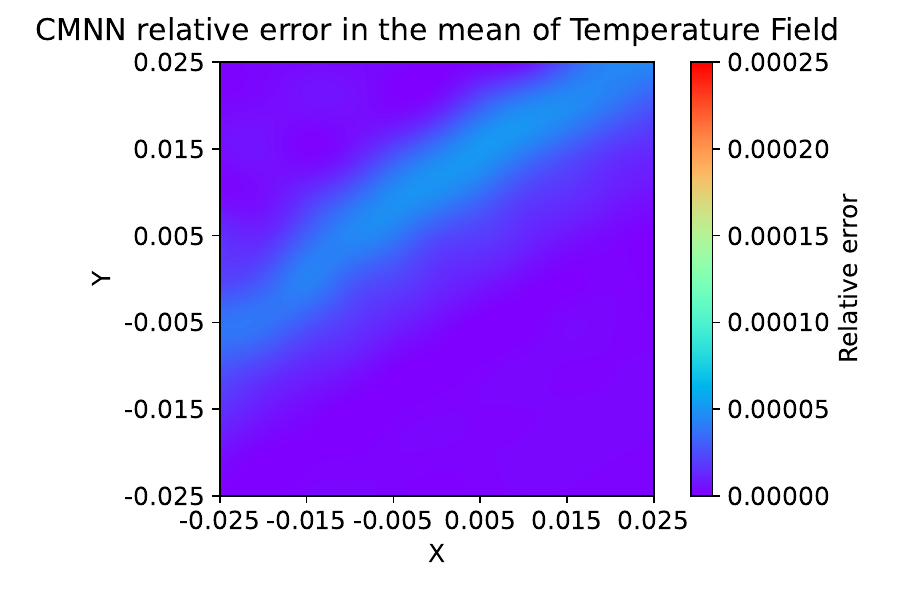}
        \caption{CMNN relative error in the mean.}
        \label{fig:CMNN_mean_rel_scl}
    \end{subfigure}   
    
    
    \begin{subfigure}[t]{0.29\textwidth}
        \includegraphics[width=\textwidth, trim=1.0cm 1.0cm 3.2cm 0.8cm,clip]{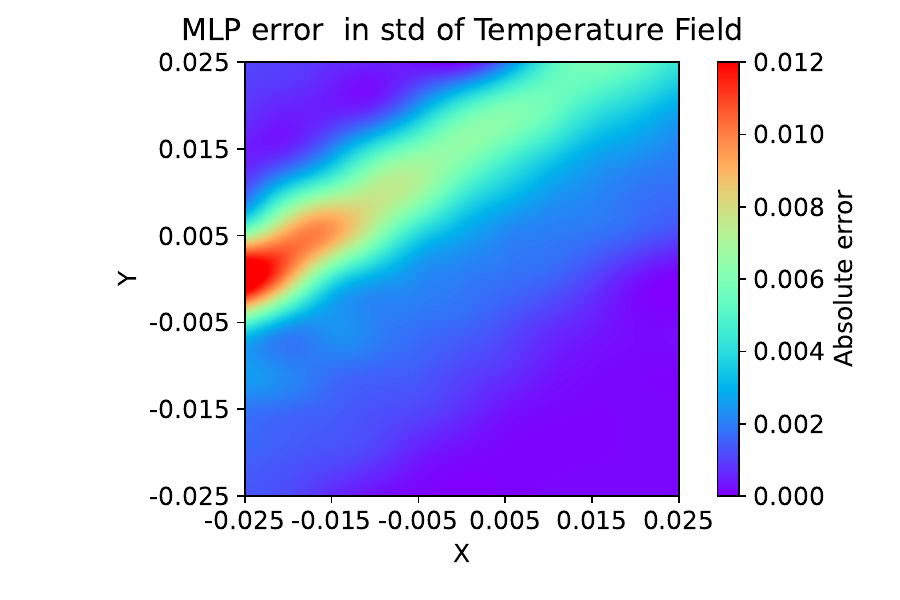}
        \caption{MLP absolute error in the standard deviation.}
        \label{fig:MLP_std_abs_scl}
    \end{subfigure}    
    \hfill
    \begin{subfigure}[t]{0.257\textwidth}        \includegraphics[width=\textwidth,trim=2.6cm 1.0cm 3.2cm 0.8cm,clip]{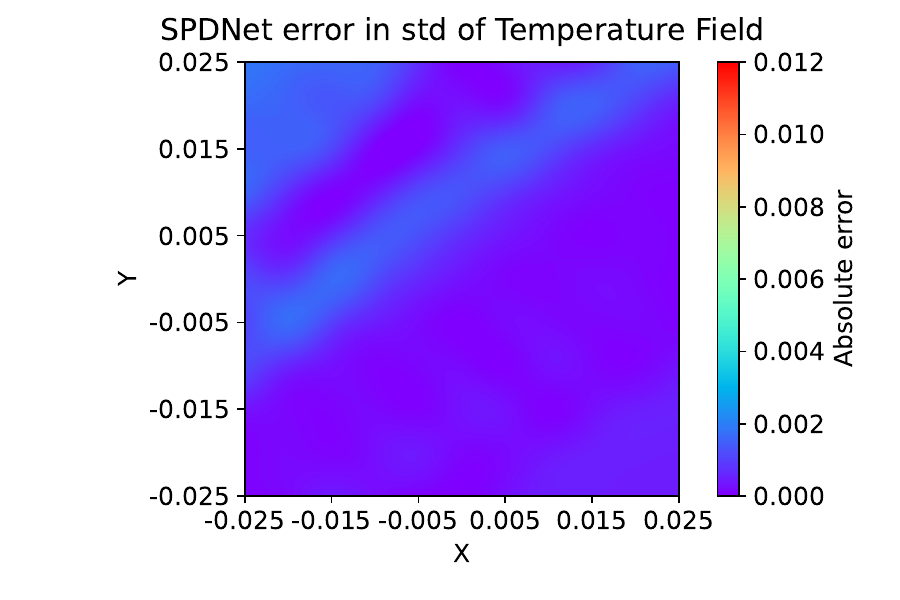}
        \caption{SPDNet absolute error in the standard deviation.}
        \label{fig:SPD_std_abs_scl}
    \end{subfigure}
    \hfill
    \begin{subfigure}[t]{0.328\textwidth}        \includegraphics[width=\textwidth,trim=2.6cm 1.0cm 0.5cm 0.8cm,clip]{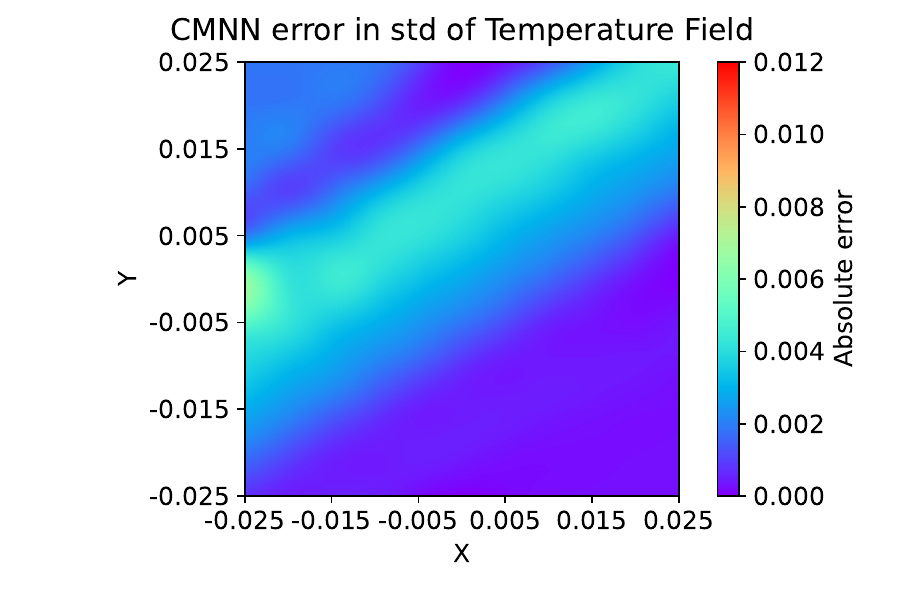}
        \caption{CMNN absolute error in the standard deviation.}
        \label{fig:CMNN_std_abs_scl}
    \end{subfigure}
    \caption{Errors in approximation of the temperature field by the neural networks with the scaling uncertainty dataset.}
    \label{fig:errors_scl_1}
    \vspace{0.5cm}
\end{figure}

\begin{figure}[!ht]               
    \captionsetup{justification=centering,margin=0cm}
    \centering
    \vspace{-0.6cm}
    \begin{subfigure}[t]{0.28\textwidth}
        \includegraphics[width=\textwidth, trim=1.0cm 1.0cm 3.2cm 0.8cm,clip]{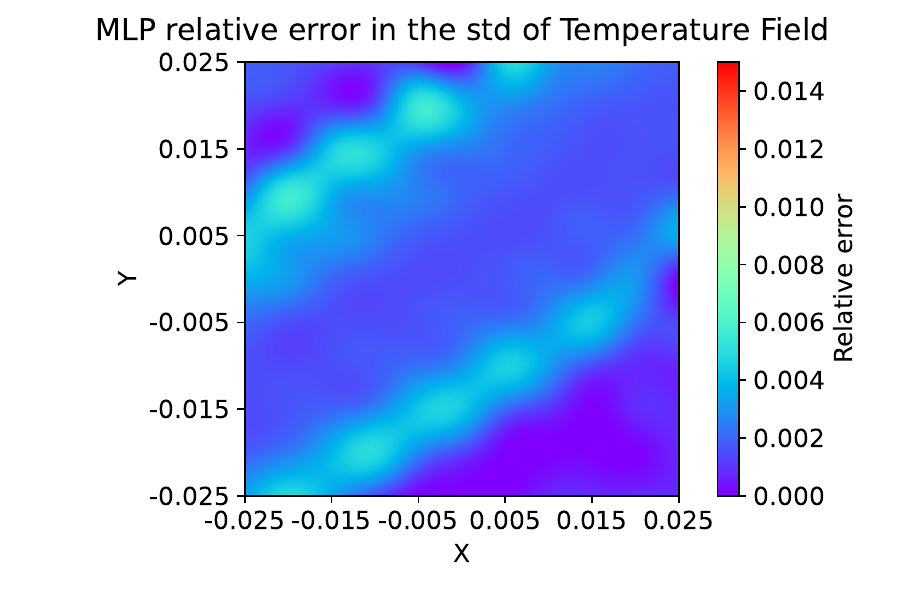}
        \caption{MLP relative error in the standard deviation.}
        \label{fig:MLP_std_rel_scl}
    \end{subfigure}    
    \hfill
    \begin{subfigure}[t]{0.247\textwidth}        \includegraphics[width=\textwidth,trim=2.3cm 1.0cm 3.2cm 0.8cm,clip]{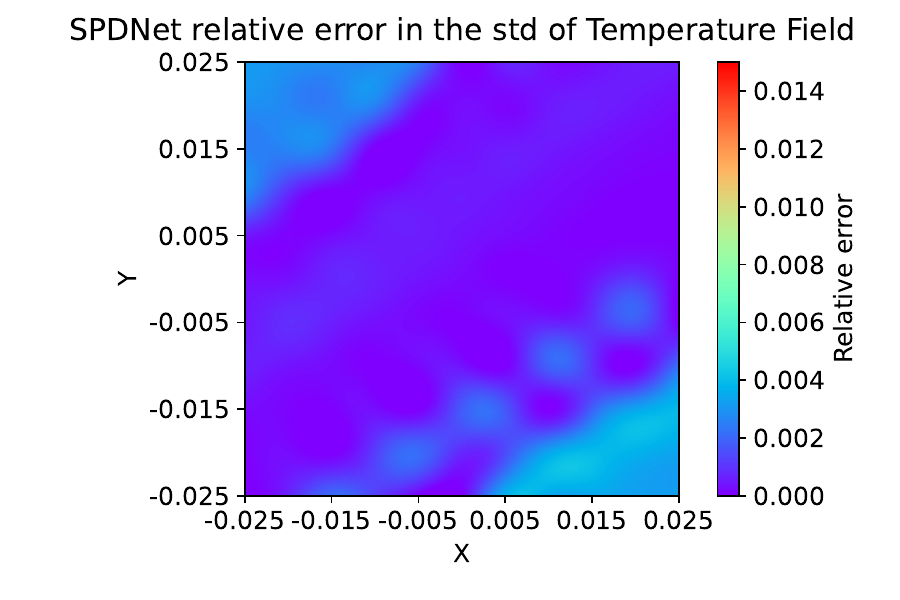}
        \caption{SPDNet relative error in the standard deviation.}
        \label{fig:SPD_std_rel_scl}
    \end{subfigure}
    \hfill
    \begin{subfigure}[t]{0.328\textwidth}        \includegraphics[width=\textwidth,trim=2.3cm 1.0cm 0.5cm 0.8cm,clip]{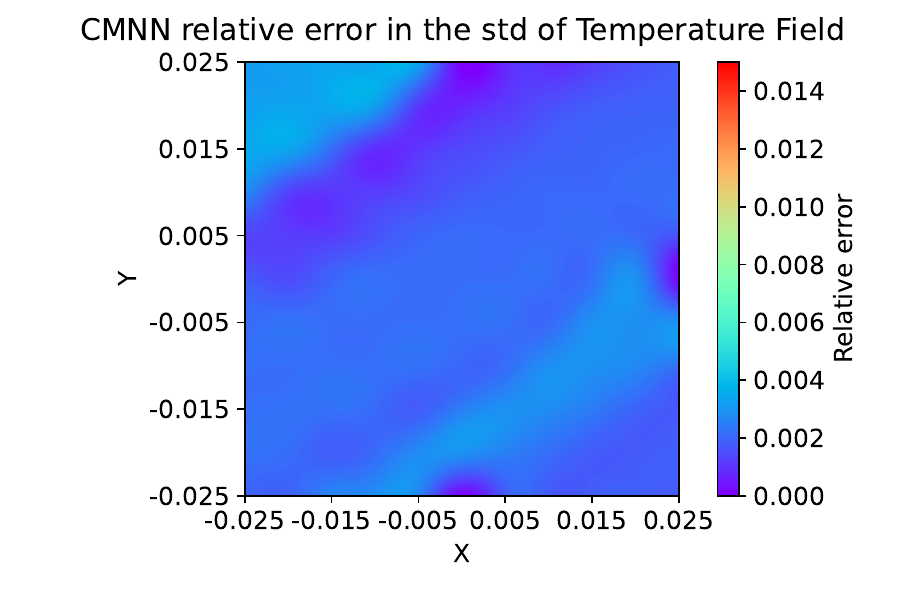}
        \caption{CMNN relative error in the standard deviation.}
        \label{fig:CMNN_std_rel_scl}
    \end{subfigure}    
    \hfill
    
    \begin{subfigure}[t]{0.28\textwidth}
        \includegraphics[width=\textwidth, trim=1.0cm 0.0cm 3.6cm 0.8cm,clip]{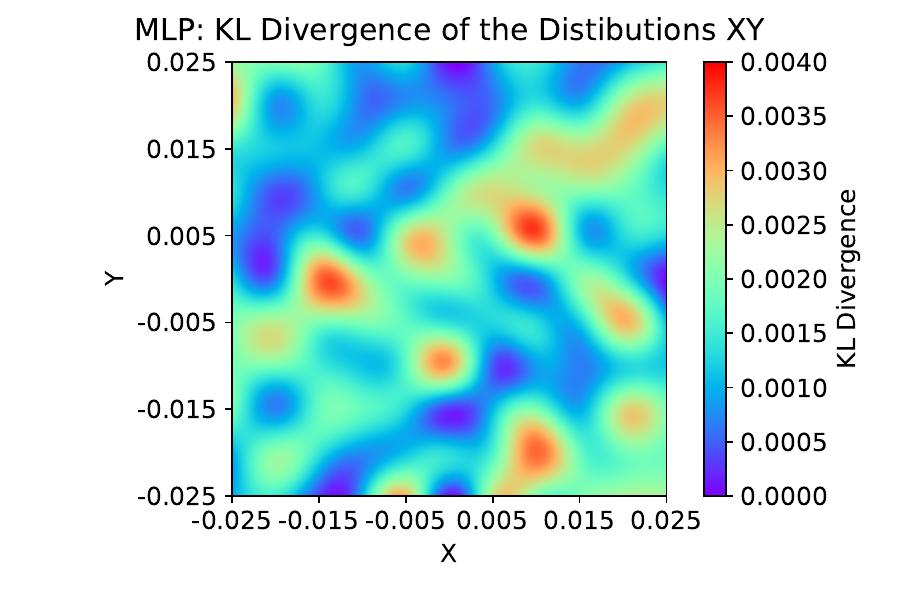}
        \caption{MLP Kullback-Leibler divergence.}
        \label{fig:MLP_KLD_scl}
    \end{subfigure}  
    \hfill
    \begin{subfigure}[t]{0.247\textwidth}        
    \includegraphics[width=\textwidth,trim=2.3cm 0.0cm 3.6cm 0.8cm,clip]{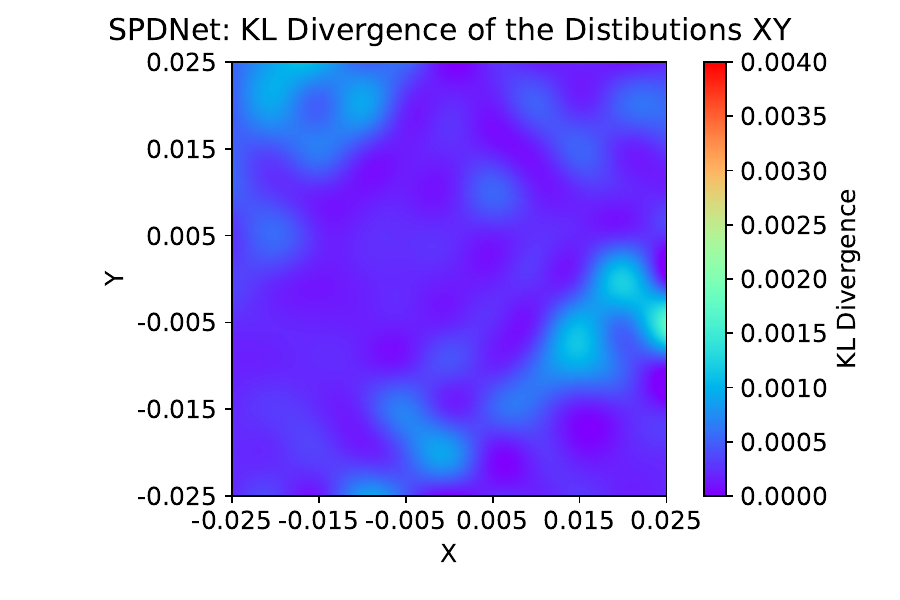}
        \caption{SPDNet Kullback-Leibler divergence.}
        \label{fig:SPD_KLD_scl}
    \end{subfigure} 
    \hfill    
    \begin{subfigure}[t]{0.328\textwidth}        \includegraphics[width=\textwidth,trim=2.3cm 0.0cm 0.5cm 0.8cm,clip]{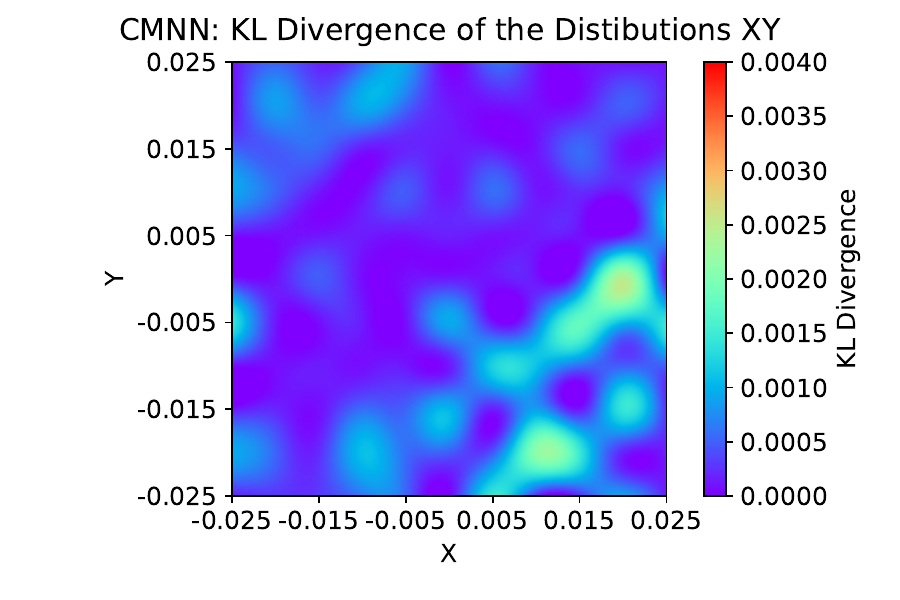}
        \caption{CMNN Kullback-Leibler divergence.}
        \label{fig:CMNN_KLD_scl}
    \end{subfigure} 
    \vspace{0.25cm}
    \caption{Errors in approximation of the temperature field by the neural networks with the scaling uncertainty dataset.}
    \label{fig:errors_scl_2}
\end{figure}


\refFIG{\ref{fig:errors_scl_1}–\ref{fig:errors_scl_2}} show the errors of the MLP, SPDNet, and CMNN architectures in approximating the temperature field under scaling uncertainty. The first row, \refFIG{\ref{fig:MLP_rel_scl}–\ref{fig:CMNN_rel_scl}}, displays the relative error in a random sample, highlights significant errors, particularly in regions with steep temperature gradients along the diagonal, where variability is highest. SPDNet demonstrates a smoother and more uniform error distribution, though localised areas of elevated errors remain. CMNN shows the lowest relative errors, with minimal and uniformly distributed relative errors across the domain. The second and third rows, \refFIG{\ref{fig:MLP_mean_abs_scl}–\ref{fig:CMNN_mean_rel_scl}}, show the absolute and relative errors in the mean, following a similar trend. Errors are highest for the MLP, concentrated near the patch at $(-0.025, 0)$. CMNN shows lower errors, mainly across the diagonal, while SPDNet achieves the smallest overall errors. The fourth and fifth rows, \refFIG{\ref{fig:MLP_std_abs_scl}–\ref{fig:MLP_std_rel_scl}}, illustrate absolute and relative errors in the standard deviation. Here, SPDNet is generally more accurate. The final row examines the KL divergence, comparing the predicted and reference temperature field distributions. The MLP exhibits scattered and high divergences, while SPDNet and CMNN show significantly lower errors, particularly in high-variance regions.

Overall, SPDNet and to a slightly lesser extent the CMNN outperform the standard MLP when working with scaling uncertainty.

\subsubsection{Surrogate model for orientation uncertainty}
 
\begin{figure}[t!]               
    \captionsetup{justification=centering,margin=0cm}
    \centering
    \includegraphics[width=0.45\textwidth, trim=5.3cm 7.5cm 0cm 4cm,clip]{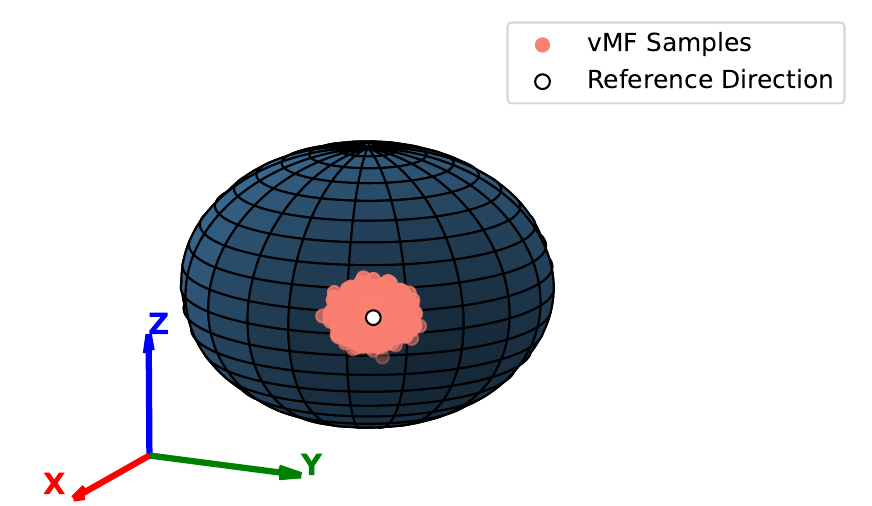}
    \caption{Von mises fisher distribution with $\varVMFcon = 200$.}
    \label{fig:VMF_sphere}
\end{figure}

\begin{figure}[b!]               
    \captionsetup{justification=centering,margin=0cm}
    \centering
    
    \begin{subfigure}[t]{0.32\textwidth}
        \includegraphics[width=\textwidth, trim=2.3cm 0.5cm 0.7cm 0.9cm,clip]{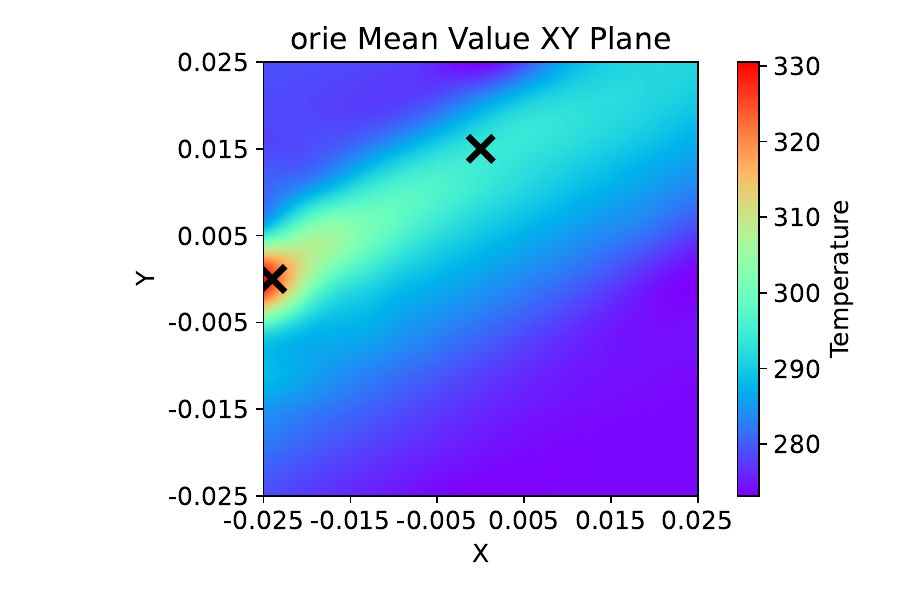}
        \caption{Mean temperature field with markers.}
        \label{fig:ref_mean_ori_YZ}
    \end{subfigure}    
    \hfill
    \begin{subfigure}[t]{0.31\textwidth}        \includegraphics[width=\textwidth, trim=2.9cm 0.5cm 0.7cm 0.9cm,clip]{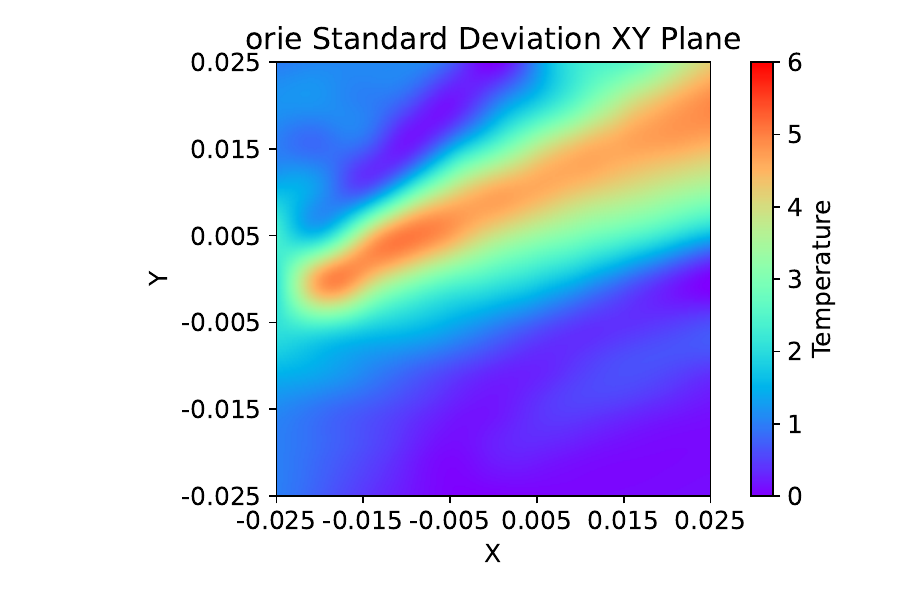}
        \caption{Standard deviation over the temperature field.}
        \label{fig:ref_std_ori_YZ}
    \end{subfigure} 
    \hfill    
    \begin{subfigure}[t]{0.33\textwidth}        
        \includegraphics[width=\textwidth, trim=0.3cm 0.0cm 0.7cm 1.3cm,clip]{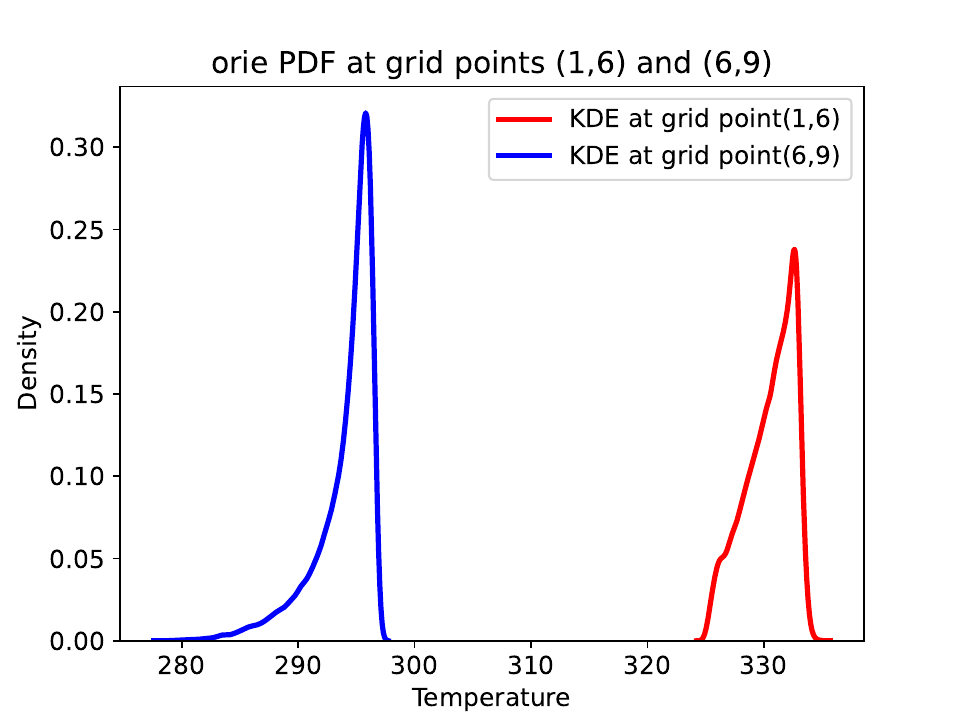}
        \caption{Probability density functions for the points marked on the mean plot.}
        \label{fig:ref_pdf_ori}
    \end{subfigure} 
    
    \begin{subfigure}[t]{0.27\textwidth}
        \includegraphics[width=\textwidth, trim=2.3cm 1.0cm 2.85cm 0.94cm,clip]{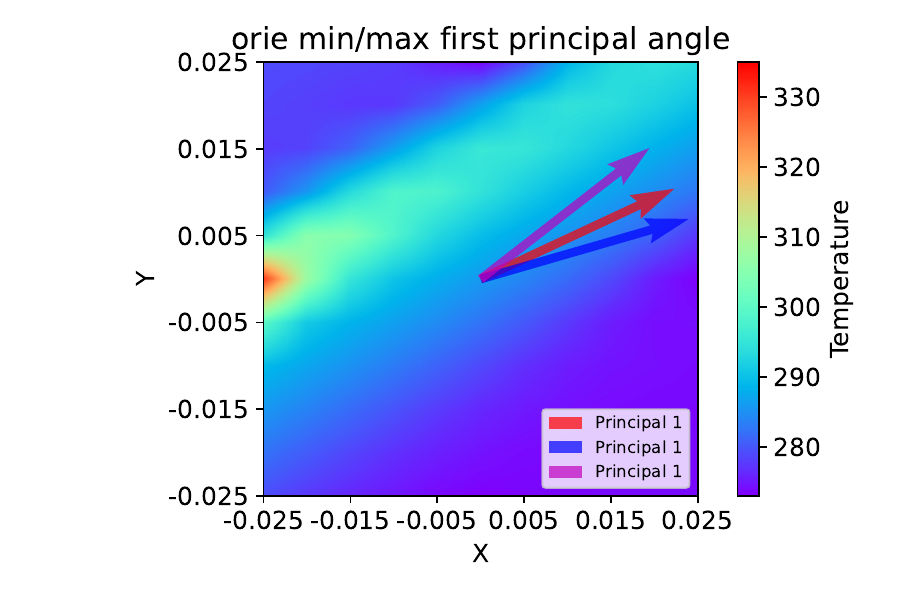}
        \caption{Lower (5\%), mean and upper (95\%) angles on the mean temperature plot.}
        \label{fig:minmax_angle_YZ}
    \end{subfigure}    
    \hfill
    \begin{subfigure}[t]{0.255\textwidth}        
        \includegraphics[width=\textwidth,trim=2.8cm 1.0cm 2.85cm 0.94cm,clip]{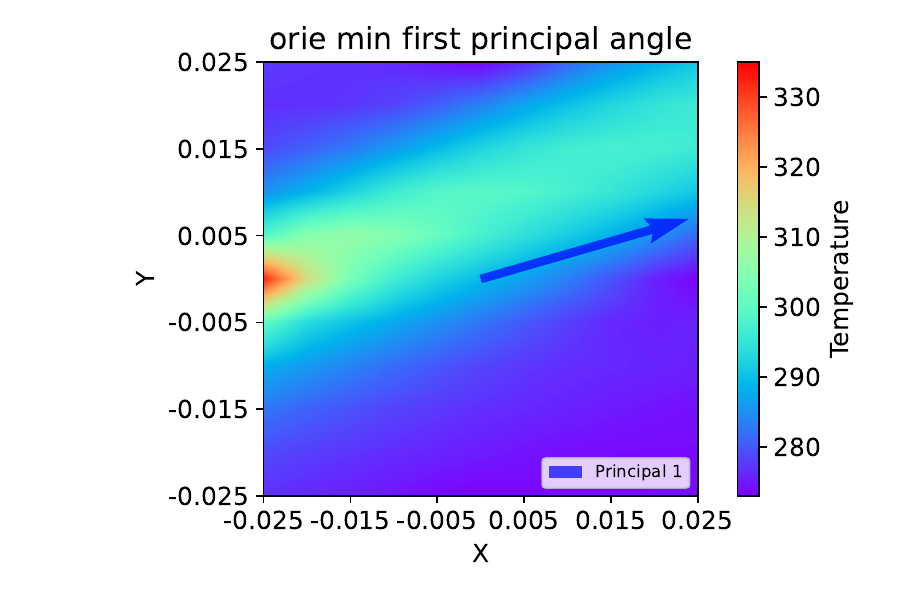}
        \caption{Lower (95\%) angle temperature field}
        \label{fig:min_angle_YZ}
    \end{subfigure} 
    \hfill    
    \begin{subfigure}[t]{0.32\textwidth}        
        \includegraphics[width=\textwidth,trim=2.8cm 1.0cm 0.5cm 0.94cm,clip]{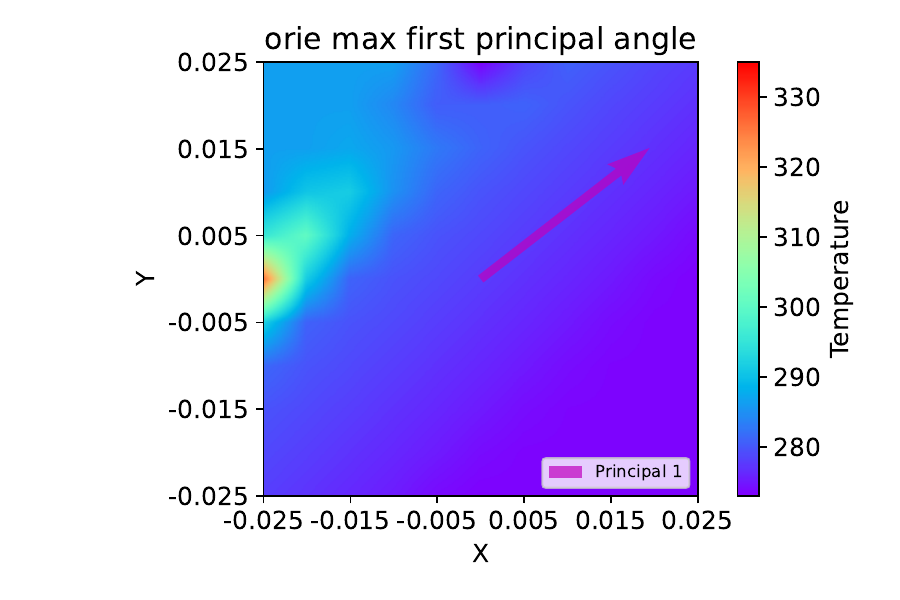}
        \caption{Upper (95\%) angle temperature field}
        \label{fig:max_angle_YZ}
    \end{subfigure} 
    \vspace{-0.25cm}
    \caption{Temperature field statistics in the XY plane with the orientation uncertainty dataset.}
    \label{fig:ref_ori}
\end{figure}

\begin{figure}[!b]
    \captionsetup{justification=centering}
    \centering
    \begin{subfigure}[b]{0.41\textwidth}
        \includegraphics[width=\textwidth, trim=0.35cm 0cm 0.25cm 1.3cm,clip]{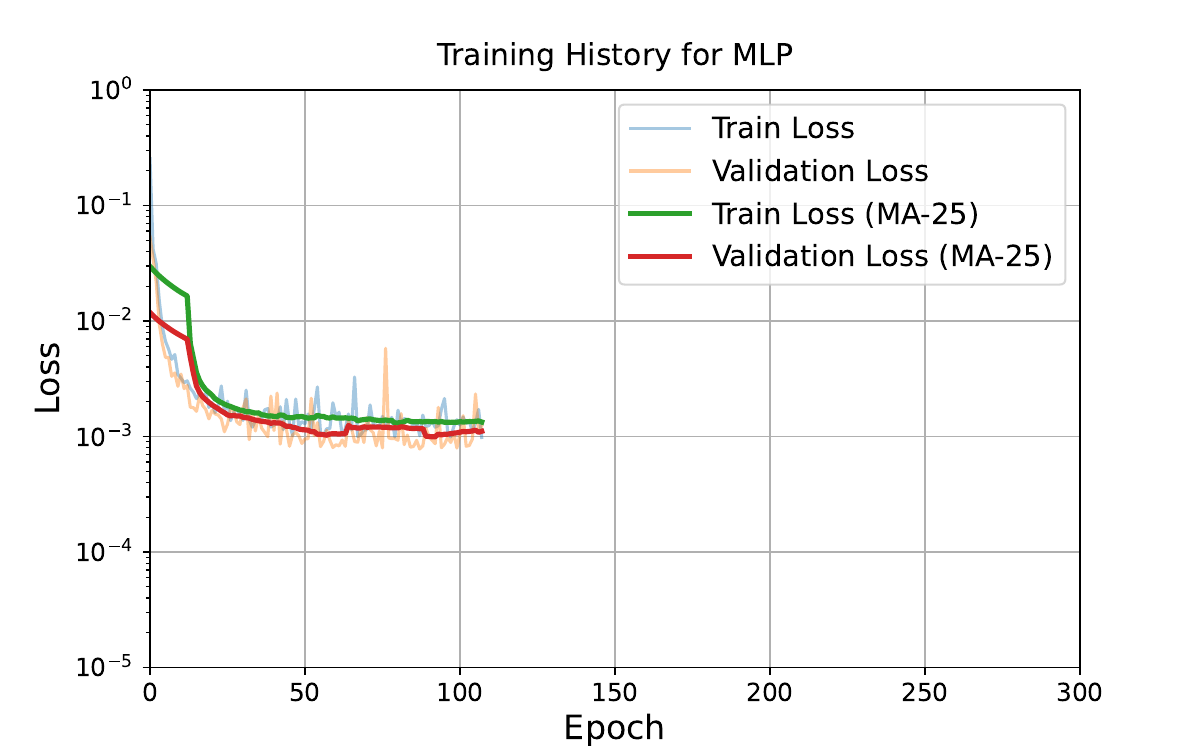}
        \caption{MLP}
        \label{fig:his_ori_vec}
    \end{subfigure}
    \begin{subfigure}[b]{0.41\textwidth}
        \includegraphics[width=\textwidth,trim=0.35cm 0cm 0.25cm 1.3cm,clip]{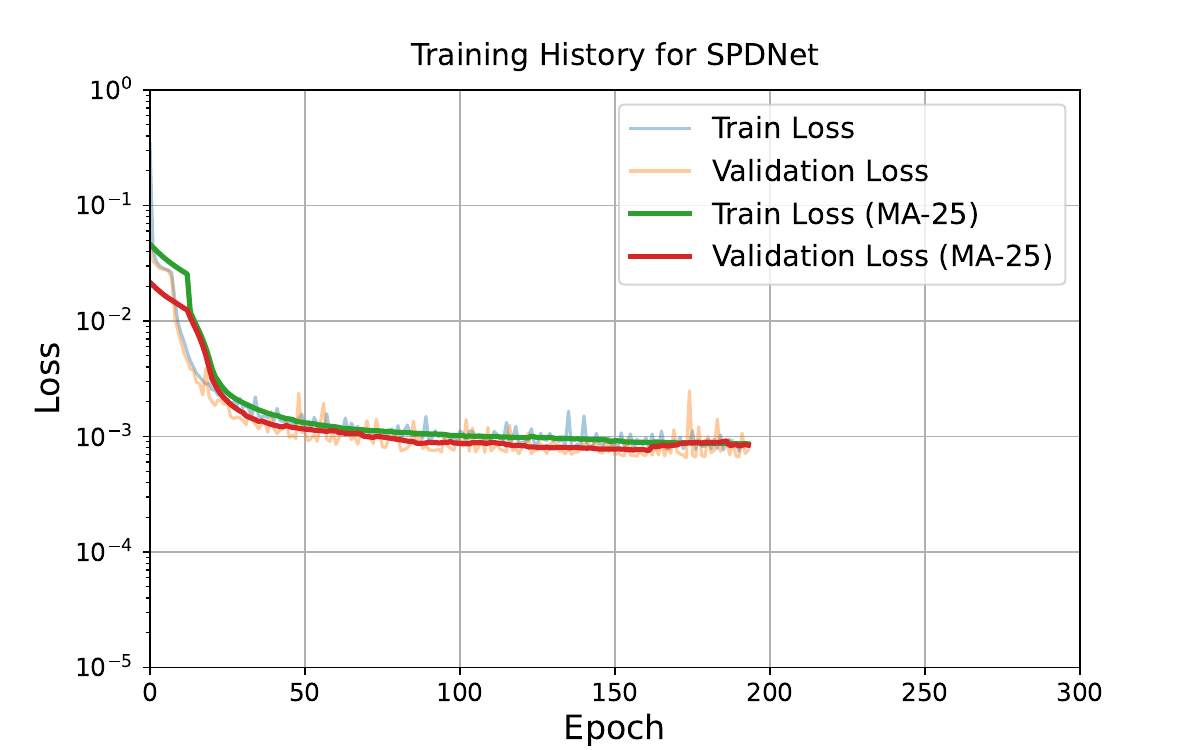}
        \caption{SPDNet}
        \label{fig:his_ori_spdnet}
    \end{subfigure}    
    \begin{subfigure}[b]{0.41\textwidth}
        \includegraphics[width=\textwidth, trim=0.35cm 0cm 0.25cm 1.3cm,clip]{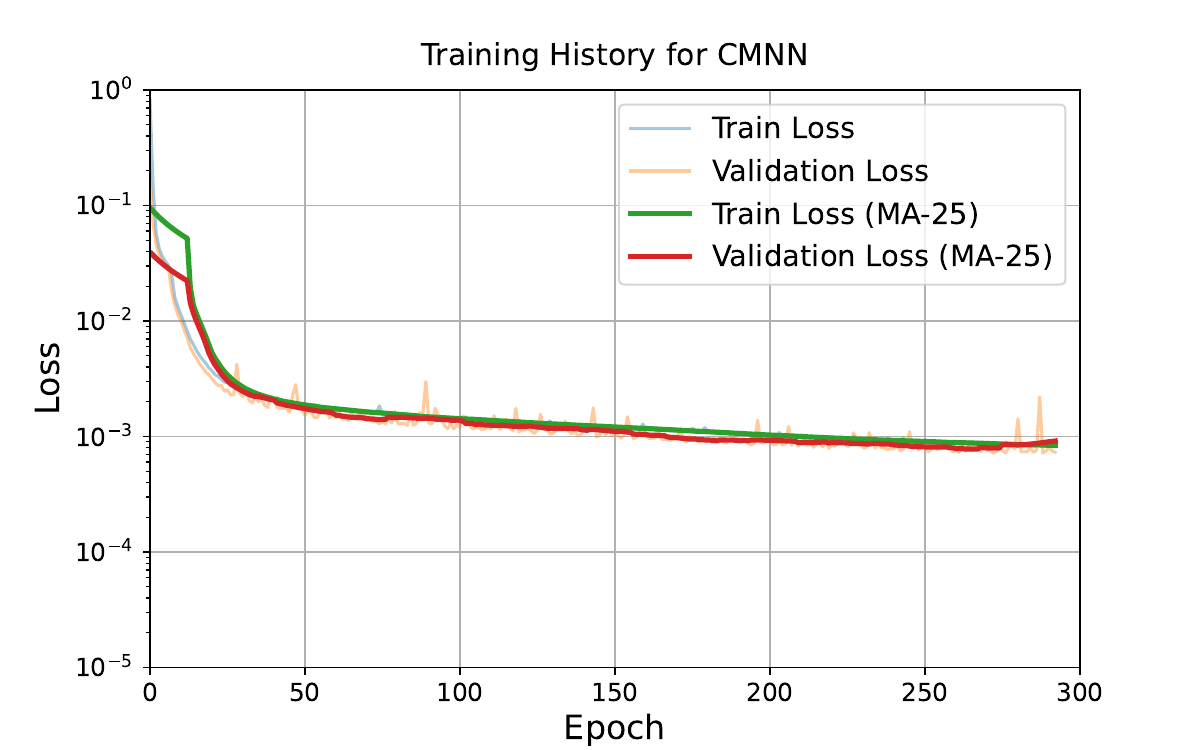}
        \caption{CMNN}
        \label{fig:his_ori_cmnn}
    \end{subfigure}        
    \vspace{-0.25cm}
    \caption{Training history of the neural networks with the orientation uncertainty dataset.}
    \vspace{-0.35cm}
    \label{fig:his_ori}
\end{figure}

\noindent
To investigate neural network performance under orientation uncertainty, we model the eigenvectors as uncertain using \refEQ{\ref{eq:stochastic_eigvec}}. The von Mises–Fisher distribution obtained from \refEQ{\ref{eq:VMF}}, is shown in \refFIG{\ref{fig:VMF_sphere}}, illustrating the directional concentration of samples around the reference vector. In this visualization, the reference vector $\boldsymbol{\mu}$ is defined as the first column of the eigenvector matrix specified in \refEQ{\ref{eq:initial_tensor}}. A concentration parameter of $\eta = 200$ is characterises the spread of the distribution

\refFIG{\ref{fig:ref_ori}} illustrates the reference solution under orientation uncertainty, as randomisation of principal directions leads to localised changes in heat flow alignment. The mean temperature field, depicted in \refFIG{\ref{fig:ref_mean_ori_YZ}}, aligns closely with the deterministic solution. The standard deviation, shown in \refFIG{\ref{fig:ref_std_ori_YZ}}, reveals lower uncertainty near the entry patch at $(-0.025,0)$, where the gradient direction is largely preserved. However, further from the entry point, uncertainty increases as heat flows toward the patches at $(0,0.025)$ and $(0.025,0)$, reflecting the effects of orientation variability.

The influence of orientation is even more pronounced in the temperature fields shown in \refFIG{\ref{fig:min_angle_YZ}} and \refFIG{\ref{fig:max_angle_YZ}}, which corresponds to the lower (5\%) and upper (95\%) angle percentile, respectively. The temperature field for the lower-angle shows clear deviations from the mean, with noticeable shifts in heat distribution. The temperature field for the upper-angle exhibits even greater changes, emphasizing the sensitivity of the temperature field to orientation variations. This behaviour is further reflected in the probability distributions presented in \refFIG{\ref{fig:ref_pdf_ori}}, where orientation uncertainty leads to significant variations in the statistical characteristics of the temperature field. Compared to scaling uncertainty, orientation-driven variability causes more spatially distributed heat transport.

\begin{table}[!b]
\centering
\caption{Training metrics of the neural networks with the orientation uncertainty dataset. Data in order of magnitude of $ e^{-4}$.}
\begin{tabularx}{\textwidth}{l|XXX|XXX}
\toprule
Layer & Best Val.\ Loss & Test Loss & Train Loss &
Mean Val.\ Loss (Std.) & Mean Test Loss (Std.) & Mean Train Loss (Std.) \\
\midrule
MLP    & 7.79 & 10.16 & 13.12 & 8.6\,(0.8)  & 14.1\,(5.8) & 12.2\,(1.7) \\
SPDnet & 6.54 &  8.79 &  8.23 & 7.1\,(0.4)  &  9.5\,(1.7) &  9.1\,(1.5) \\
CMNN   & 7.18 &  8.37 &  7.75 & 8.0\,(0.8)  &  9.3\,(2.0) &  9.5\,(1.3) \\
\bottomrule
\end{tabularx}
\label{tab:ori_results}
\end{table}

\begin{table}[!b]
\centering
\caption{Normalised and per-sample norms of the neural networks with the orientation uncertainty dataset.}
\begin{tabularx}{\textwidth}{l|XXX|XXX|XXX}
\toprule
Model & L1 Sample $e^{-4}$ & L2 Sample $e^{-5}$ & L$\infty$ Sample $e^{-3}$ & 
        L1 Mean $e^{-4}$ & L2 Mean $e^{-5}$ & L$\infty$ Mean $e^{-3}$ & 
        L1 \; Std $e^{-3}$ & L2 \; Std $e^{-5}$ & L$\infty$ Std $e^{-2}$ \\
\midrule
MLP    & 5.02 & 2.22 & 6.78 & 3.54 & 1.44 & 1.71 & 1.84 & 7.03 & 1.15 \\
SPDNet & 2.34 & 1.03 & 2.97 & 1.32 & 0.63 & 1.85 & 1.08 & 4.67 & 1.24 \\
CMNN   & 1.76 & 0.85 & 2.69 & 1.12 & 0.54 & 1.13 & 0.63 & 2.90 & 0.67 \\
\bottomrule
\end{tabularx}
\label{tab:ori_norm_results}
\end{table}

\begin{figure}[h!]               
    \captionsetup{justification=centering,margin=0cm}
    \centering
    \vspace{-0.6cm}
    \begin{subfigure}[t]{0.28\textwidth}
        \includegraphics[width=\textwidth, trim=1.0cm 1.0cm 3.4cm 0.8cm,clip]{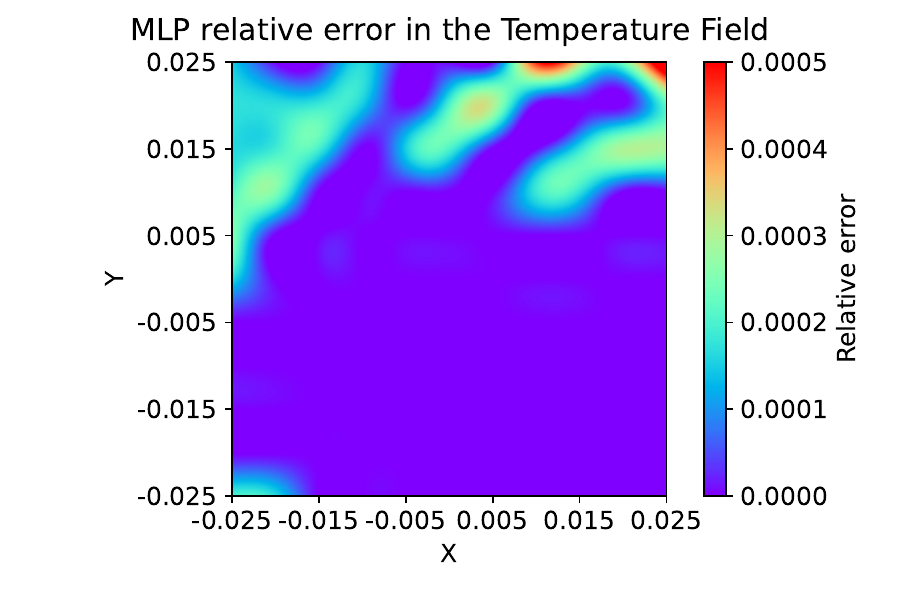}
        \caption{MLP relative error in random sample.}
        \label{fig:MLP_rel_ori}
    \end{subfigure}        
    \hfill
    \begin{subfigure}[t]{0.246\textwidth}
        \includegraphics[width=\textwidth,trim=2.3cm 1.0cm 3.4cm 0.8cm,clip]{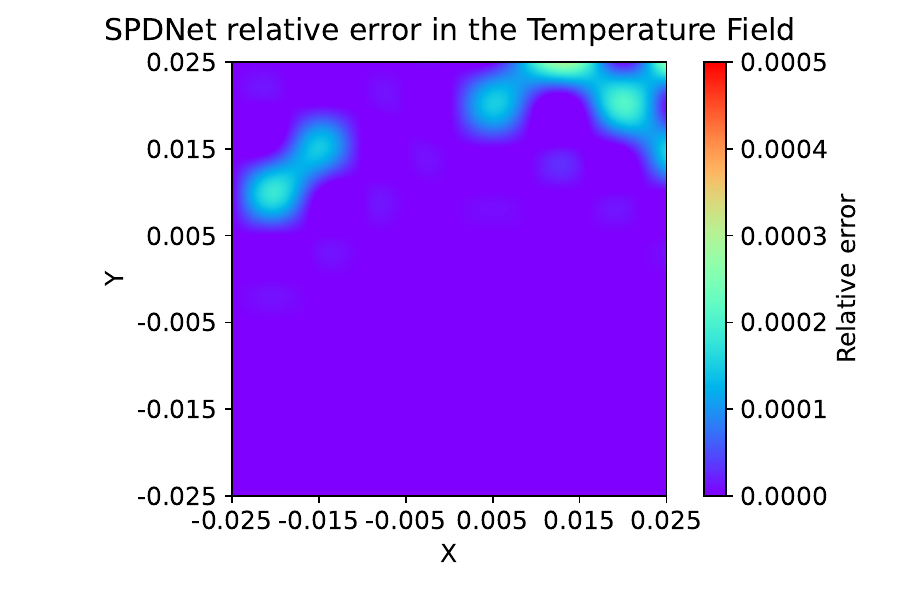}
        \caption{SPDNet relative error in random sample}
        \label{fig:SPD_rel_ori}
    \end{subfigure}    
    \hfill
    \begin{subfigure}[t]{0.328\textwidth}
        \includegraphics[width=\textwidth,trim=2.3cm 1.0cm 0.5cm 0.8cm,clip]{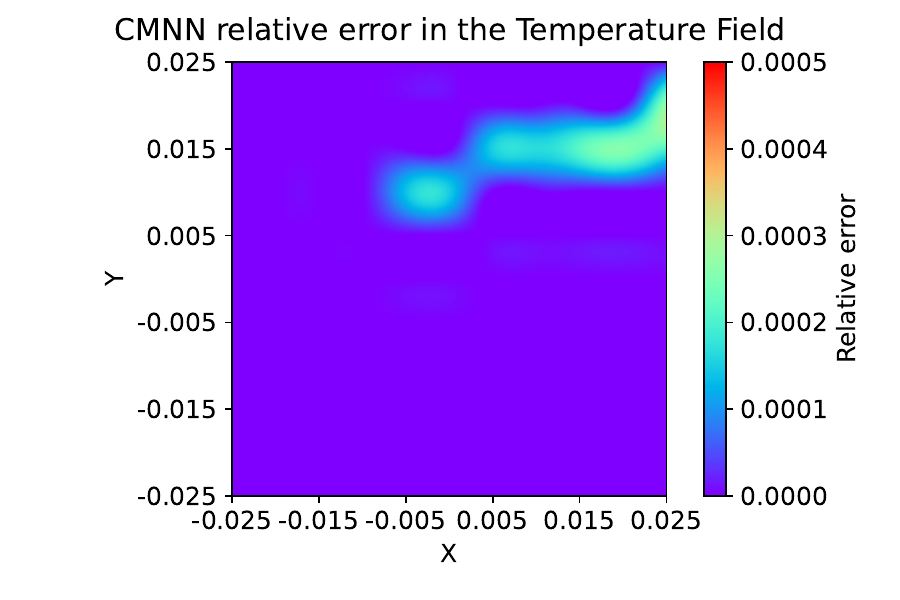}
        \caption{CMNN relative error in random sample.}
        \label{fig:CMNN_rel_ori}
    \end{subfigure}    
    \hfill
    
    \begin{subfigure}[t]{0.3\textwidth}        
        \includegraphics[width=\textwidth, trim=1.0cm 1.0cm 2.95cm 0.8cm,clip]{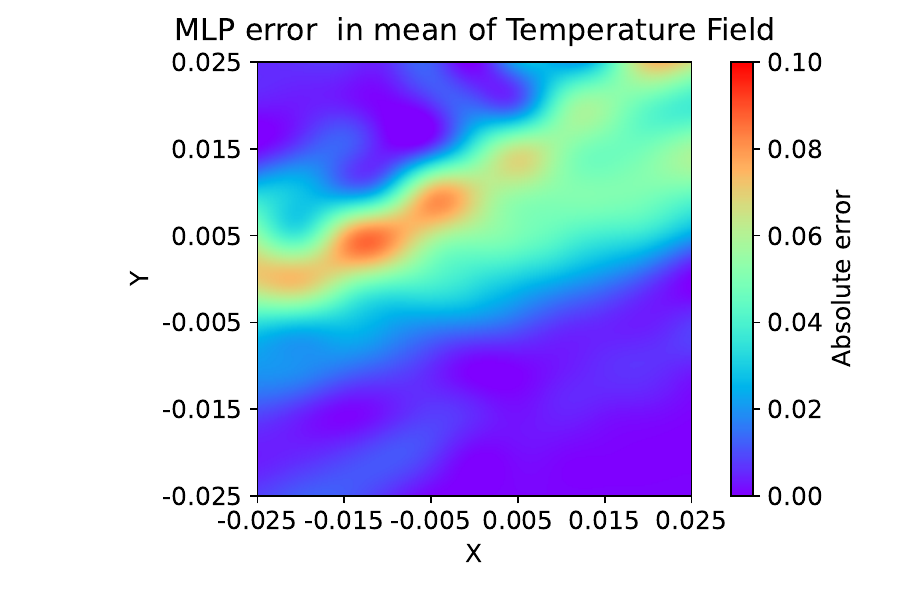}
        \caption{MLP absolute error in the mean.}
        \label{fig:MLP_mean_abs_ori}
    \end{subfigure}
    \hfill
    \begin{subfigure}[t]{0.257\textwidth}        
        \includegraphics[width=\textwidth,trim=2.6cm 1.0cm 2.95cm 0.8cm,clip]{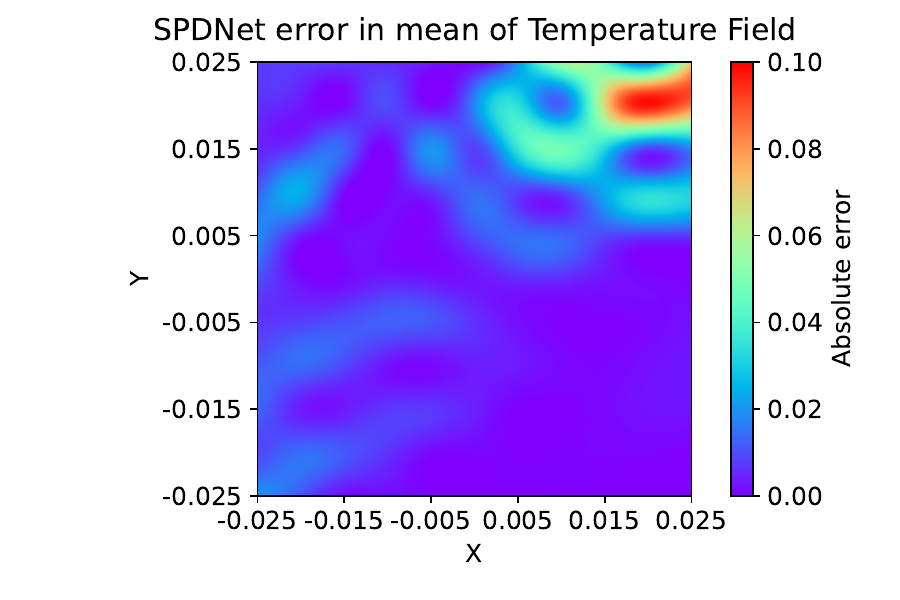}
        \caption{SPDNet absolute error in mean.}
        \label{fig:SPD_mean_abs_ori}
    \end{subfigure}    
    \hfill
    \begin{subfigure}[t]{0.33\textwidth}
        \includegraphics[width=\textwidth,trim=2.6cm 1.0cm 0.5cm 0.8cm,clip]{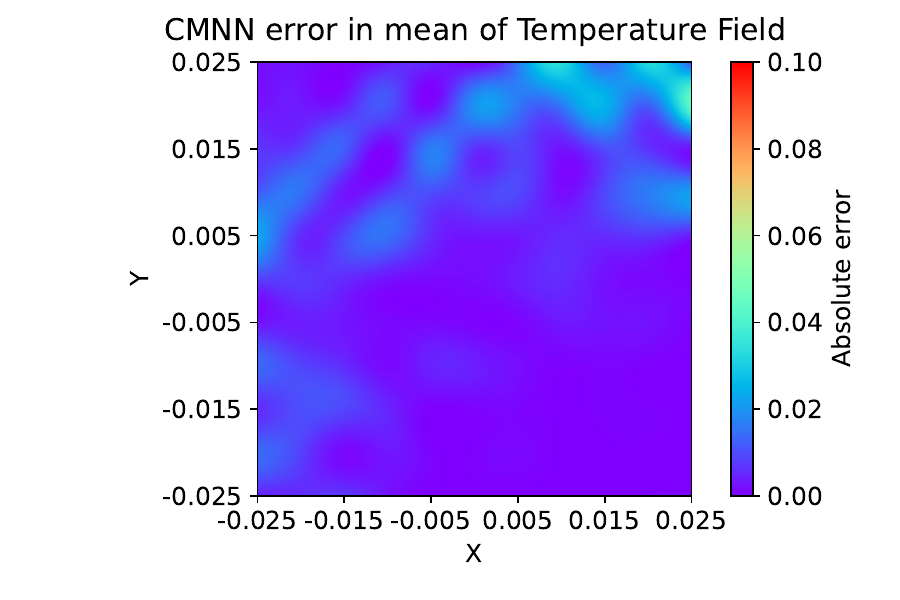}
        \caption{CMNN absolute error in mean.}
        \label{fig:CMNN_mean_abs_ori}
    \end{subfigure}      
    \hfill
    
    \begin{subfigure}[t]{0.27\textwidth}
        \includegraphics[width=\textwidth, trim=1.0cm 1.0cm 3.85cm 0.8cm,clip]{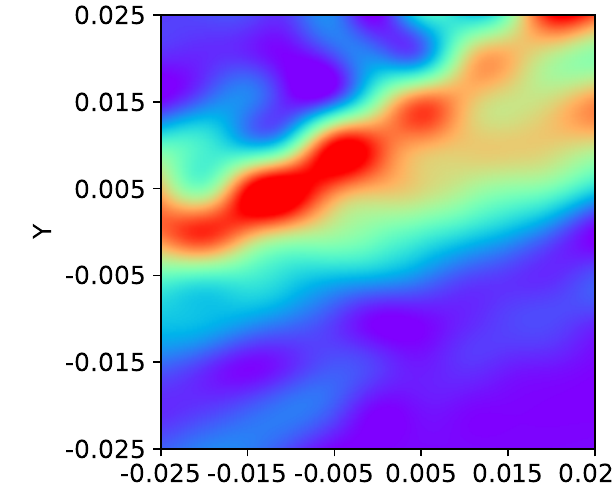}
        \caption{MLP relative error in the mean.}
        \label{fig:MLP_mean_rel_ori}
    \end{subfigure}    
    \hfill
    \begin{subfigure}[t]{0.247\textwidth}        \includegraphics[width=\textwidth,trim=2.0cm 1.0cm 3.85cm 0.8cm,clip]{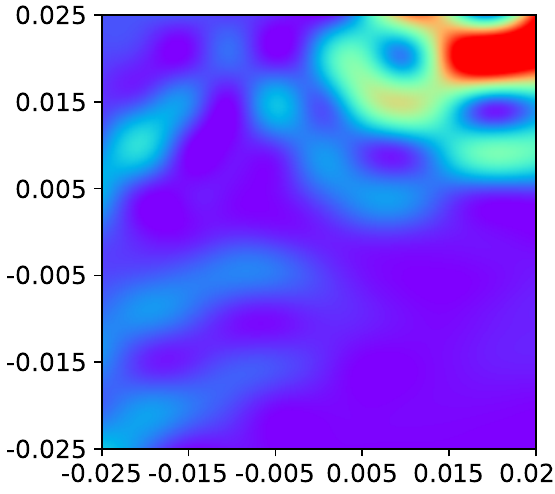}
        \caption{SPDNet relative error in the mean.}
        \label{fig:SPD_mean_rel_ori}
    \end{subfigure}
    \hfill
    \begin{subfigure}[t]{0.328\textwidth}        \includegraphics[width=\textwidth,trim=2.0cm 1.0cm 0.5cm 0.8cm,clip]{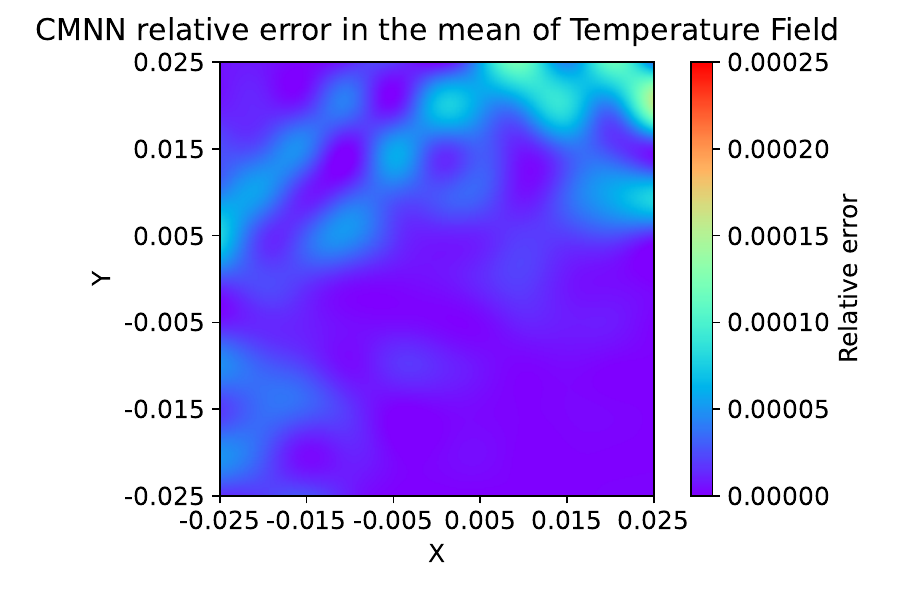}
        \caption{CMNN relative error in the mean.}
        \label{fig:CMNN_mean_rel_ori}
    \end{subfigure}   
    
    
    \begin{subfigure}[t]{0.28\textwidth}        
        \includegraphics[width=\textwidth, trim=1.0cm 1.0cm 3.4cm 0.8cm,clip]{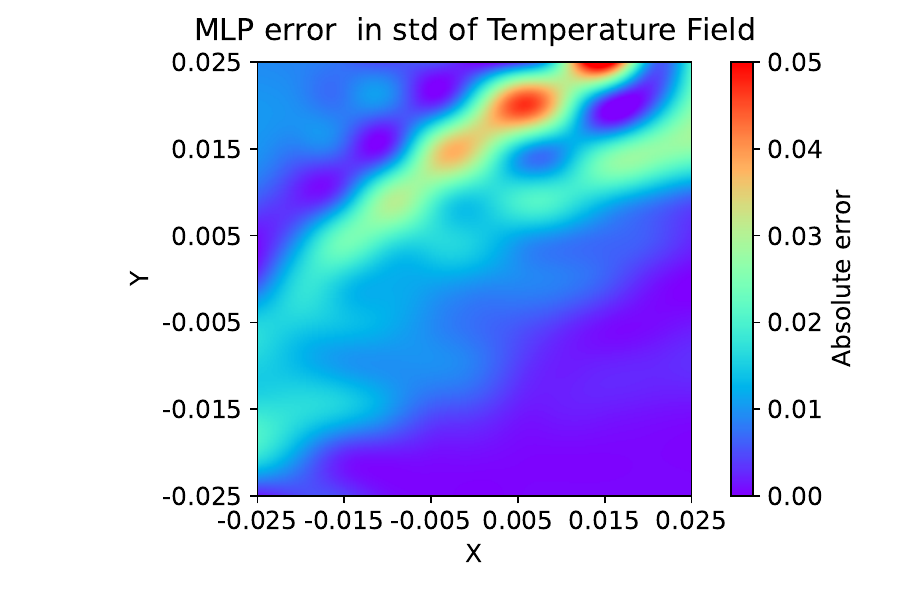}
        \caption{MLP absolute error in the standard deviation.}
        \label{fig:MLP_std_abs_ori}
    \end{subfigure}
    \hfill
    \begin{subfigure}[t]{0.247\textwidth}        
        \includegraphics[width=\textwidth,trim=2.3cm 1.0cm 3.4cm 0.8cm,clip]{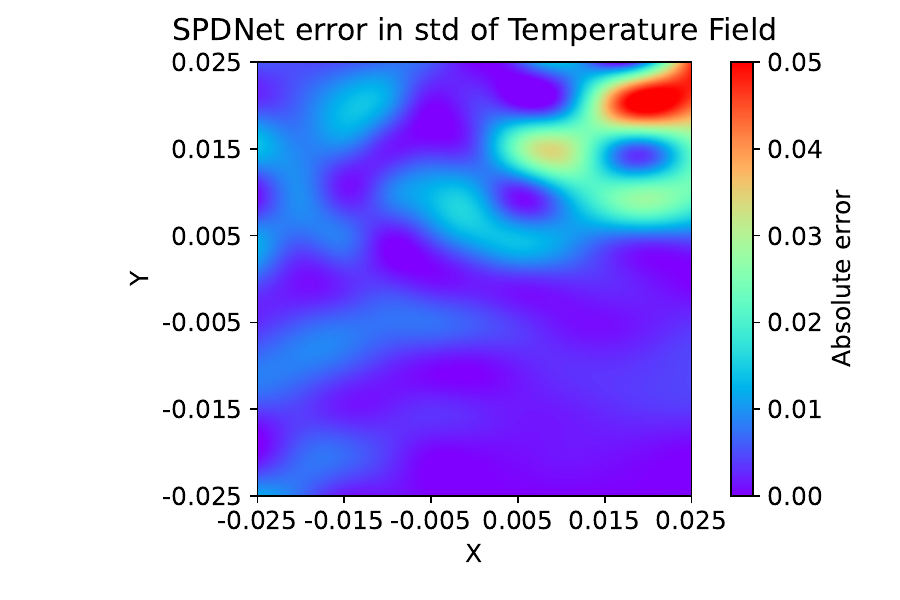}
        \caption{SPDNet absolute error in the standard deviation.}
        \label{fig:SPD_std_abs_ori}
    \end{subfigure}    
    \hfill
    \begin{subfigure}[t]{0.328\textwidth}        
        \includegraphics[width=\textwidth,trim=2.3cm 1.0cm 0.5cm 0.8cm,clip]{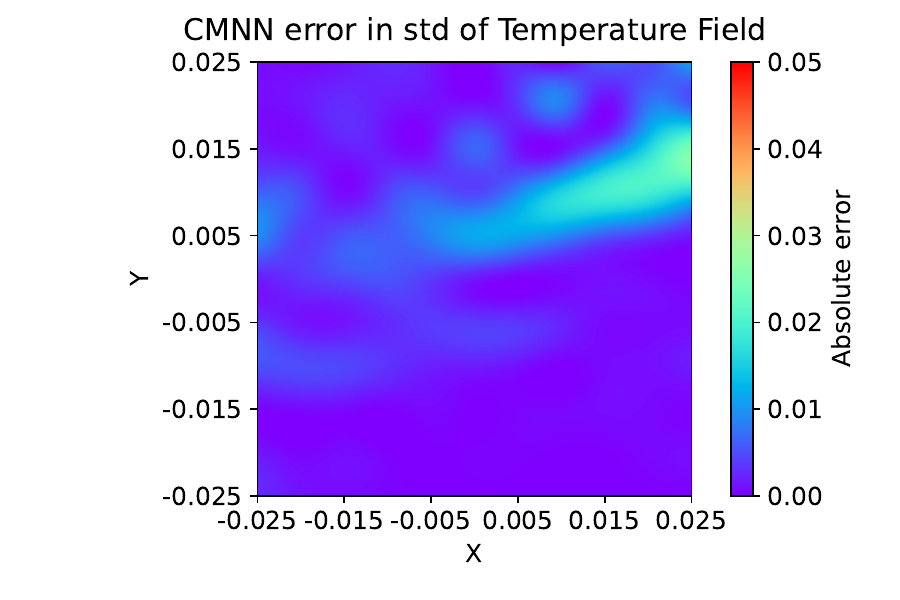}
        \caption{CMNN absolute error in the standard deviation.}
        \label{fig:CMNN_std_abs_ori}
    \end{subfigure}
    \vspace{0.75cm}
    \caption{Errors in approximation of the temperature field by the neural networks with the orientation uncertainty dataset.}
    \label{fig:errors_ori_1}
\end{figure}


To train the networks, the optimised hyper-parameters were 500 epochs, a batch size of 32, hidden layers of sizes 32 and 8, a sigmoid activation function, and a learning rate of $7.5\times e^{-3}$. Table \ref{tab:ori_results} summarises the training performance of the MLP, SPDNet and CMNN architectures on the random-orientation data set. Although the performance gaps are smaller than in the scaling-uncertainty case, the MLP again shows slightly lower consistency and accuracy. SPDNet exhibits the smallest variation, particularly in the validation loss, whereas CMNN attains the lowest test and training losses. All three models perform consistently across the ten repeated experiments. The normalised per-sample norms in Table \ref{tab:ori_norm_results} show that CMNN consistently outperforms SPDNet and, in particular, the MLP. In \refFIG{\ref{fig:his_ori}}, a similar phenomenon to that seen in the scaling-uncertainty training histories appears: the MLP displays the most erratic behaviour, SPDNet is smoother, and CMNN remains the most stable, although it requires more epochs to converge.

\begin{figure}[t!]               
    \captionsetup{justification=centering,margin=0cm}
    \centering
    \vspace{-0.6cm}
    \begin{subfigure}[t]{0.28\textwidth}
        \includegraphics[width=\textwidth, trim=1.0cm 1.0cm 3.4cm 0.8cm,clip]{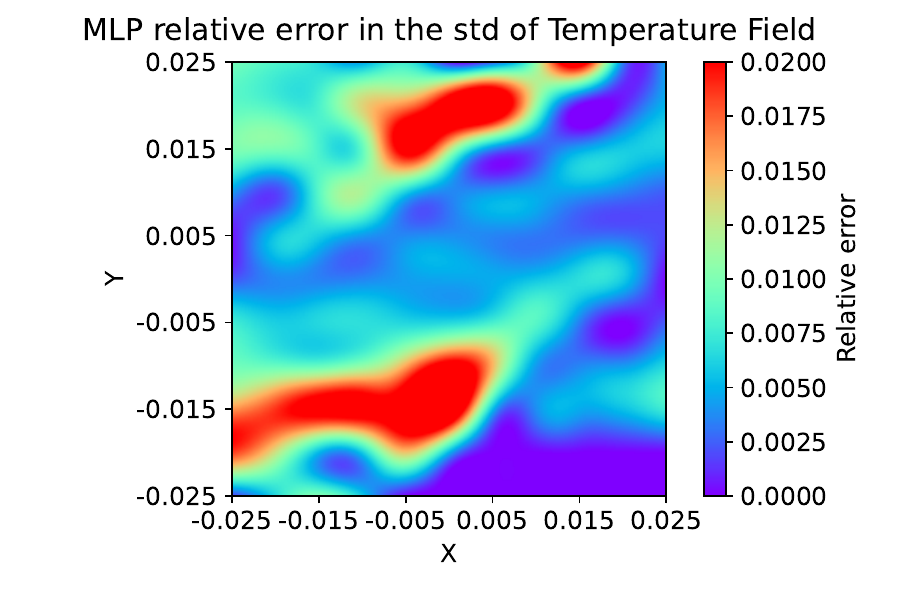}
        \caption{MLP relative error in the standard deviation.}
        \label{fig:MLP_std_rel_ori}
    \end{subfigure}    
    \hfill
    \begin{subfigure}[t]{0.247\textwidth}        \includegraphics[width=\textwidth,trim=2.3cm 1.0cm 3.4cm 0.8cm,clip]{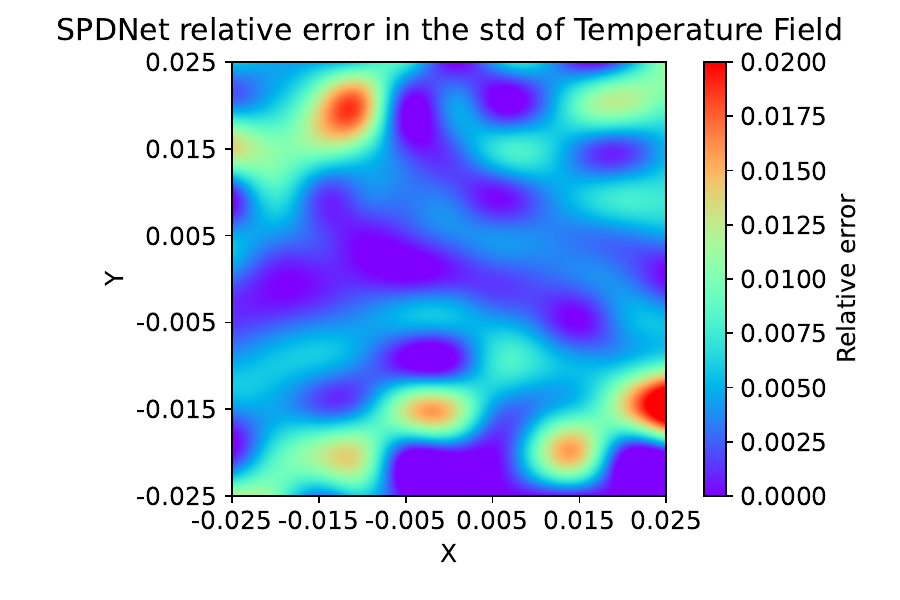}
        \caption{SPDNet relative error in the standard deviation.}
        \label{fig:SPD_std_rel_ori}
    \end{subfigure}
    \hfill
    \begin{subfigure}[t]{0.328\textwidth}        \includegraphics[width=\textwidth,trim=2.3cm 1.0cm 0.5cm 0.8cm,clip]{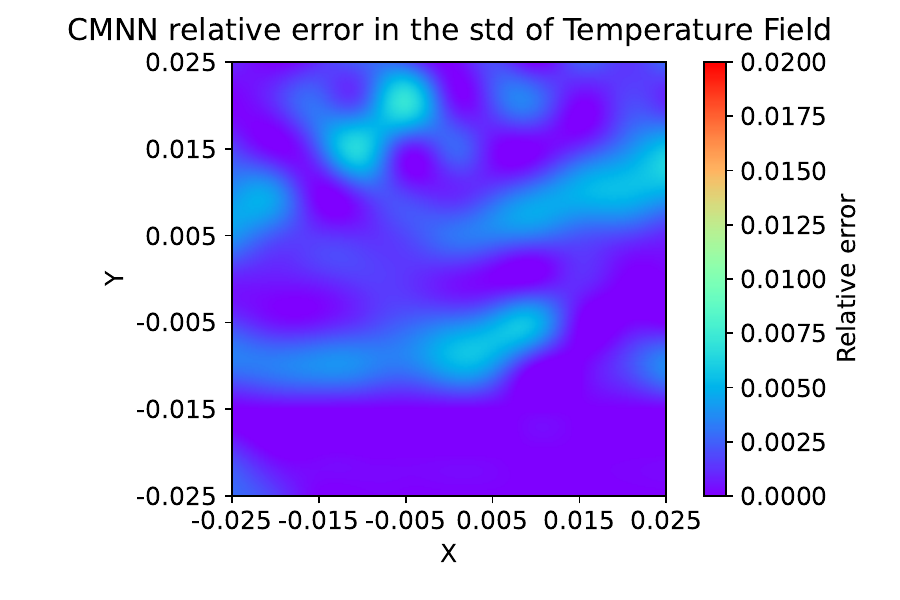}
        \caption{CMNN relative error in the standard deviation.}
        \label{fig:CMNN_std_rel_ori}
    \end{subfigure}    
    \hfill
    
    \begin{subfigure}[t]{0.285\textwidth}
        \includegraphics[width=\textwidth, trim=1.5cm 0.0cm 2.95cm 0.8cm,clip]{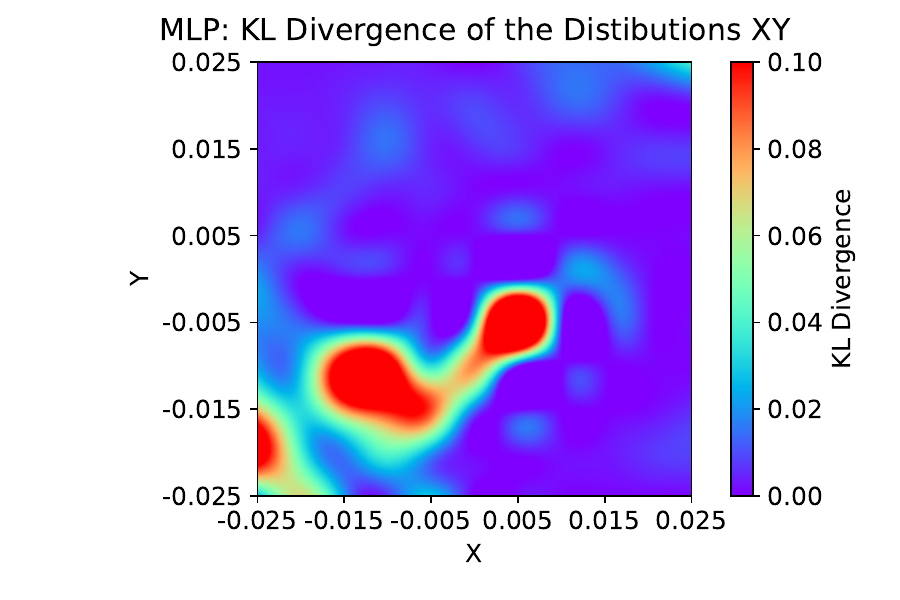}
        \caption{KLD MLP}
        \label{fig:MLP_KLD_ori}
    \end{subfigure}    
    \hfill
    \begin{subfigure}[t]{0.247\textwidth}        
    \includegraphics[width=\textwidth,trim=2.8cm 0.0cm 2.95cm 0.8cm,clip]{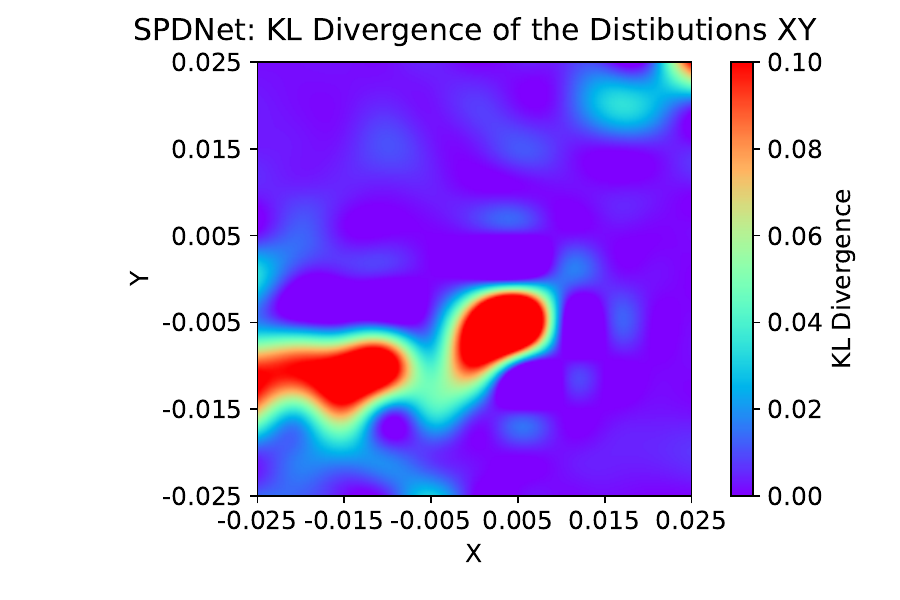}
        \caption{KLD SPDNet}
        \label{fig:SPD_KLD_ori}
    \end{subfigure} 
    \hfill    
    \begin{subfigure}[t]{0.31\textwidth}        \includegraphics[width=\textwidth,trim=2.8cm 0.0cm 0.5cm 0.8cm,clip]{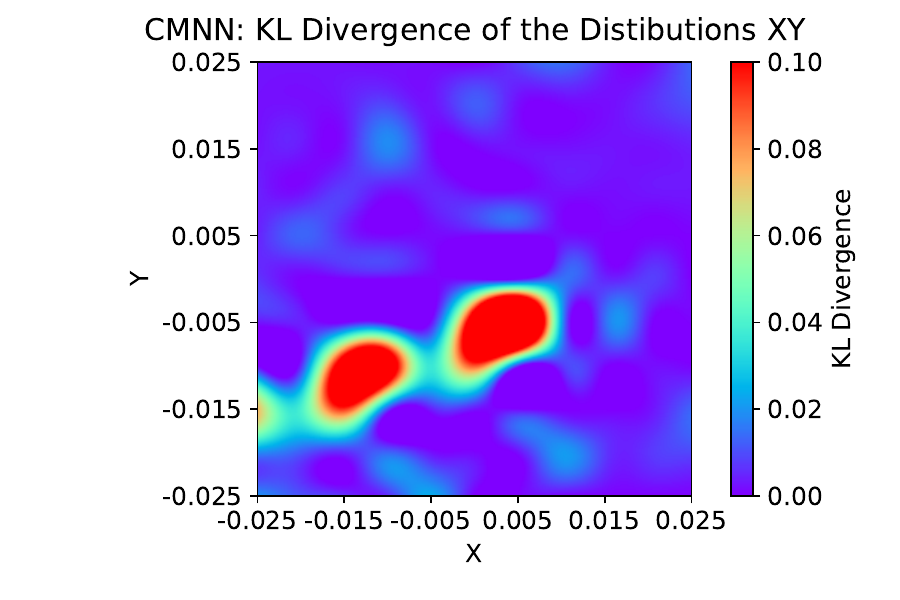}
        \caption{KDL CMNN}
        \label{fig:CMNN_KLD_ori}
    \end{subfigure} 
     \vspace{-0.25cm}    
    \caption{Errors in approximation of the temperature field by the neural networks with the orientation uncertainty dataset.}   
    \label{fig:errors_ori_2}
     \vspace{1cm}
\end{figure}

\refFIG{\ref{fig:errors_ori_1}-\ref{fig:errors_ori_2}} illustrates the errors in approximating the temperature field under orientation uncertainty for the MLP, SPDNet and CMNN architectures. The relative errors in \refFIG{\ref{fig:MLP_rel_ori}} remain generally low for all three models, with only minor local variations. The absolute and relative errors in the mean temperature field—shown in \refFIG{\ref{fig:MLP_mean_abs_ori}–\ref{fig:CMNN_mean_rel_ori}}—and in the standard deviation—depicted in \refFIG{\ref{fig:MLP_std_abs_ori}–\ref{fig:CMNN_std_rel_ori}}—reveal that the MLP produces the largest errors along the high-temperature diagonal, while SPDNet exhibits noticeable mean and standard-deviation errors in the upper-right corner. The final row, focusing on the Kullback–Leibler divergence in \refFIG{\ref{fig:MLP_KLD_ori}–\ref{fig:CMNN_KLD_ori}}, shows elevated values for all models in regions where the probability-density functions have high kurtosis, as seen in \refFIG{\ref{fig:max_angle_YZ}}.

Overall, CMNN displays the most stable and consistent training behaviour under orientation uncertainty, followed by SPDNet and, lastly, the MLP.

\subsubsection{Surrogate model for the combined scaling-orientation uncertainty}

\begin{figure}[!t] 
\vspace{-1cm}
    \captionsetup{justification=centering,margin=1cm}
    \centering
    \hfill
    \begin{subfigure}[t]{0.3\textwidth}
        \includegraphics[width=\textwidth, trim=2.3cm 0.5cm 0.7cm 0.9cm,clip]{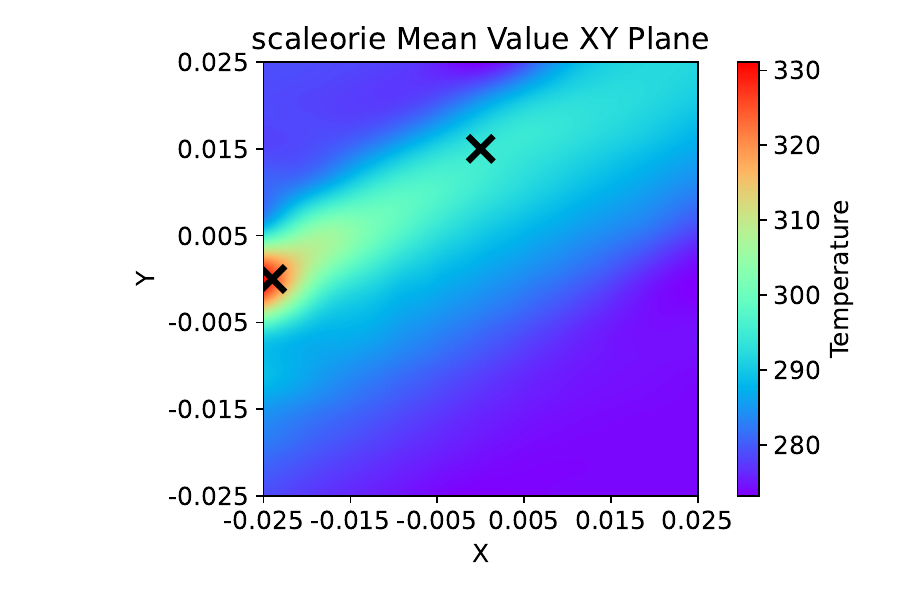}
        \caption{Mean temperature field with markers.}
        \label{fig:ref_mean_sclori_YZ}
    \end{subfigure}    
    \hfill
    \begin{subfigure}[t]{0.3\textwidth}        
        \includegraphics[width=\textwidth, trim=2.3cm 0.5cm 0.7cm 0.9cm,clip]{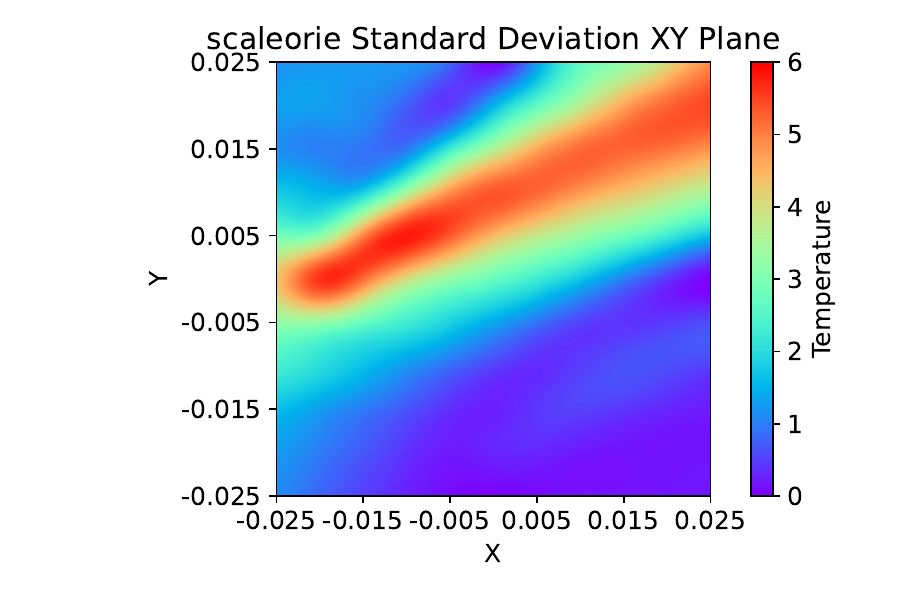}
        \caption{Standard deviation over the temperature field.}
        \label{fig:ref_std_sclori_YZ}
    \end{subfigure} 
    \hfill    
    \begin{subfigure}[t]{0.33\textwidth}       
        \includegraphics[width=\textwidth, trim=0.3cm 0.5cm 0.7cm 1.3cm,clip]{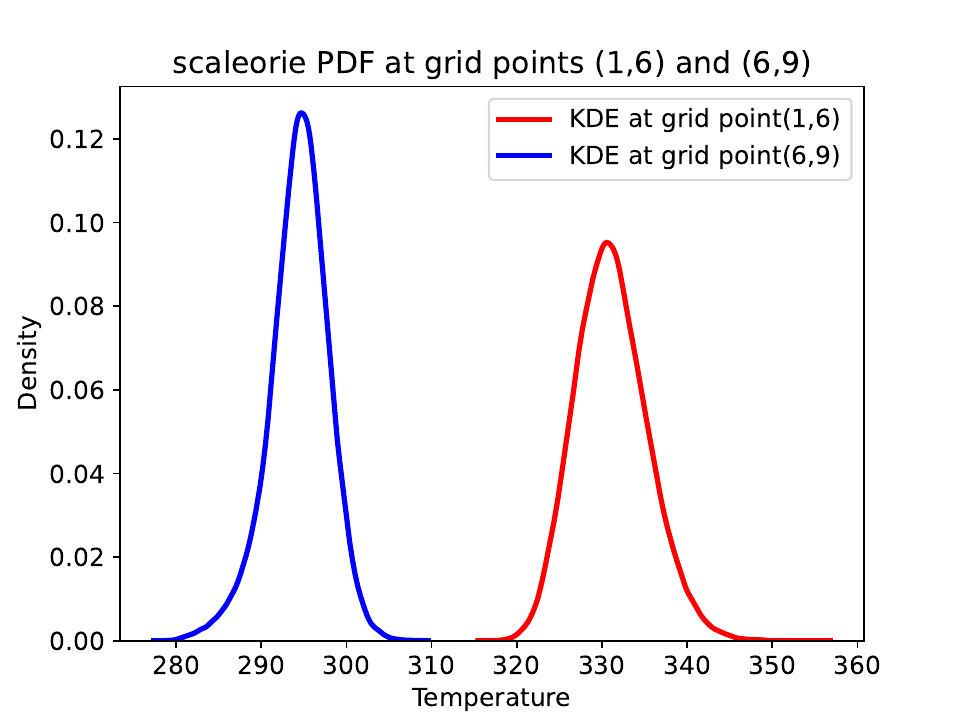}
        \caption{Probability density functions for the points marked on the mean plot.}
        \label{fig:ref_dist_sclori_YZ}
    \end{subfigure} 
    \hfill    
    \vspace{0.25cm}
    \caption{Temperature field statistics in the XY plane with the combined scaling-orientation uncertainty dataset.}
    \label{fig:ref_sclori}
\end{figure}

\begin{figure}[!t]    
    \captionsetup{justification=centering}
    \centering
    \begin{subfigure}[b]{0.41\textwidth}
        \includegraphics[width=\textwidth,height = 0.5\textwidth, trim=0.35cm 0cm 0.25cm 1.3cm,clip]{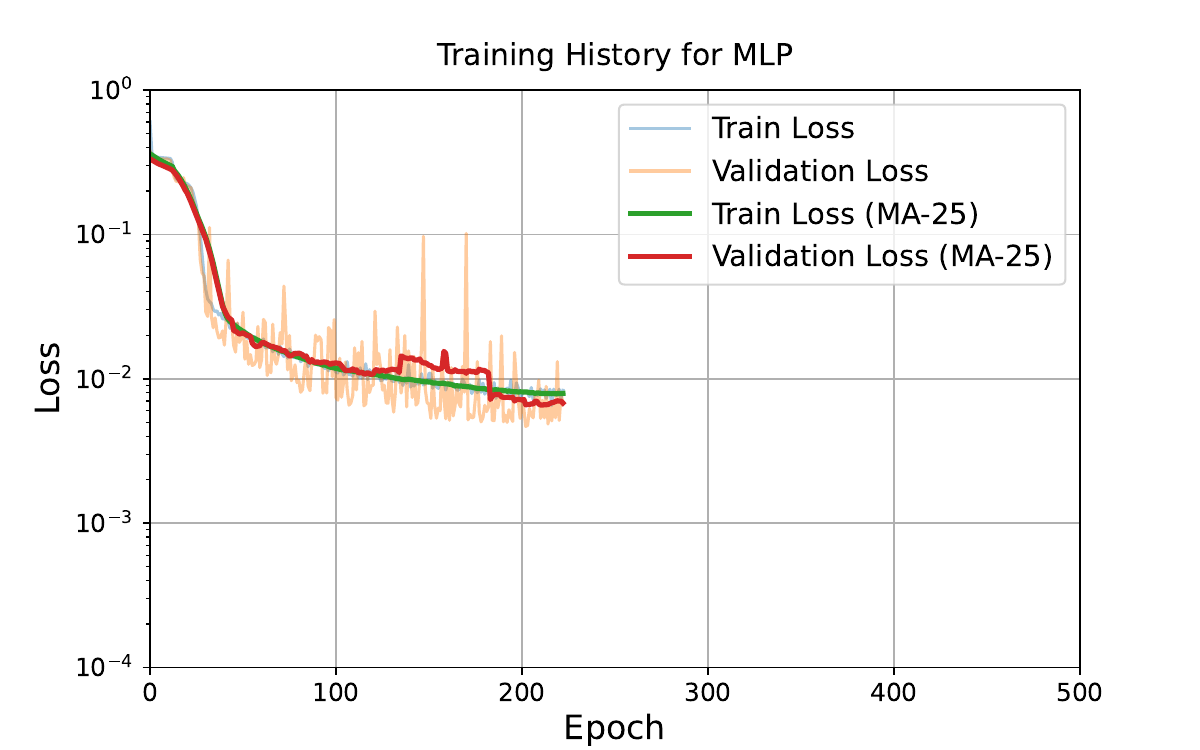}
        \caption{MLP}
        \label{fig:his_sclori_MLP}
    \end{subfigure}
    \begin{subfigure}[b]{0.41\textwidth}
        \includegraphics[width=\textwidth,height = 0.5\textwidth,trim=0.35cm 0cm 0.25cm 1.3cm,clip]{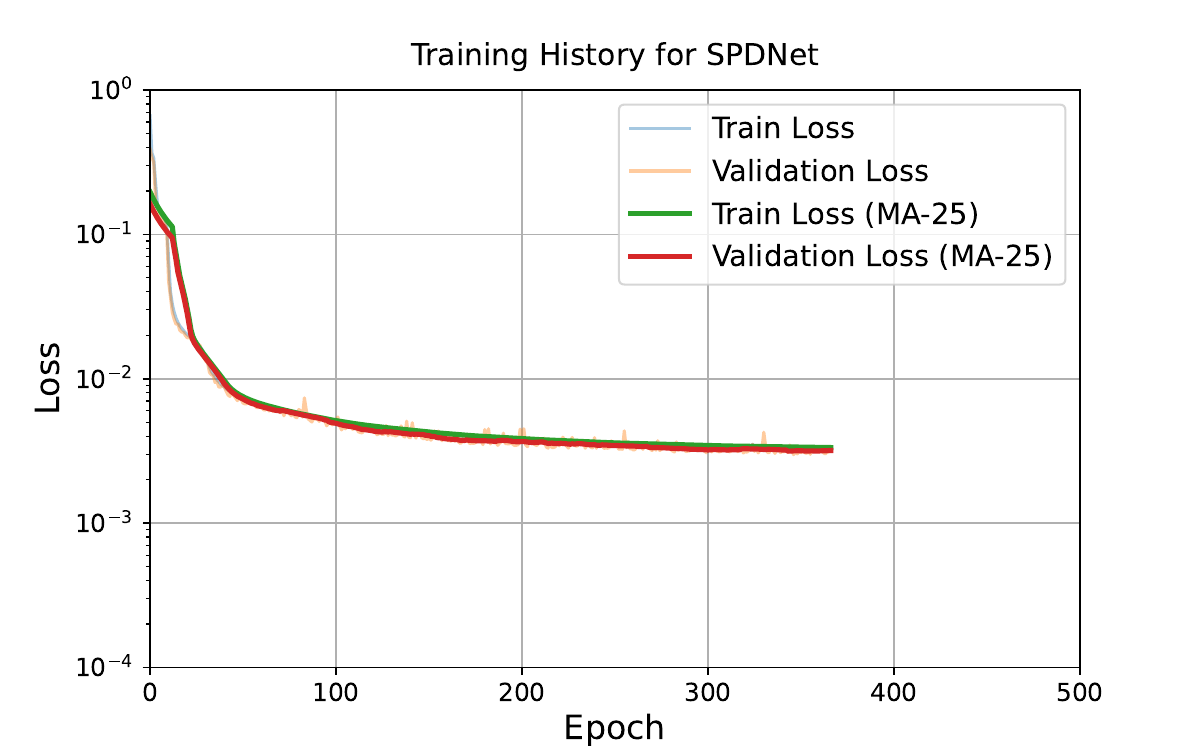}
        \caption{SPDNet}
        \label{fig:his_sclori_SPD}
    \end{subfigure}
    \vspace{0.25cm}    
    
    \begin{subfigure}[b]{0.41\textwidth}
        \includegraphics[width=\textwidth,height = 0.5\textwidth,trim=0.35cm 0cm 0.25cm 1.3cm,clip]{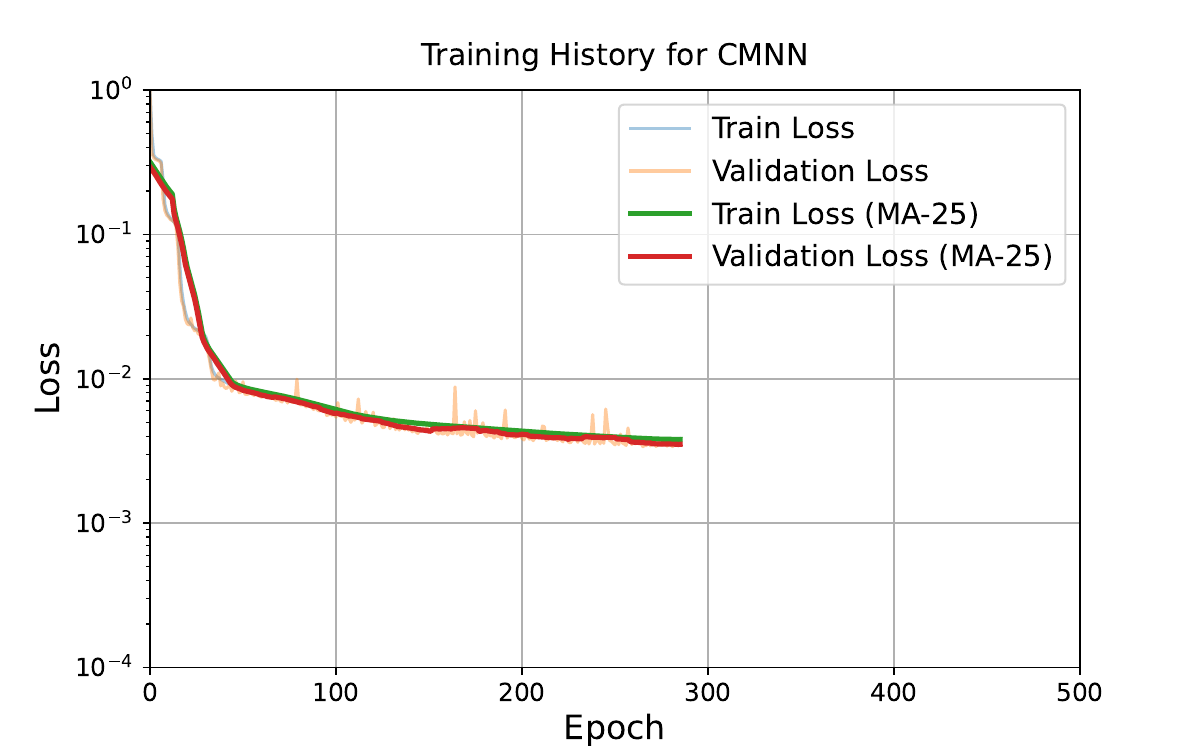}
        \caption{CMNN}
        \label{fig:his_sclori_CMNN}
    \end{subfigure}    
    \caption{Training history of the neural networks with the combined scaling-orientation uncertainty dataset.}
    \vspace{0.35cm}
    \label{fig:his_sclori}
\end{figure}

\begin{table}[!b]
\centering
\caption{Training metrics of the neural networks with the combined scaling-orientation uncertainty dataset. Data in order of magnitude of $ e^{-3}$.}
\begin{tabularx}{\textwidth}{l|XXX|XXX}
\toprule
Layer & Best Val.\ Loss & Test Loss & Train Loss &
Mean Val.\ Loss (Std.) & Mean Test Loss (Std.) & Mean Train Loss (Std.) \\
\midrule
MLP    & 4.68 & 7.02 & 8.17 & 6.46\,(1.15) & 8.69\,(2.27) & 9.35\,(1.16) \\
SPDnet & 2.98 & 3.42 & 3.34 & 3.43\,(0.21) & 3.45\,(0.27) & 3.47\,(0.23) \\
CMNN   & 3.38 & 3.77 & 3.89 & 3.74\,(0.33) & 3.77\,(0.25) & 3.83\,(0.21) \\
\bottomrule
\end{tabularx}
\label{tab:sclori_results}
\end{table}

\begin{table}[!b]
\centering
\caption{Normalised and per-sample norms of the neural networks with the combined scaling-orientation uncertainty dataset.}
\begin{tabularx}{\textwidth}{l|XXX|XXX|XXX}
\toprule
Model &
L1 Sample $e^{-3}$ & L2 Sample $e^{-5}$ & L$\infty$ Sample $e^{-3}$ &
L1 Mean $e^{-4}$ & L2 Mean $e^{-5}$ & L$\infty$ Mean $e^{-3}$ &
L1 \; Std  $e^{-3}$ & L2 \; Std $e^{-5}$ & L$\infty$ Std $e^{-3}$ \\
\midrule
MLP    & 2.42 & 9.3  & 19.1 & 11.5 & 4.5  & 7.1  & 6.57 & 23.7 & 25.9 \\
SPDNet & 1.83 & 7.5  & 15.7 & 2.94 & 1.1  & 1.86 & 0.94 & 3.5  & 7.36 \\
CMNN   & 2.54 & 10.2 & 18.7 & 2.85 & 1.1  & 1.42 & 1.29 & 5.7  & 14.2 \\
\bottomrule
\end{tabularx}
\label{tab:sclori_norm_results}
\end{table}

\noindent
To test the neural networks under combined scaling and orientation uncertainty, we combine the random-scaling and random-orientation parameters from \refEQ{\ref{eq:random_both}}. These six random variables are detailed in Table \ref{tab:lognormal_variables} and visualised in \refFIG{\ref{fig:VMF_sphere}}. The reference results appear in \refFIG{\ref{fig:ref_sclori}}. The mean temperature field in \refFIG{\ref{fig:ref_mean_sclori_YZ}} closely matches the deterministic solution. The standard deviation in \refFIG{\ref{fig:ref_std_sclori_YZ}} reflects the combined effects of scaling and orientation. Its pattern resembles the sum of the standard deviations from the pure-orientation case (\refFIG{\ref{fig:ref_std_ori_YZ}}) and the pure-scaling case (\refFIG{\ref{fig:ref_std_scl_YZ}}). Finally, \refFIG{\ref{fig:ref_dist_sclori_YZ}} shows the probability-density functions at two marked points; these distributions have greater spread and are nearer to a normal shape, since the combined uncertainty reduces the influence of any single source and mitigates skewness.

\begin{figure}[h!]               
    \captionsetup{justification=centering,margin=0cm}
    \centering
    \vspace{-0.6cm}
    \begin{subfigure}[t]{0.3\textwidth}
        \includegraphics[width=\textwidth, trim=1.0cm 1.0cm 3.4cm 0.8cm,clip]{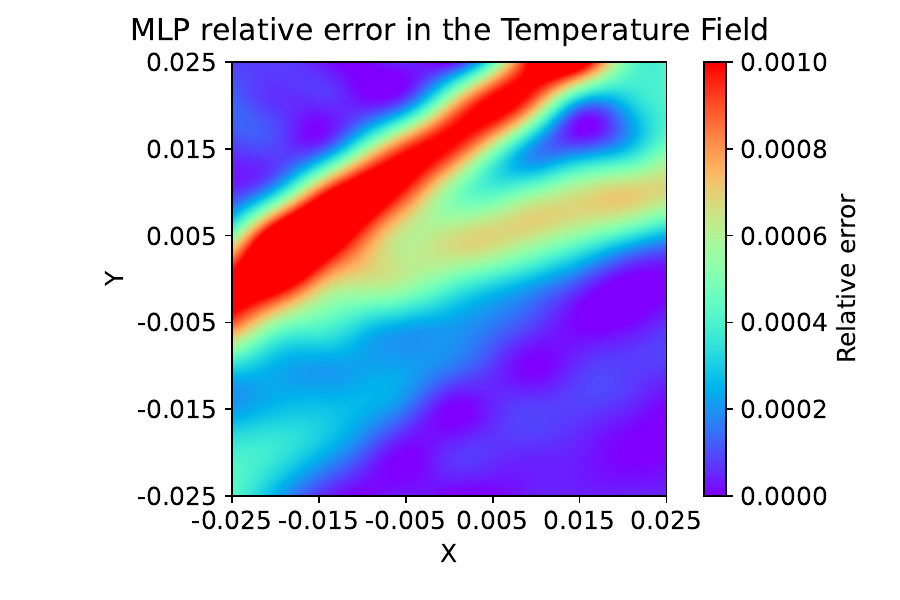}
        \caption{MLP relative error in a random sample.}
        \label{fig:MLP_rel_sclori}
    \end{subfigure}    
    \hspace{0.25cm}
    \begin{subfigure}[t]{0.258\textwidth}        
    \includegraphics[width=\textwidth,trim=2.3cm 1.0cm 3.6cm 0.8cm,clip]{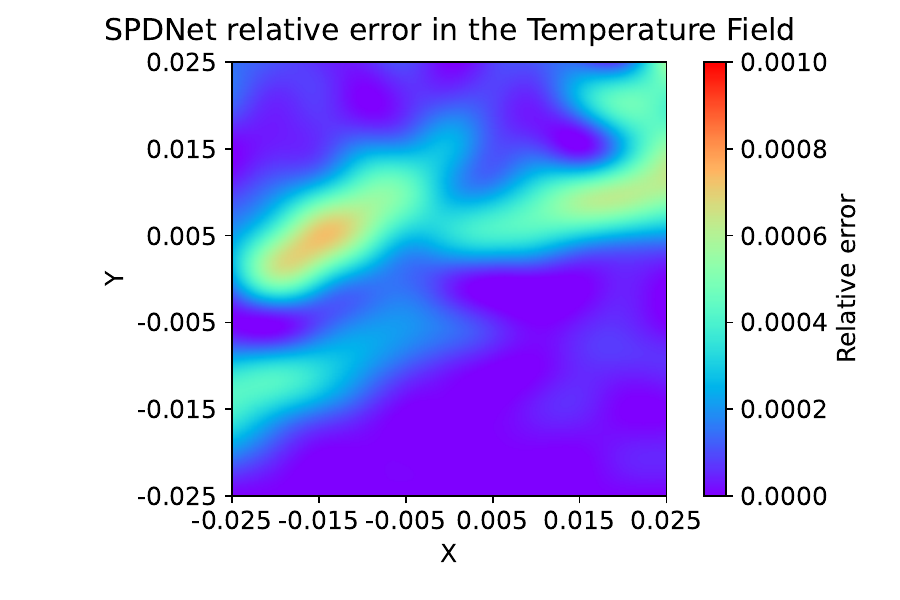}
        \caption{SPDNet relative error in a random sample.}
        \label{fig:SPD_rel_sclori}
    \end{subfigure}
     \hspace{0.25cm}
    \begin{subfigure}[t]{0.34\textwidth}        \includegraphics[width=\textwidth,trim=2.3cm 1.0cm 0.5cm 0.8cm,clip]{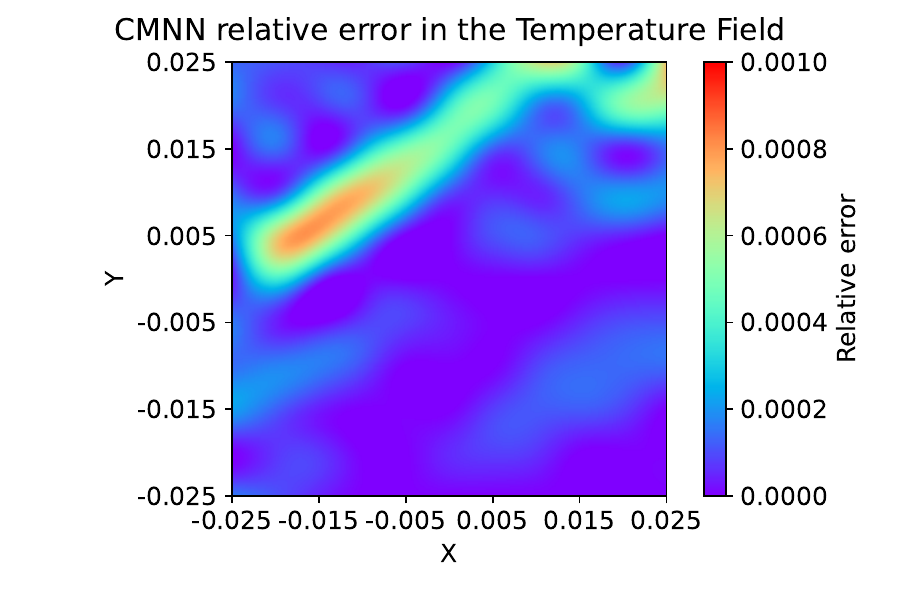}
        \caption{CMNN relative error in a random sample.}
        \label{fig:CMNN_rel_sclori}
    \end{subfigure}    
    \hfill
    
    \begin{subfigure}[t]{0.3\textwidth}
        \includegraphics[width=\textwidth, trim=1.45cm 1.0cm 2.95cm 0.8cm,clip]{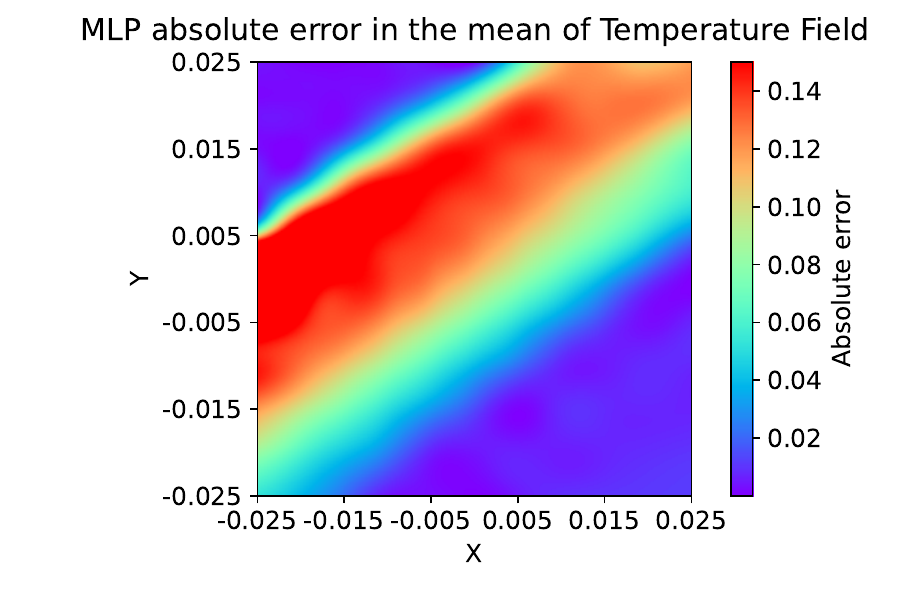}
        \caption{MLP absolute error in the mean.}
        \label{fig:MLP_mean_abs_sclori}
    \end{subfigure}    
    \hspace{0.25cm}
    \begin{subfigure}[t]{0.26\textwidth}        \includegraphics[width=\textwidth,trim=2.6cm 1.0cm 2.95cm 0.8cm,clip]{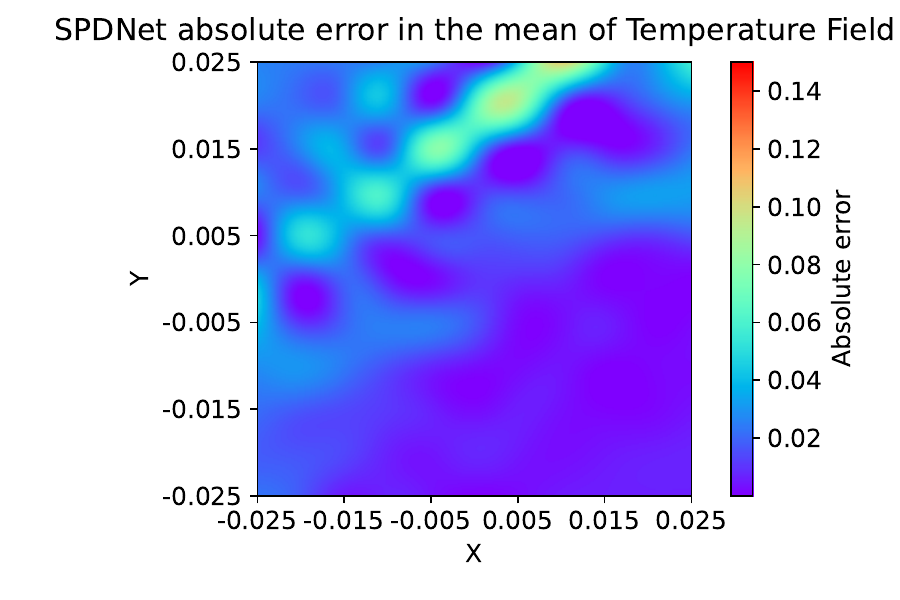}
        \caption{SPDNet absolute error in the mean.}
        \label{fig:SPD_mean_abs_sclori}
    \end{subfigure}
     \hspace{0.25cm}
    \begin{subfigure}[t]{0.34\textwidth}        \includegraphics[width=\textwidth,trim=2.6cm 1.0cm 0.5cm 0.8cm,clip]{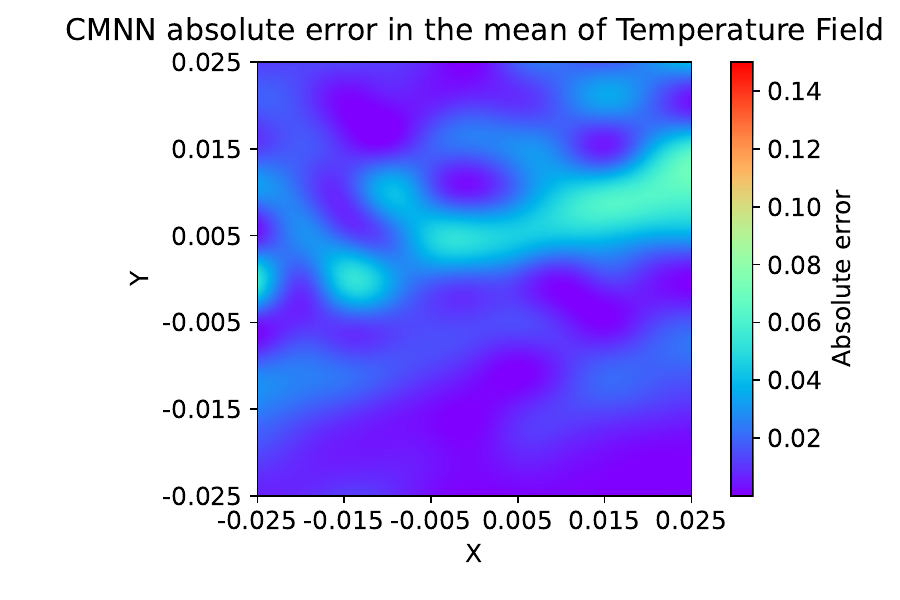}
        \caption{CMNN absolute error in the mean.}
        \label{fig:CMNN_mean_abs_sclori}
    \end{subfigure}    
    \hfill
    
    \begin{subfigure}[t]{0.3\textwidth}
        \includegraphics[width=\textwidth, trim=0.9cm 1.0cm 3.6cm 0.8cm,clip]{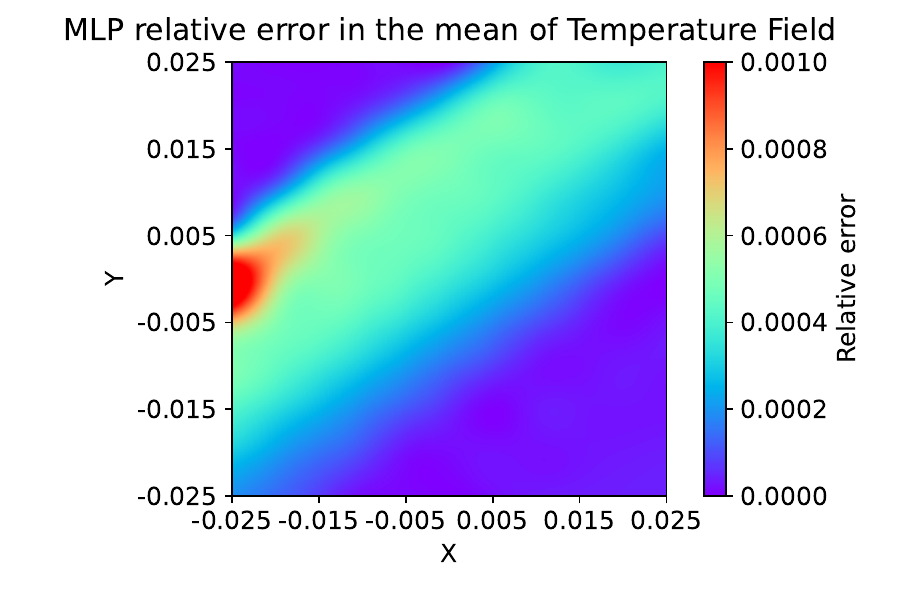}
        \caption{MLP relative error in the mean.}
        \label{fig:MLP_mean_rel_sclori}
    \end{subfigure}    
    \hspace{0.25cm}
    \begin{subfigure}[t]{0.26\textwidth}        \includegraphics[width=\textwidth,trim=2.2cm 1.0cm 3.65cm 0.8cm,clip]{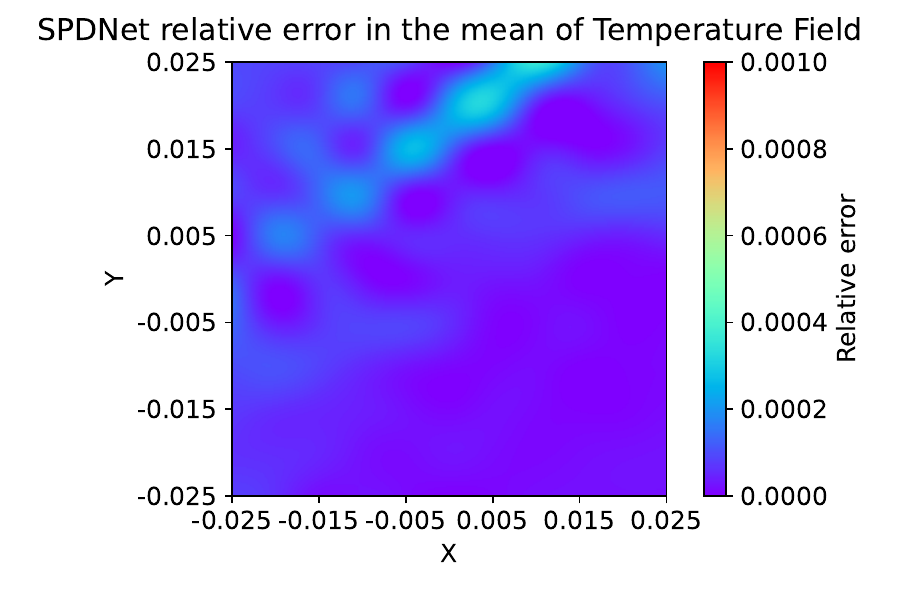}
        \caption{SPDNet relative error in the mean.}
        \label{fig:SPD_mean_rel_sclori}
    \end{subfigure}
     \hspace{0.25cm}
    \begin{subfigure}[t]{0.345\textwidth}        \includegraphics[width=\textwidth,trim=2.2cm 1.0cm 0.5cm 0.8cm,clip]{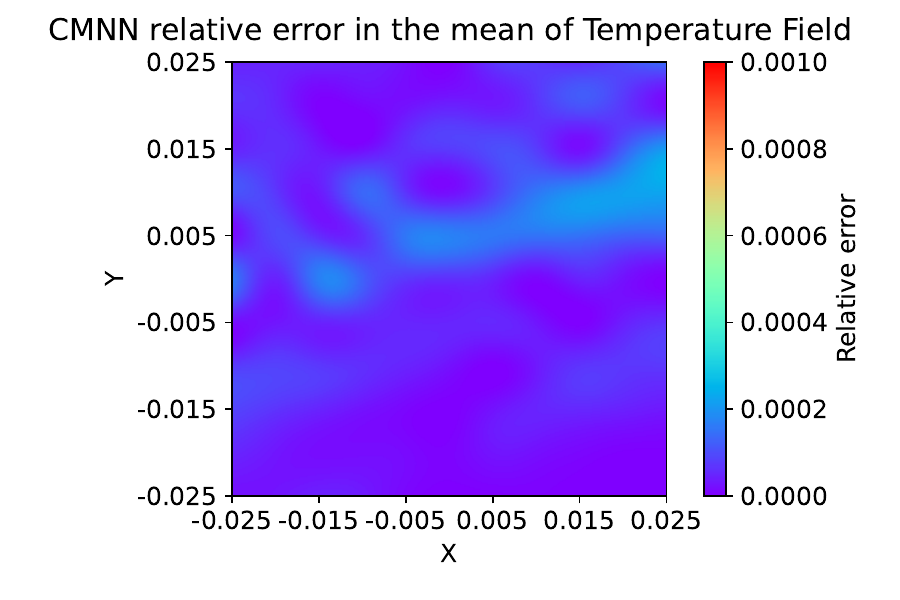}
        \caption{CMNN relative error in the mean.}
        \label{fig:CMNN_mean_rel_sclori}
    \end{subfigure} 
    \hfill
    
    \begin{subfigure}[t]{0.3\textwidth}
        \includegraphics[width=\textwidth, trim=1.3cm 1.0cm 3.0cm 0.8cm,clip]{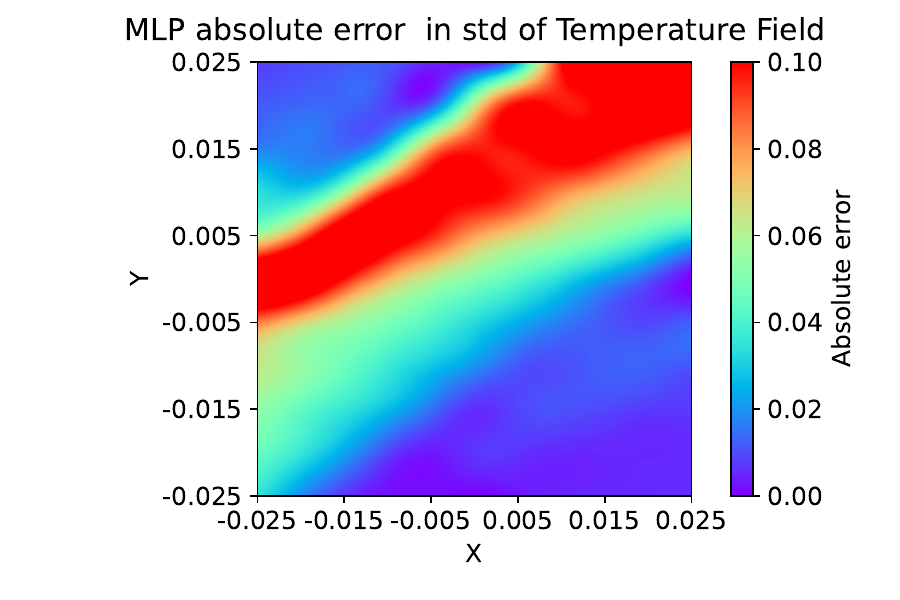}
        \caption{MLP absolute error in the standard deviation.}
        \label{fig:MLP_std_abs_sclori}
    \end{subfigure}    
    \hspace{0.25cm}
    \begin{subfigure}[t]{0.26\textwidth}        \includegraphics[width=\textwidth,trim=2.5cm 1.0cm 3.0cm 0.8cm,clip]{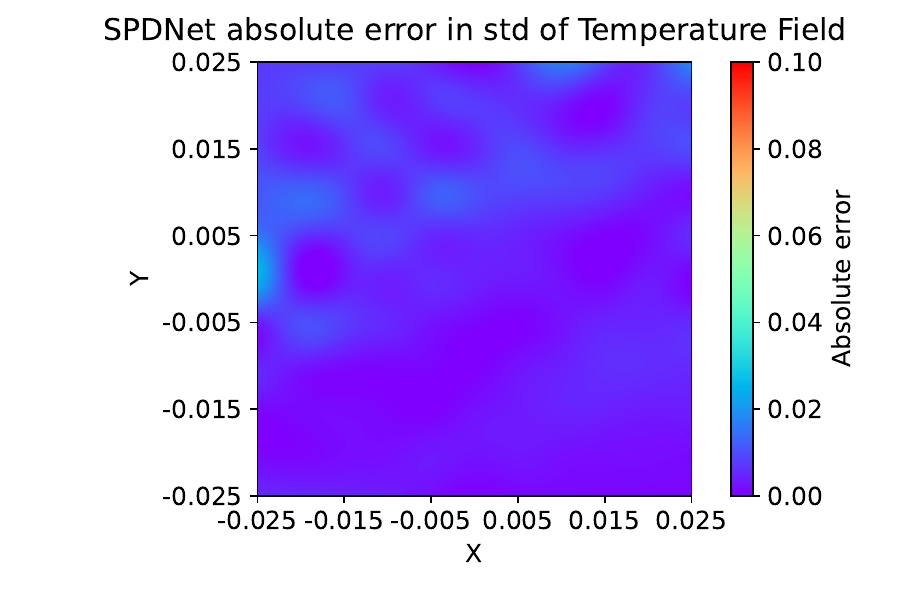}
        \caption{SPDNet absolute error in the standard deviation.}
        \label{fig:SPD_std_abs_sclori}
    \end{subfigure}
     \hspace{0.25cm}
    \begin{subfigure}[t]{0.335\textwidth}        \includegraphics[width=\textwidth,trim=2.4cm 1.0cm 0.5cm 0.8cm,clip]{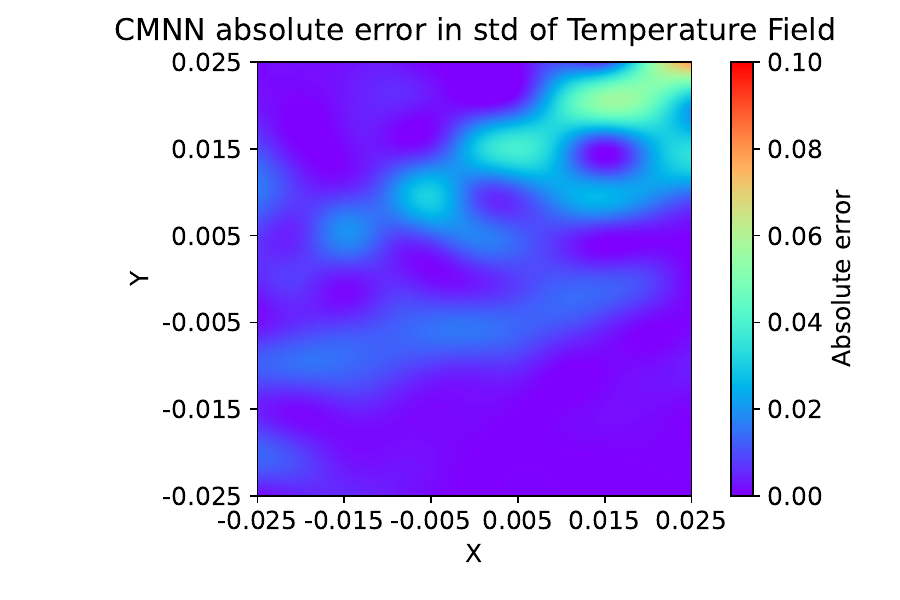}
        \caption{CMNN absolute error in the standard deviation.}
        \label{fig:CMNN_std_abs_sclori}
    \end{subfigure}
    \hfill
    \caption{Errors in approximation of the temperature field by the neural networks with the combined scaling-orientation uncertainty dataset.}
    \label{fig:errors_sclori_1}
\end{figure}

\begin{figure}[h!]               
    \captionsetup{justification=centering,margin=0cm}
    \centering
    \vspace{-0.6cm}
    \begin{subfigure}[t]{0.28\textwidth}
        \includegraphics[width=\textwidth, trim=1.0cm 1.0cm 3.2cm 0.8cm,clip]{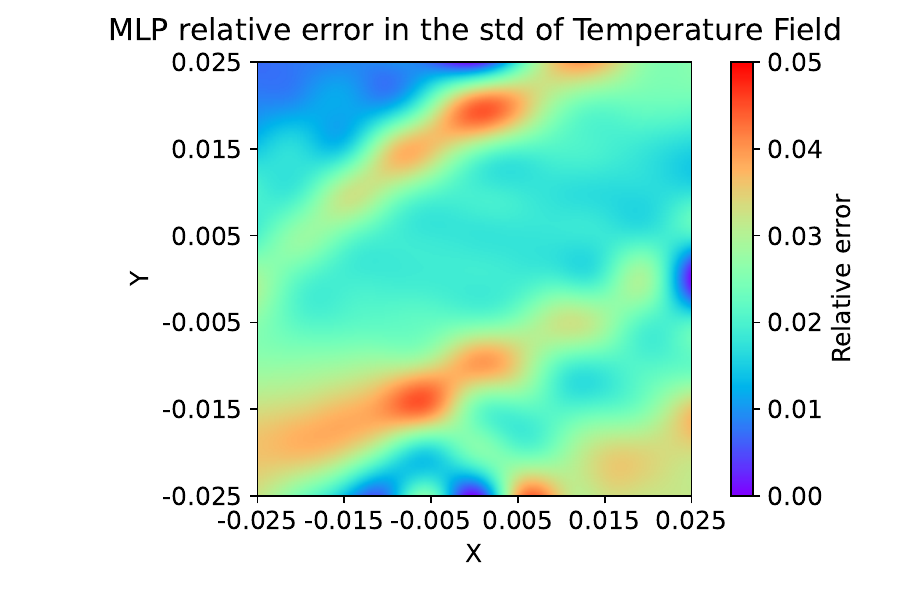}
        \caption{MLP relative error in the standard deviation.}
        \label{fig:MLP_std_rel_sclori}
    \end{subfigure}    
    \hfill
    \begin{subfigure}[t]{0.247\textwidth}        \includegraphics[width=\textwidth,trim=2.3cm 1.0cm 3.2cm 0.8cm,clip]{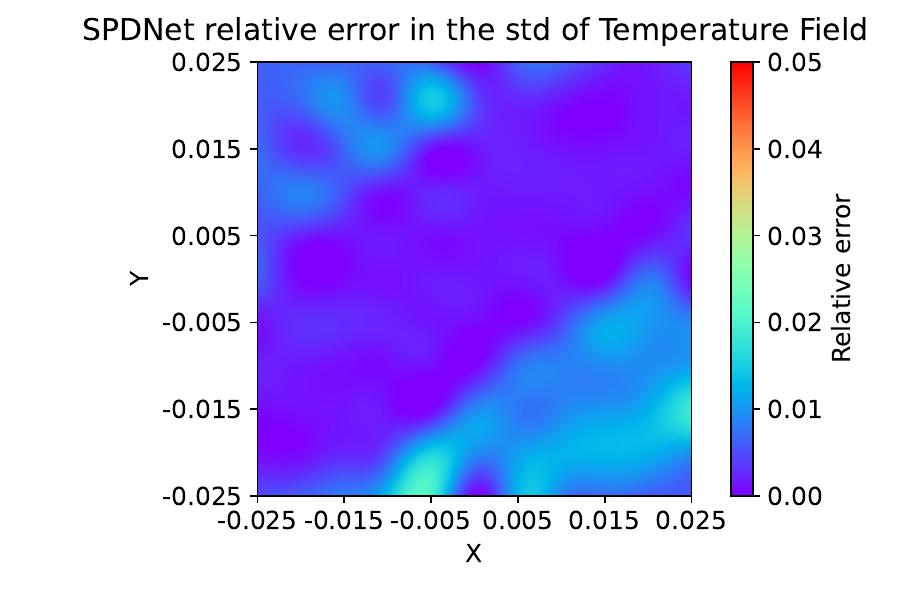}
        \caption{SPDNet relative error in the standard deviation.}
        \label{fig:SPD_std_rel_sclori}
    \end{subfigure}
    \hfill
    \begin{subfigure}[t]{0.328\textwidth}        \includegraphics[width=\textwidth,trim=2.3cm 1.0cm 0.5cm 0.8cm,clip]{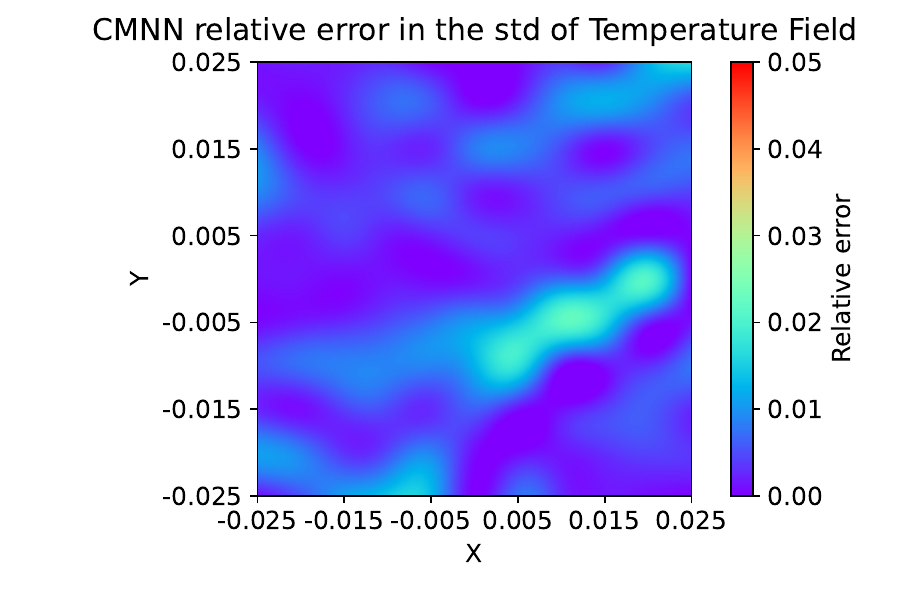}
        \caption{CMNN relative error in the standard deviation.}
        \label{fig:CMNN_std_rel_sclori}
    \end{subfigure}    
    \hfill
    
    \begin{subfigure}[t]{0.28\textwidth}
        \includegraphics[width=\textwidth, trim=1.0cm 0.0cm 3.6cm 0.8cm,clip]{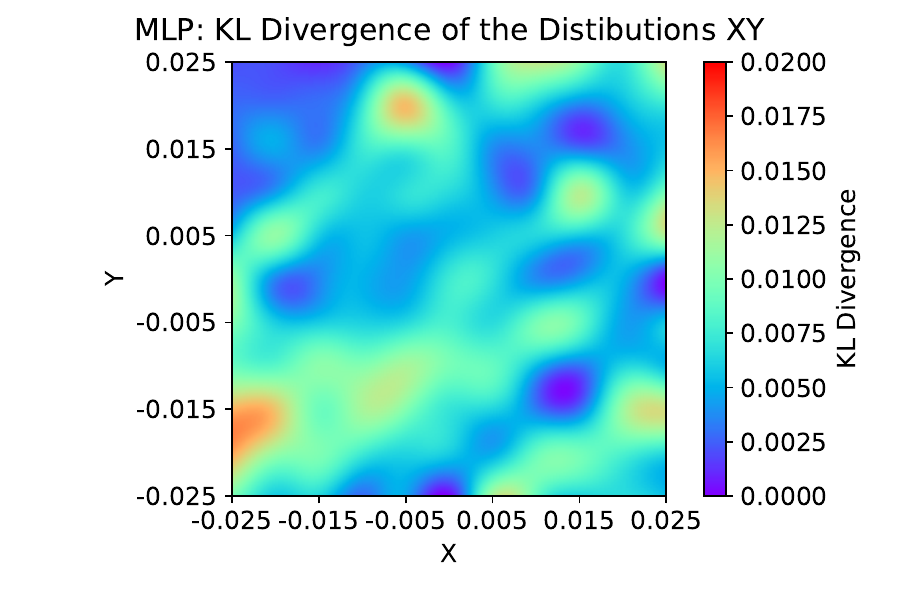}
        \caption{MLP Kullback-Leibler divergence.}
        \label{fig:MLP_KLD_sclori}
    \end{subfigure}  
    \hfill
    \begin{subfigure}[t]{0.247\textwidth}        
    \includegraphics[width=\textwidth,trim=2.3cm 0.0cm 3.6cm 0.8cm,clip]{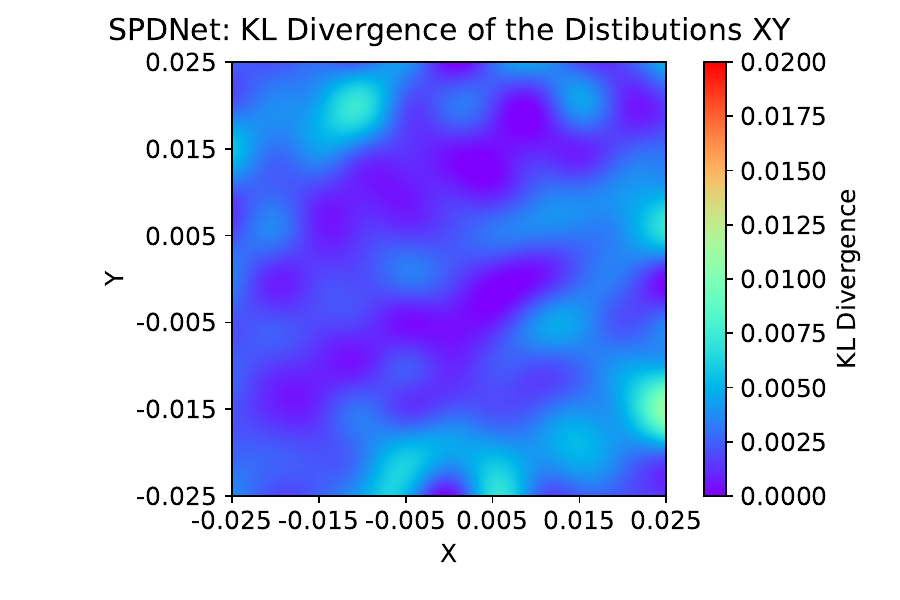}
        \caption{SPDNet Kullback-Leibler divergence.}
        \label{fig:SPD_KLD_sclori}
    \end{subfigure} 
    \hfill    
    \begin{subfigure}[t]{0.328\textwidth}        \includegraphics[width=\textwidth,trim=2.2cm 0.0cm 0.5cm 0.8cm,clip]{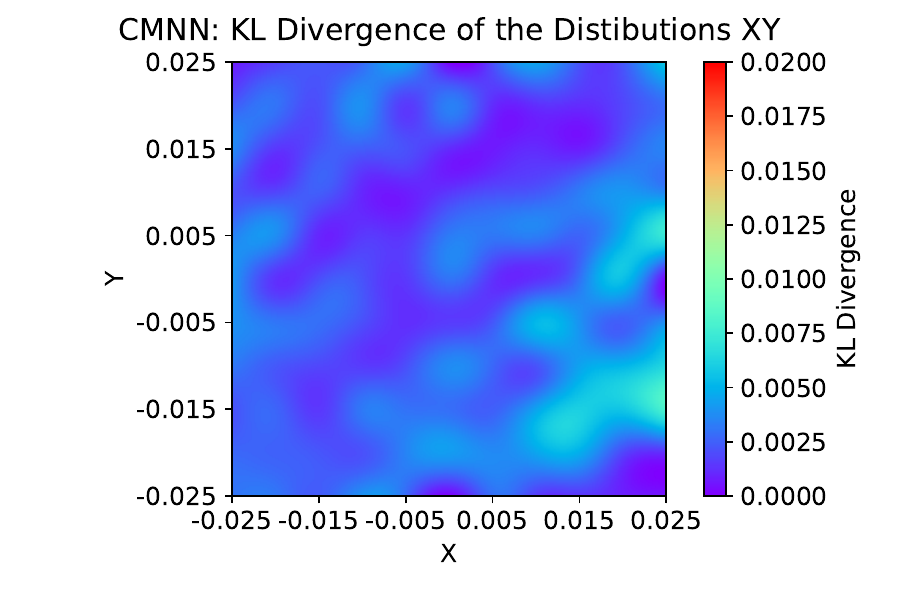}
        \caption{CMNN Kullback-Leibler divergence.}
        \label{fig:CMNN_KLD_sclori}
    \end{subfigure} 
    \vspace{0.1cm}
    \caption{Errors in approximation of the temperature field by the neural networks with the combined scaling-orientation uncertainty dataset.}
    \label{fig:errors_sclori_2}
\end{figure}

For the MLP, SPDNet and CMNN architectures the optimised hyper-parameters are 300 epochs, a batch size of 16, three hidden layers of sizes 48, 16 and 8, a sigmoid activation function, a learning rate of $2.5\times e^{-3}$ and the Adam optimiser. Table~\ref{tab:sclori_results} shows their training, validation and test losses on the combined scaling-and-orientation data set. Note that these losses lie in the order of $e^{-3}$ rather than $e^{-4}$.

The MLP exhibits the highest variability and poorest performance, with a best validation loss of $4.68\times e^{-3}$, a test loss of $7.02\times e^{-3}$ and a training loss of $8.17\times e^{-3}$. Its mean validation and test losses, $6.46\times e^{-3}$ and $8.69\times e^{-3}$ respectively, and their standard deviations underscore its inconsistency. Between CMNN and SPDNet the differences are smaller, with SPDNet slightly outperforming CMNN across all metrics. As in the scaling-uncertainty case (Table~\ref{tab:scl_norm_results}), SPDNet edges out in estimating the standard deviation.

The training history for the neural networks under combined scaling and orientation uncertainty is shown in \refFIG{\ref{fig:his_sclori}}. The MLP, \refFIG{\ref{fig:his_sclori_MLP}}, exhibits noticeable instability during training, with significant fluctuations in both training and validation losses. In contrast, SPDNet shown in \refFIG{\ref{fig:his_sclori_SPD}} and CMNN  shown in \refFIG{\ref{fig:his_sclori_CMNN}} display considerably more stable convergence, with smoother loss curves throughout training.

\refFIG{\ref{fig:errors_sclori_1}–\ref{fig:errors_sclori_2}} presents the errors in approximating the temperature field under combined scaling and orientation uncertainty for the MLP, SPDNet, and CMNN architectures. In a random sample, the relative errors shown in \refFIG{\ref{fig:MLP_rel_sclori} - \ref{fig:CMNN_rel_sclori}} for the MLPare noticeably higher compared to SPDNet and CMNN, particularly in regions with high temperature gradients. This trend persists in both the absolute and relative errors of the mean seen in \refFIG{\ref{fig:MLP_mean_abs_sclori}–\ref{fig:CMNN_mean_rel_sclori}} and the absolute and relative errors of the standard deviation in \refFIG{\ref{fig:MLP_std_abs_sclori}–\ref{fig:CMNN_std_rel_sclori}}, where the MLP exhibits greater inaccuracies. The Kullback–Leibler (KL) divergence shown in \refFIG{\ref{fig:MLP_KLD_sclori} - \ref{fig:CMNN_KLD_sclori}} further highlights these discrepancies, with SPDNet showing localised areas of higher divergence, likely due to inaccuracies in approximating the mean in those regions. In contrast, CMNN and MLP exhibit slightly higher divergence in specific regions; however, CMNN maintains better overall consistency compared to both the MLP and SPDNet.

In summary, both geometry-aware models surpass the standard MLP on the combined-uncertainty dataset. SPDNet achieves the lowest validation errors, and CMNN matches it closely on the test set while retaining smooth training behaviour.

\subsubsection{Case study discussion}
\noindent
Across the three case studies, one recurring issue is the reliability of the validation loss as an indicator of generalisation performance. When the validation set happens to align closely with patterns the network has learned, the loss can be over-optimistic and lower than the true error on the full data set. Selecting the model with the smallest validation loss therefore amounts to cherry-picking. Part of the mismatch also stems from using the mean-squared-error loss, which does not always capture meaningful differences for outputs that reside in stochastic spaces. For comparing predicted distributions, divergence measures such as the Kullback–Leibler divergence or the Wasserstein distance better reflect discrepancies in statistical structure.

Using one set of shared hyper-parameters for all architectures may have limited individual performance. The Bayesian optimisation was run with equal weighting across the MLP, SPDNet and CMNN, and it tended to favour the weakest model rather than unlocking the full potential of each. Separate hyper-parameter searches would probably improve results for individual networks. Additional techniques such as drop-out or batch normalisation could further stabilise training, but they were excluded here to keep the comparison focused on the three input maps: flattening for the MLP, LogEig for SPDNet and StrAng for CMNN.

The training curves in \refFIG{\ref{fig:his_scl}}, \refFIG{\ref{fig:his_ori}} and \refFIG{\ref{fig:his_sclori}} where a good indicator of generalisation performance. CMNN and SPDNet show smooth, stable convergence under every uncertainty type, whereas the MLP fluctuates markedly. Together with the norm tables and field-wise error plots, these histories provide a clear picture of global and local behaviour. From these sources it can be concluded that CMNN performed best with orientation uncertainty and SPDNet performed best with scaling uncertainty, both clearly benefitting from manifold-aware preprocessing. The standard MLP lagged behind, especially in approximating probability distributions.


\section{Conclusions}
\noindent
This work investigates three neural network architectures for physics-based problems involving uncertain symmetric positive-definite tensors. The proposed methods differ in how they preserve the relationship between samples in the input space while mapping from the tensor manifold to the Euclidean space in which neural networks operate. These architectures are employed to approximate the function defined in \refEQ{\ref{eq:stochastic_heat_pde}}, where the stochastic tensor model is specified by \refEQ{\ref{eq:random_both}}. In the case study, MLP, SPDNet, and CMNN are evaluated for their ability to approximate this function under scaling, orientation, and combined uncertainties. Both the newly formulated CMNN and the SPDNet, which is introduced here in the context of physics-based problems, outperform the MLP in terms of training stability and accuracy, as demonstrated in the previous section. These results highlight the potential of CMNN and SPDNet in effectively handling SPD material properties for function approximation tasks, making them promising candidates for uncertainty-aware machine learning in physics-based applications.

The heat transfer case study focuses on a relatively simple physical problem with predominantly linear behaviour. In more complex scenarios involving nonlinear physics, such as multi-physics interactions or heterogeneous materials, the advantages of architectures like SPDNet and CMNN could become even more pronounced. These methods may offer greater robustness in capturing intricate dependencies that standard MLPs struggle to describe effectively. Additionally, the observed differences between SPDNet and CMNN are relatively minor in this study, with the CMNN performing slightly better on uncertainty propagation for orientation uncertainty, possibly due to the matrix logarithm used for the eigenvectors. 

While neural networks offer scalability and efficiency, they also face inherent challenges such as limited interpretability, sensitivity to hyperparameter choices, and occasional instability, as seen for the MLP network. These limitations highlight the continued importance of numerical solvers for solving physics problems, as the approximation of the function can only be as good as the underlying data from which it originates.

This work demonstrates that embedding geometric representations of SPD tensors into neural networks leads to more accurate and robust uncertainty quantification. By extending beyond traditional Euclidean mappings, this framework contributes to more robust surrogate modelling of stochastic materials and offers new opportunities for computational material science.


\section{Acknowledgements}
This research was carried out as part of the project ENLIGHTEN \cite{noauthor_enlighten_nodate} (project number N21010 - m) in the framework of the Partnership Program of the Materials Innovation institute M2i (www.m2i.nl) and the
Netherlands Organization for Scientific Research (www.nwo.nl).

\bibliographystyle{ieeetr}
\bibliography{references}

\end{document}